\documentclass[12pt,twocolumn,tightenlines,superscriptaddress,nofootinbib]{revtex4}
\usepackage{graphicx}
\usepackage{xcolor}
\usepackage{caption}
\usepackage{subfig}
\newcommand{\bea}{\begin{eqnarray}}
\newcommand{\eea}{\end{eqnarray}}
\usepackage{multirow}
\usepackage{amsmath,amssymb}  
\usepackage{bm}
\usepackage{float}
\usepackage{graphics}
\usepackage{slashed}
\usepackage{comment}
\usepackage[normalem]{ulem}
\usepackage{verbatim}

\begin{document}

\title{Quark mass dependence of the $T_{cc}(3875)^+$ pole}

\author{F. Gil-Dom\'inguez}
\email{Fernando.Gil@ific.uv.es}
\affiliation{Departamento de F\'{\i}sica Te\'orica and IFIC,
Centro Mixto Universidad de Valencia-CSIC, Parc Científic UV, C/ Catedrático José Beltrán, 2, 46980 Paterna, Spain}

\author{A. Giachino} 
\email{Alessandro.Giachino@ific.uv.es}
\affiliation{Departamento de F\'{\i}sica Te\'orica and IFIC,
Centro Mixto Universidad de Valencia-CSIC, Parc Científic UV, C/ Catedrático José Beltrán, 2, 46980 Paterna, Spain}  

\author{R. Molina}
\email{Raquel.Molina@ific.uv.es}
\affiliation{Departamento de F\'{\i}sica Te\'orica and IFIC,
Centro Mixto Universidad de Valencia-CSIC, Parc Científic UV, C/ Catedrático José Beltrán, 2, 46980 Paterna, Spain}

\begin{abstract}

     Recently, several LQCD simulations have proven that the interaction in the isoscalar channel in $DD^*$ scattering is attractive. This channel is naturally connected to the $T_{cc}(3875)^+$ which is observed in the $D^0D^0\pi^+$ invariant mass distribution. However, it remains an open question whether the virtual bound state found in these several LQCD simulations is actually linked to the LHCb experimental observation. In this article we perform an EFT-based analysis of the LQCD data and demonstrate that a proper chiral extrapolation leads to a $T_{cc}$ pole compatible with experiment. At the physical pion mass, we find a virtual bound state with a binding energy $\Delta E=-0.06 \left(^{+1.30}_{-2.20}\right) \left(^{+0.50}_{-1.11}\right)$. Moreover, we extract from a global analysis both the light and heavy quark mass dependence of the $T_{cc}$ pole, and study the role of the $\rho$- and $\pi$-meson exchanges.

\end{abstract}
   
\maketitle
\section{Introduction}

As one of the first flavor exotic states observed in the charm sector, the $T_{cc}(3875)^+$ has attracted much attention since its discovery in 2021 by the LHCb~\cite{LHCb:2021vvq,LHCb:2021auc}. It was predicted by several theoretical approaches as a $DD^*$ bound state~\cite{Janc:2004qn,Yang:2009zzp,Carames:2011zz,Ohkoda:2012hv,Li:2012ss,Liu:2019stu,Liu:2020nil} and also as a compact tetraquark~\cite{Ader:1981db,Zouzou:1986qh,Heller:1986bt,Silvestre-Brac:1993zem,Navarra:2007yw,Ebert:2007rn,Karliner:2017qjm,Yang:2019itm,Tang:2019nwv}. See also Refs.~\cite{Dong:2021bvy,Chen:2022asf} for recent reviews. One of its intriguing properties is its tiny binding energy, $\delta E=-360\pm 40^{+4}_{-0}$~keV. Its mass is located extremely close to the $D^{*+}D^0$ threshold, and its state has a width as small as $\Gamma=48\pm2^{+0}_{-14}$~keV. The minimum components of $T_{cc}(3875)^+$ are four quarks, $cc\bar{u}\bar{d}$, and its properties are qualitatively similar to those of one of the most controversial exotic states, $X(3872)$~\cite{Belle:2003nnu}. Whether or not these two states are of similar nature is a subject of current debate~\cite{Wang:2023ovj}. 

Remarkably, the $DD^*$ scattering has recently been studied by LQCD, and it was concluded that the interaction is attractive in the isoscalar channel~\cite{Padmanath:2022cvl,Chen:2022vpo,Lyu:2023xro,Collins:2024sfi,Whyte:2024ihh}. Most of these simulations obtain a virtual pole at unphysical pion masses which is associated with the $T_{cc}$, being closer to the $DD^*$ threshold as the pion mass approaches its physical value. In particular, the HALQCD Collaboration performed a simulation for a pion mass of $146$~MeV and encountered a virtual state with a binding energy of $\delta E=-59^{+53}_{-99}(^{+2}_{-67})$~keV.

Since $T_{cc}$ has a three-body decay, three-body interactions could, in principle, play a role. Experimentally, $T_{cc}$ is seen in the $D^0D^0\pi^+$ invariant mass distribution, with most of the events (~$90$~\%) coming from the $D^{*+}$ decay~\cite{LHCb:2021auc}. In fact, theoretical analysis of the experimental data that do not include three-body effects and are merely based on a short-range interaction lead to a good description of the line shape and a pole position close to the experimental one~\cite{Feijoo:2021ppq,Albaladejo:2021vln}. See also Ref.~\cite{Ling:2021bir}. Similar results for these observables are obtained in a more complete theoretical analysis; the results do not neglect these effects, considering explicitly the dynamics from the pion exchange and momentum-dependent $D^*$ decay width~\cite{Du:2021zzh}, needed to guarantee a self-consistent three-body unitary formalism~\cite{Aaron:1968aoz,Mai:2017vot}. This is an indication that three-body dynamics might have a minor role in the physical pion mass. Indeed, NLO calculations of the strong decay of $T_{cc}$ including pion exchange and rescattering effects also find good agreement with the experimental line shape and show that the contact interactions are dominant in a similar way as for $X(3872)$~\cite{Dai:2023mxm,Fleming:2007rp,Dai:2019hrf}. However, for pion masses larger than the physical one, such that $m_\pi>m_{D^*}-m_D$, as in the present LQCD simulations, there is currently a debate, since the left-hand cut (lhc) caused by the consideration of the pion exchange becomes closer to the pole. This is the case for the recent LQCD simulation~\cite{Padmanath:2022cvl,Collins:2024sfi} that uses $m_\pi\simeq 280$~MeV, for which the L\"uscher method is not applicable~\cite{Luscher:1986pf,Luscher:1990ux}. A comprehensive explanation of this problem is given in Refs.~\cite{Du:2023hlu,Meng:2023bmz}, where an alternative method based on EFT with an interaction that consists of a contact term plus the pion exchange term is used to extract the pole position of the $T_{cc}$ from the energy levels. See also other works on possible extensions of the L\"uscher method including the lhc~\cite{Raposo:2023oru,Bubna:2024izx,Hansen:2024ffk,Du:2024snq}. 

Still, the nature of the attraction in $T_{cc}$ is a matter of debate. In Ref.~\cite{Abolnikov:2024key}, it is argued that the role of the pion might become relevant for pion masses larger than $230$~MeV, where its effect would lead to a virtual resonance instead of a virtual bound state as found in the LQCD simulations~\cite{Padmanath:2022cvl,Lyu:2023xro,Whyte:2024ihh} when the lhc is properly taken into account. For pion masses near the physical point, the HALQCD simulation with $m_\pi = 146$ MeV extracts a local potential where the interaction is dominated by an attractive short range plus a two-pion exchange term~\cite{Lyu:2023xro}. In Ref.~\cite{Chen:2022vpo}, the authors analyze the different contributions of the diagrams in the isoscalar $DD^*$ scattering on the lattice and infer that the $\rho$-meson exchange is dominant and responsible for the attractive interaction. In a recent LQCD simulation of $DD^*-D^*D^*$ scattering, performed at $m_\pi=391$~MeV, the authors also obtain a virtual bound state~\cite{Whyte:2024ihh}, as in Refs.~\cite{Padmanath:2022cvl,Lyu:2023xro}; however, they use a parametrization for the $K$-matrix that neglects the lhc contribution in the extraction of the phase shifts. Nevertheless, in this work, an excellent description of the energy levels is obtained within a thorough partial-wave analysis including both the $DD^*$ and $D^*D^*$ channels, indicating that the effect of the lhc might not be that relevant for this pion mass. In this simulation, a virtual state associated with $T_{cc}$ and a resonance below the $D^*D^*$ threshold are found. This finding might be related to the state predicted in Ref.~\cite{Molina:2010tx} as an isocalar $D^*D^*$ molecule and updated in the recent work~\cite{Dai:2021vgf}. 

In a recent analysis, the LQCD data from Ref.~\cite{Padmanath:2022cvl} and the experimental data~\cite{LHCb:2021vvq,LHCb:2021auc} are combined to provide a prediction of the light quark mass dependence of $T_{cc}$~\cite{Abolnikov:2024key}. However, the data from other LQCD simulations are not considered. The heavy quark mass dependence possibly inferred from Ref.~\cite{Collins:2024sfi} has not been extracted either. 

While there are hints of an attractive interaction from the $DD^*$ scattering in the isoscalar channel from the different LQCD simulations~\cite{Padmanath:2022cvl,Chen:2022vpo,Lyu:2023xro,Collins:2024sfi,Whyte:2024ihh}, the compatibility between these simulations and with the experimental data~\cite{LHCb:2021vvq,LHCb:2021auc} has not been studied yet; nor has the role of the possible $\rho$-meson exchange. The quark mass dependence, light and heavy for $T_{cc}$, has not yet been extracted from an analysis of the existing LQCD simulations. 

This article is structured as follows: In Section \ref{sec:formalism}, we present the theoretical formalism, including the infinite volume formalism (Sec. \ref{subs:hga}), the finite volume formalism used to fit energy levels (Sec. \ref{sec:finvol}), and a discussion of the LQCD data analyzed (Sec. \ref{subs:ld}). Section \ref{sec:results} discusses the results, covering the fit to energy levels, the extraction of the light and heavy quark mass dependence of the $T_{cc}$ pole position (Sec. \ref{Energy_levels_fit}) and the impact of off-shell effects and pion exchange on the scattering phase shifts and binding energies (Sec. \ref{infinite_volume}). Section \ref{sec:conclusions} concludes with a summary of the findings and their implications for the nature of $T_{cc}(3875)^+$. The appendixes provide additional technical details: the $D^*D\pi$ coupling vertex determination (Appendix \ref{pion-form-factor}), the treatment of polarization vectors (Appendix \ref{app:polvec}), some plots of the on-shell potentials for various LQCD data sets (Appendix \ref{onshelV}), and an analysis of possible higher-order effects (Appendix \ref{mpi4}).

\section{Formalism }~\label{sec:formalism}

In this section, we present the infinite volume formalism in Sec.~\ref{subs:hga}, the finite volume formalism used to fit the energy levels in Sec.~\ref{sec:finvol}, and a discussion about the lattice data analyzed in Sec.~\ref{subs:ld}.

\subsection{$T_{cc}$ in the infinite volume}~\label{subs:hga}

The interaction between a vector meson ($V^\mu$) and a pseudoscalar meson ($P$) and also between vector mesons, can be evaluated from the local hidden gauge Lagrangians  \cite{Bando:1987br,Harada:2003jx,Meissner:1987ge,Nagahiro:2008cv}:
\begin{equation}
\begin{array}{l}
\mathcal{L}_{V P P}=-i g\left\langle\left[P, \partial_\mu P\right] V^\mu\right\rangle \\
\mathcal{L}_{V V V}=i g\left\langle\left(V^\nu \partial_\mu V_\nu-\partial_\mu V^\nu V_\nu\right) V^\mu\right\rangle \, .
\end{array}
\label{eq:lagrangians}
\end{equation}
Here, the symbol $\langle\rangle$ denotes the trace in the SU(3) flavor space; $P$ and $V$ stand for the  pseudoscalar and vector-meson nonet matrices, respectively; and the coupling $g=m_\rho/2f_\pi$. This formalism was extended to SU(4) in Refs.~\cite{Molina:2009eb,Molina:2009ct}, where one can write the matrices
\begin{widetext}
\begin{equation}
P=\left(\begin{array}{cccc}
\frac{\pi^0}{\sqrt{2}}+\frac{\eta}{\sqrt{3}}+\frac{\eta^{\prime}}{\sqrt{6}} & \pi^{+} & K^{+} & \bar{D}^0 \\
\pi^{-} & -\frac{\pi^0}{\sqrt{2}}+\frac{\eta}{\sqrt{3}}+\frac{\eta^{\prime}}{\sqrt{6}} & K^0 & D^{-} \\
K^{-} & \bar{K}^0 & -\frac{\eta}{\sqrt{3}}+\sqrt{\frac{2}{3}} \eta^{\prime} & D_s^{-} \\
D^0 & D^{+} & D_s^{+} & \eta_c
\end{array}\right) 
\label{eq:p}
\end{equation}
and
\begin{equation}
    V_\mu=\left(\begin{array}{cccc}\frac{\rho^0}{\sqrt{2}}+\frac{\omega}{\sqrt{2}} & \rho^{+} & K^{*+} & \bar{D}^{* 0} \\ \rho^{-} & -\frac{\rho^0}{\sqrt{2}}+\frac{\omega}{\sqrt{2}} & K^{* 0} & D^{*-} \\ K^{*-} & \bar{K}^{* 0} & \phi & D_s^{*-} \\ D^{* 0} & D^{*+} & D_s^{*+} & J / \psi\end{array}\right)_\mu\ .
    \label{eq:v}
\end{equation}
\end{widetext}

\begin{figure*}
     \centering
     \begin{minipage}{0.4\textwidth}
     \centering
     \includegraphics[width=1.\linewidth]{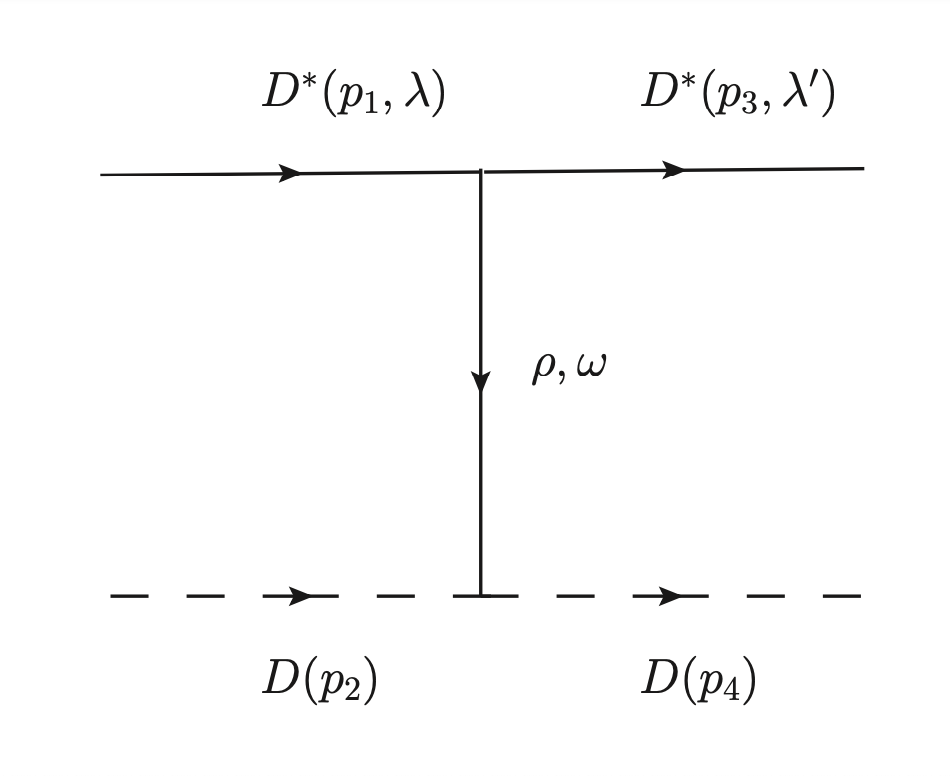}
   \end{minipage}
   \begin{minipage}{0.4\textwidth}
     \centering
     \includegraphics[width=1.\linewidth]{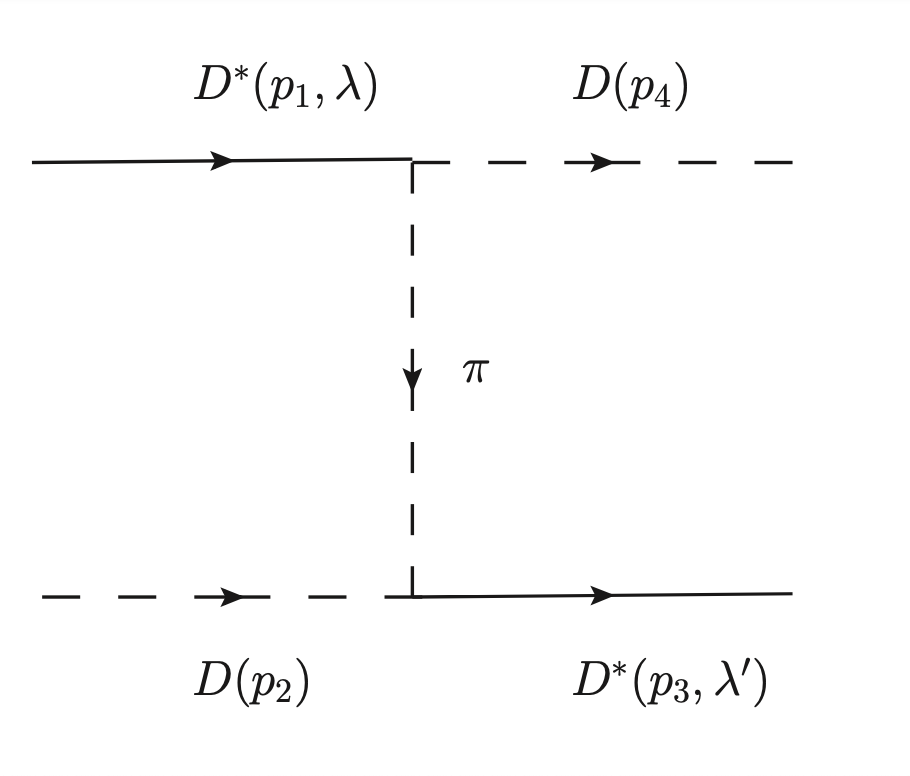}
   \end{minipage}
   \caption{Feynman diagrams for
the rho exchange (left-hand side) and  the pion exchange (right-hand side)  
interactions in the center-of-mass frame of the  $D D^{*}$ scattering process.}
   \label{fig:feyndiagrams}
\end{figure*} 

In this work, we consider vector- and pseudoscalar-meson exchange. The main Feynman diagrams are depicted in Fig.~1\label{fig:feyndiagrams}. Note that the use of SU(4) in Eqs.~(\ref{eq:p}) and~(\ref{eq:v}) is merely formal since, in the vertices of these diagrams, the light-exchanged particle dominates the interaction in such a way that the $c$ quark is acting as a spectator, the symmetry is broken to SU(3), and indeed, the form of the interaction term given by Eq.~(\ref{eq:lagrangians}) satisfies heavy quark symmetry~\cite{Xiao:2013yca}. Notice also that we are extending the approach of Ref.~\cite{Feijoo:2021ppq}, which only considers vector-meson exchange.
We consider the isospin limit case for $I=0$, where the interaction is attractive~\cite{Molina:2010tx}.
After projecting the interaction in isospin ($I$), for $I=0$ and considering only the light-vector-meson exchange, $i.e.$, the $\rho$, $\omega$, and  $\pi$ mesons, as displayed in Fig. \ref{fig:feyndiagrams} the $I=0$ tree-level scattering amplitudes for $D^*(p_1)D(p_2)\to D^*(p_3)D(p_4)$ derived from the effective Lagrangians of Eq. (\ref{eq:lagrangians}) are given by

\begin{widetext}
\begin{eqnarray}\label{treeampli_rho}
V^{\rho (I=0)}_{\lambda,\lambda'} (p,p')&=&-g^2 \frac{(p_1+p_3)_\mu(p_2+p_4)^\mu}{t-m_{\rho}^2}
\, \epsilon_{\lambda, \nu}(p_{1}) \epsilon^{* ~\nu}_{\lambda'}(p_{3})
\\
V^{\pi (I=0)}_{\lambda,\lambda'} (p,p') &=&\frac{3}{4}  \, g^2_{D^*D\pi}\,\frac{e^{u/\Lambda^2}  }{u-m_{\pi}^2}
\, (2p_4 -  p_1 )_{\mu} \epsilon^{~\mu}_\lambda(p_1)\;
(2p_2 -  p_3 )_{\nu} \epsilon^{*~\nu}_{\lambda'}(p_3)\,
\label{eq:treeampli_pi}
\end{eqnarray}
\end{widetext}

where we have approximated $m_\omega\simeq m_{\rho}$, leading to an exact cancellation between the $\rho^0$ and $\omega$ exchange~\cite{Feijoo:2021ppq}. In the above equation, $p_{1}$ and $p_{3}$ are the initial and final $D^*$-meson four-momenta, and $p_{2}$ and $p_{4}$ are the incoming and outgoing $D$-meson four-momenta, so $p^0\equiv p_1^0$ and $p'\,^0\equiv p_3^0$, $\vec{p}\equiv \vec{p}_1=-\vec{p}_2$, $\vec{p}\,'\equiv\vec{p}_3=-\vec{p}_4$, in the center-of-mass frame; $\epsilon_{h}(p)$ is the polarization vector of the vector meson with momenta $p$ and helicity $h$ (see Appendix \ref{app:polvec}); $s$ is the squared total energy of the $D^*D$-meson system in the center-of-mass reference frame; and $t$, $u$ are the Mandelstam variables defined as $t=(p_{1}-p_{3})^2$ and $u=(p_{1}-p_{4})^2$.
The pion form factor $e^{u/\Lambda^2}$ takes into account that the pion can be off shell. We take $\Lambda=1300$~MeV~\cite{Molina:2020hde,Molina:2022jcd}. To obtain the experimental decay width of $\Gamma(D^{*+}\to D\pi)$, we must fix $g_{D^*D\pi}=g_{D^*D\pi}^\mathrm{exp}=8.33\pm 0.1$, as explained in Appendix~\ref{pion-form-factor}. 
 
It is convenient to expand the tree-level scattering amplitudes of Eqs. (\ref{treeampli_rho}) and (\ref{eq:treeampli_pi}) in partial waves \cite{Sadasivan:2021emk}:

\begin{align}
\label{eq:VmatrixPartial}
A^{J}_{\lambda\lambda'}(p,p')=
\frac{1}{2} \int\limits_{0}^{\pi} \sin(\theta) d \theta  \, d^{\,J}_{\lambda,\lambda'}(\theta) V_{\lambda\lambda'}(\vec{p},\vec{p}\,') \,,
\end{align}
where $\theta$ is the scattering angle between $\vec{p}$ and $\vec{p}\,'$ in the center-of-mass frame, and $d^{\,J}_{\lambda,\lambda'}(\theta)$
are the Wigner $d$ functions.
The spin projection of the amplitude of Eq. (\ref{eq:VmatrixPartial}) is given by \cite{Chung:1971ri}
\begin{align}
\label{eq:transformation}
V^{J}_{LL'}(p,p')=U_{L\lambda}A^{J}_{\lambda\lambda'}(p,p') U_{\lambda' L'}\,,
\end{align}
where the matrix
\begin{align}
U_{L\lambda }&=\sqrt{\frac{2L+1}{2J+1}}\langle L01\lambda|J\lambda \rangle
\langle 1\lambda00|1\lambda \rangle
\end{align}
accounts for the spin and the initial and final angular momenta, $L(L')=0,2$, as required by momentum conservation. This implies that, in the $T_{cc}$ system the possible angular-momentum couplings are $S$-$S$ ($L = L'=0$) and $D$-$D$ ($L = L'=2$), and $S$-$D$ ($L=0, L'=2$). 

For $J=1$, the matrix $U$ can be explicitly expressed as
\begin{equation}
    U_{L\lambda}=\left( \begin{array}{ccc}
   \frac{1}{\sqrt{3}}    & \frac{1}{\sqrt{3}}& \frac{1}{\sqrt{3}}  \\
   \frac{1}{\sqrt{6}}      & -\sqrt{\frac{2}{3}} & \frac{1}{\sqrt{6}}
    \end{array}
    \right) \,.
\end{equation}
On the one hand, the $\pi$-meson exchange leads to the mixing of $S-D$ waves. Nevertheless, we have found that the effect of $S$-$D$ mixing is really small and can be safely neglected. On the other hand, Eq.~(\ref{eq:transformation}) leads to the appearance of a left-hand cut starting at a branch point $p^2_{lhc}=\frac{1}{4}(p_\pi^{0\,2}-m_\pi^2)$, with $p_\pi^0=p_1-p_4\simeq m_{D^*}-m_D$~\cite{Collins:2024sfi}. This lhc makes us question the validity of the Lüscher formula to extract the infinite volume amplitudes from the finite volume energy levels~\cite{Meng:2023bmz}. In the present work, we study the effect of the one-pion exchange in the infinite volume case.

The partial-wave-projected tree-level interactions obtained from the $\rho$- and $\pi$-meson exchange provide the kernel $V^{1}_{0,0}(p,p')=V^{1, \rho}_{0,0}(p,p')+ V^{1, \pi}_{0,0}(p,p')$ of the Bethe-Salpeter equation. In the following, we omit the $L,L'$ and ``$1$'' indices in the interaction, $V\equiv V^{1}_{0,0}$. The Bethe-Salpeter equation in its off-shell form reads

\begin{widetext}
\begin{eqnarray}
T(p,p')&=&V(p,p')+ \frac{4\pi}{(2\pi)^3} \int_0^{q_{\text{max}}} q^2 dq \,  V(p,q) I(q) T(q,p')
\label{LS}
\end{eqnarray}
In Eq.~(\ref{LS}), $I(q)$
is defined as
\begin{eqnarray}\label{Iq}
 I(q)=\frac{\omega_1(q)+\omega_2(q)}{2\omega_1(q)\omega_2(q)\left[s-(\omega_1(q)+\omega_2(q))^2 +i m_{D^*} \Gamma(M_\mathrm{inv}) +i\epsilon\right]}.
\end{eqnarray}
\end{widetext}
Note that this equation reduces to the on-shell Bethe-Salpeter equation when $V(p,q)$ is taken out of the integral with the final momenta being the on-shell momenta.\\
In Eq.~(\ref{Iq}), we have included the decay width $\Gamma(M_\mathrm{inv})$ to account for the $D^{*}\to D + \pi$ decay that happens in the physical limit. 
We observe that, since the $D^{*}$ meson is also in the loop, it is convenient to introduce an energy-dependent width $\Gamma(M_\mathrm{inv})$ to account for the fact that the $D^{*}$ meson is off shell in the loop \cite{Bayar:2023azy}:
\begin{eqnarray}
\Gamma(M_\mathrm{inv})=
\Gamma[D^*\to D\pi]
 \frac{m^2_{D^*}}{M^2_\mathrm{inv}} \left(\frac{p_\mathrm{on}(M_\mathrm{inv})}{p_\mathrm{on}(m_{D^*})}\right)^3
 \label{eq:width}
\end{eqnarray}
with $\Gamma[D^*\to D\pi]$ being the partial width for the decay process $D^*\to D\pi$, where the total widths for $D^{*+}$ and $D^{*0}$ are 83.4 keV and 56.2 keV~\cite{ParticleDataGroup:2022pth,Albaladejo:2021vln}. The branching ratios are obtained from the PDG ~\cite{ParticleDataGroup:2022pth}, $p_\mathrm{on}(M)=\frac{{\lambda^{1/2}(M^2,m^2_{D},m^2_{\pi})}}{2M}$.
Here, $\lambda(a,b,c)=a^2+b^2+c^2-2 ab- 2 ac - 2bc$ is the usual Källén function, and 
$M^2_\mathrm{inv}=(\sqrt{s}-E_{D})^2-q^2$, taking $E_D=\sqrt{m_D^2+p'\,^2}$, with $p'=|\vec{p}\,'|$ and $q=|\vec{q}\,|$ the momenta in the center-of-mass frame. 

Note that in Eq.~(\ref{LS}) we use the off shell Bethe-Salpeter equation including both $\pi$- and $\rho$-meson exchange. 
\\ \indent
We follow the normalization convention of Ref. \cite{Doring:2012eu}, in which the relation between the on-shell scattering amplitude $T(p,p)$  and  the phase shift $\delta$ is given by 
\begin{equation}
    \frac{1}{\cot{\delta} -i }=-\frac{p}{8 \pi \sqrt{s} \; T(p,p)}.
    \label{EqPS}
\end{equation}
\\

\subsection{Formalism in the finite volume}~\label{sec:finvol}

The formalism in the finite volume can be easily constructed, for instance, by assuming dominance of the $S$-wave interaction and then substituting the integral in Eq.~(\ref{LS}) by a discrete sum over the momenta $\vec{q}=\frac{2\pi}{L}\vec{n}$. However, if the vector-meson exchange is dominant, it might be possible to describe the energy levels with only vector-meson exchange while some small effects can be absorbed in the free parameters. On the other hand, we want to compare this with the phase shifts obtained by the Lüscher approach, where the effect of the lhc is neglected. Along this line, we proceed in the following way. We consider only the $\rho$-meson exchange for the energy-level analysis, and after extracting the phase shifts from the formalism presented below, we also include the pion in the infinite volume limit. In this way, we can compute the size of the effect of the pion exchange in the phase shifts.

Thus, here we consider only $V_\rho=V^{1,\rho}_{00}$ from Eqs.~(\ref{treeampli_rho}) and (\ref{eq:transformation}). Note that Eq.~(\ref{treeampli_rho}) depends on the coupling $g$, which, in principle, can vary with the pion mass. We introduce a quadratic mass dependence of the type,
\begin{equation}\label{g_rho}
    g=g_0+g_2m_\pi^2
\end{equation}
and collect $g_0,g_2$ as fitting parameters in the energy-level analysis. In principle, we could also consider some extra dependence with the lattice spacing; however, we observe that this extra term is not significant and practically does not affect the fitting results. For this reason, we adhere to Eq.~(\ref{g_rho}). Hence, considering the usual on-shell factorization of the interacting kernel, the scattering equation is reduced to~\cite{Gil-Dominguez:2023huq}

\begin{equation}
 \widetilde{T}^{-1}=V^{-1}_\rho-\widetilde{G}\ ,\label{eq:bethe}
\end{equation}
where $\widetilde{T}$ is the scattering matrix in the finite volume and $\widetilde{G}$ is the two-meson loop function in the box \cite{MartinezTorres:2011pr},
\begin{eqnarray}
 \widetilde{G}(P,\alpha )=G^{DR}(s,\alpha)+\mathrm{lim}_{q_\mathrm{max}'\to\infty}\Delta G\ .\nonumber\\\label{eq:gt}
\end{eqnarray}
In the above equation, $P\equiv P^\mu=(P^0,\vec{P}\ )$ is the full four-momentum of the two-meson system. The Mandelstam variable $s$ is related to the momentum as $s=P_0^2-\vec{P}^2=P_0^2$. 

The first term of Eq. (\ref{eq:gt}) is the two-meson loop function in the infinite volume, which can be evaluated using dimensional regularization,
\begin{widetext}
\begin{align}\label{GDM}
G^{DR}(s,\alpha)=&\frac{1}{16\pi^2}\left\{\alpha+\log\left(\frac{m_1^2}{\mu^2}\right)+\frac{p}{\sqrt{s}}\left[\log\left(\frac{s-m_1^2+m_2^2+2p\sqrt{s}}{-s+m_1^2-m_2^2+2p\sqrt{s}}\right) \right.  \right. \nonumber \\
& \left.\left. +\log\left(\frac{s-m_1^2+m_2^2+2p\sqrt{s}}{-s+m_1^2-m_2^2+2p\sqrt{s}}\right)\right] +\frac{s-m_1^2+m_2^2}{2s}\log\left(\frac{m^2_2}{m_1^2}\right) \right\},
\end{align}

where $m_1$ and $m_2$ are the masses of the two mesons, $\mu=1500$ MeV is the regularization scale, and $\alpha\equiv\alpha(q_\mathrm{max})$ is the subtraction constant, which can written as a function of the cutoff $q_\mathrm{max}$ as follows \cite{Oller:2019opk}:
\begin{align}
    \alpha(q_\mathrm{max})=&-\frac{2}{m_1+m_2}\left[ m_1\log\left( 1+\sqrt{1+\frac{m_1^2}{q_\mathrm{max}^2}} \right) + m_2\log\left( 1+\sqrt{1+\frac{m_2^2}{q_\mathrm{max}^2}}\right) \right] +2\log\left(\frac{\mu}{q_\mathrm{max}}\right).\label{eq:cut}
\end{align}
\end{widetext}

In particular, in the center-of-mass frame, where $\vec{P}=\vec{0}$, the second term in Eq. (\ref{eq:gt}), $\Delta G$, can be written as~\cite{MartinezTorres:2011pr}
\begin{eqnarray}
\Delta G=\frac{1}{L^3}\sum_{q_i}^{q'_\mathrm{max}}I(q_i)-\int_{0}^\mathrm{q'_\mathrm{max}} \frac{d^3 q}{(2\pi)^3}I(q)\ ,\nonumber\\\label{eq:gt1}
\end{eqnarray}
with $\vec{q}$ the momentum in the center-of-mass frame, which takes discrete values in the finite box, $\vec{q}_i=\frac{2\pi}{L}\vec{n}_i,\,\vec{n}_i\in\mathbb{Z}^3$, and $L$ is the spatial extent of the box. The integrand is given by
\begin{eqnarray}
 I(q )=\frac{\omega_1+\omega_2}{2\omega_1\omega_2\left[s-(\omega_1+\omega_2)^2+i\epsilon\right]},
\end{eqnarray}
with $\omega_i=\sqrt{q^{2}+m_i^2}$. For moving frames, we refer to the procedure in Ref. \cite{doringframe}. Notice that, in Eq.~(\ref{eq:gt1}), a large cutoff $q'_\mathrm{max}$ has to be used to evaluate the difference $\Delta G$~\cite{MartinezTorres:2011pr}. Finally, the energy levels are determined by
\begin{equation}
\mathrm{det}(\delta_{ll'}\delta_{mm'}-V_l\tilde{G}_{lm,l'm'})=0\ ,\label{eq:det}
\end{equation}
where $V_l$ and $\tilde{G}_{lm,l'm'}$ stand for the projection of $V$ and $\tilde{G}$ in partial waves, the latter being a nondiagonal matrix due to the partial-wave mixing~\cite{doringframe}. \\
Phase shifts are extracted in the energy-level analysis using the on-shell factorization of the potential in Eq.~(\ref{LS}). 

\subsection{Lattice data sets}\label{subs:ld}

We consider the LQCD simulations for $DD^*$ scattering from Refs.~\cite{Padmanath:2022cvl, Collins:2024sfi, Chen:2021vhg,Whyte:2024ihh,Lyu:2023xro}, where we analyze the energy levels provided in Refs.~\cite{Padmanath:2022cvl, Collins:2024sfi, Chen:2021vhg,Whyte:2024ihh}. Below, we summarize these simulations:
\begin{itemize}
    \item References~\cite{Padmanath:2022cvl, Collins:2024sfi}: These simulations are based on the CLS ensembles for $m_\pi\simeq 280$~MeV utilizing the nonperturbative Wilson-clover action with $u/d,s$ dynamical quarks. The simulation of $DD^*$ scattering was first completed for two different charm quark masses in Ref.~\cite{Padmanath:2022cvl} and later extended to five different charm quark masses in Ref.~\cite{Collins:2024sfi}, which vary from $m_D\simeq 1.7-2.4$~GeV. The lattice spacing corresponds to $a=0.08636(98)(40)$, and the spatial extents are $L=2.07, 2.76$~fm. The pole analyses are completed in two different ways. The first way is the effective range expansion (ERE), and the second one the effective field theory (EFT) with one pion. In the first case, the analysis is restricted to the levels above the threshold because of the possible effect of lhc, but the results are consistent with the fit, including energies below this threshold. In all cases, a virtual pole identified with $T_{cc}$ is found; its distance to threshold varies from $-8.3$ to $-1.2$~MeV when increasing the charm quark mass.\footnote{The real part of the pole obtained is compatible within errors in both approaches. The EFT should be more accurate. See Table 2 of Ref.~\cite{Collins:2024sfi}.}
    \item Reference~\cite{Chen:2022vpo}: The $I=0,1$ $DD^*$ scattering is simulated in $N_f=2$ by employing the Clover action in a box at a fixed volume $L=2.4$~fm. The temporal lattice spacing is set to $a^{-1}_t = 6.894(51)$~GeV, and the spatial lattice spacing is $a_s=0.152(1)$~fm. The simulation corresponds to a pion mass $m_\pi\simeq 348.5(1.0)$~MeV, and the charm quark mass is tuned to its physical value through the spin average charmonia mass. The pole is not evaluated, but the scattering lengths are obtained through the ERE. The interaction turns out to be attractive for $I=0$ and repulsive for $I=1$. The diagrams at the quark level which are responsible for the dominant attractive interaction between the $D$ and $D^*$ mesons are attributed to the $\rho$-meson exchange. 
    \item Reference~\cite{Whyte:2024ihh}: The $I=0$ $DD^*-D^*D^*$ scattering is studied for a pion mass $m_\pi\simeq 391$~MeV. The Wilson-clover action with $N_f=2 + 1$ flavors of dynamical quarks in the fermion sector is used. The strange and charm quarks are tuned to their physical values, in the latter case through the $\eta_c$  mass which turns out to be slightly lower than the experimental value, $m_{\eta_c} \simeq 2965$ MeV. Three boxes of different volumes within the range $1.9-2.9$~fm are used. The temporal lattice spacing is used to improve the energy resolution, with $a_t^{-1}=5.667$~GeV. A thorough partial-wave mixing is presented. The positions of the poles in the complex plane are extracted through the $K$-matrix parametrization with an interaction expanded in powers of $s$. A virtual bound state related to $DD^*$ and a resonance below the $D^*D^*$ threshold are obtained. The effects of the lhc are not accounted for when extracting the phase shifts from the L\"uscher approach but the amplitudes obtained are consistent with the finite volume spectrum. The effect of the $^3S_1-^3D_1$ mixing is found to be negligible, but this is not the case for the $DD^*-D^*D^*$ coupling in the amplitude analysis, which could be more relevant for the energy levels close to or above the $D^*D^*$ threshold. 
    \item Reference~\cite{Lyu:2023xro}: An $N_f=2+1$ simulation with a pion mass near its physical value, $m_\pi= 146$~MeV, is studied. The charm quark mass is set to its physical value using the spin average mass of charmonia. The size of the box is much large than in the previous cases, $L\sim 8$~fm, and the lattice spacing is $a=0.0846$~fm. While Refs.~\cite{Padmanath:2022cvl,Collins:2024sfi,Chen:2022vpo,Whyte:2024ihh} use the L\"uscher approach to connect energy levels at different volumes with phase shifts, the HALQCD method is based on calculating the $I=0;J^P=1^+$ $S$-wave $DD^*$ nonrelativistic effective local potential, which is derived from the hadronic spacetime correlation function. The pole position is obtained by solving the Schrödinger equation and the scattering length is evaluated through the ERE. An attractive interaction is found, also leading to a virtual bound state but this time much closer to the $DD^*$ threshold and evolving to a bound state when a chiral extrapolation is completed. 
\item Reference~\cite{Ikeda:2013vwa}: This work studies a $2+1$ flavor full QCD gauge configuration generated by the PACS-CS Collaboration on a $32^3 \times 64$ lattice with the renormalization group improved  Wilson gauge action and using a lattice spacing $a=$ $0.0907(13) \mathrm{fm}$, leading to the spatial lattice volume $L\simeq 2.9 \, \mathrm{fm}$. In this case, the pole is not evaluated, but the scattering lengths are obtained through the ERE.
\end{itemize}

All the above simulations consider $DD^*$ operators to extract either the energy levels or potential, except for Ref.~\cite{Whyte:2024ihh}, which also considers $D^*D^*$ operators. Exploratory simulations with diquark-antidiquark interpolators have become available recently~\cite{Vujmilovic:2024snz}. Here, we do not consider this other type of operator. In addition, we stay below the $D^*D^*$ threshold in our analysis.

The relevant data of these lattice simulations are summarized in Tables~\ref{tab:latdat} and \ref{tab:latdatPad}. In the latter, we show the charmed meson spin average mass of the sets $1-5$ used in Ref.~\cite{Collins:2024sfi}.

We use the interaction described by Eq. (\ref{treeampli_rho}), that is, considering just vector-meson exchange interactions. This analysis involves solving Eq. (\ref{eq:bethe}), which provides the lattice energy spectrum adjusted by a fit to the lattice data. 
\begin{table*}
\centering
 \setlength{\tabcolsep}{0.5em}
{\renewcommand{\arraystretch}{1.6}
\begin{tabular}{|c|c|c|c|c|c|}
\hline
 Collaboration & $a$ & $L$ & $m_\pi$ & $M^{c\bar{c}}_{avg}$ \\
 \hline
Padmanath24 \cite{Padmanath:2022cvl}\cite{Collins:2024sfi}  & $0.086$ & $2.07 - 2.76$ & $280$ & $3103$\\
 \hline
CLQCD22
 \cite{Chen:2022vpo} & $0.152$ & $2.4$ & $349$ & $3069$\\
 \hline
 HSC24 \cite{Whyte:2024ihh} & $0.120$ & $1.9-2.9$ & $391$ & $3024$ \\
 \hline
HALQCD23 \cite{Lyu:2023xro} & $0.0846$ & $8$ & $146$ & $3097$ \\
 \hline
 HALQCD14 \cite{Ikeda:2013vwa} & $0.0907$ & $2.9$ & $411$ & $3070$ \\
 \hline
 \end{tabular}}
\caption{Comparison of the charm quark mass settings of the different collaborations with $DD^*$ scattering data. Here, $L$ and $a$ are given in fm, and the masses are in units of MeV. In the last column, $M^{c\bar{c}}_{avg}$ stands for $M^{c\bar{c}}_{avg}= 1/4(m_{\eta_c} + 3m_{J\Psi}) $. }
\label{tab:latdat}
\end{table*}

\begin{table*}
\centering
 \setlength{\tabcolsep}{0.5em}
{\renewcommand{\arraystretch}{1.6}
\begin{tabular}{|c|c|c|c|}
\hline
 Collaboration & $m_\pi$ & Set & $\bar{m}_D$ \\
 \hline
\multirow{5}{*}{Padmanath24 \cite{Padmanath:2022cvl}\cite{Collins:2024sfi}} & \multirow{5}{*}{$280$} & 1 & 1864 \\
\cline{3-4}
 & & 2 & 2019 \\
\cline{3-4}
  & & 3 & 2148\\
\cline{3-4}
  & & 4 & 2269 \\
\cline{3-4}
  & & 5 & 2484\\
 \hline
 \end{tabular}}
\caption{Charmed meson masses for Padmanath24 \cite{Padmanath:2022cvl}\cite{Collins:2024sfi} data. Masses are given in MeV.}
\label{tab:latdatPad}
\end{table*}

\section{Results and discussion}\label{sec:results}
In Sec.~\ref{Energy_levels_fit}, we present the results of the energy-level fit including $\rho$-meson exchange, as well as the result for the quark mass dependence of the $T_{cc}$ binding energy-light and heavy. Then, in Sec.~\ref{infinite_volume}, we study off-shell effects from the LS equation and the role of the pion exchange in the phase shifts.

\subsection{Energy-level fit}\label{Energy_levels_fit}

First, we show the results of the energy-level fit from Refs.~\cite{Padmanath:2022cvl, Collins:2024sfi} and Ref.~\cite{Whyte:2024ihh} individually. Then, we perform a global fit, including the first two energy levels from Refs.~\cite{Padmanath:2022cvl, Collins:2024sfi,Whyte:2024ihh,Chen:2022vpo} with data points below the $D^*D^*$ threshold, together with the scattering length data from Refs.~\cite{Lyu:2023xro} and \cite{Ikeda:2013vwa}. Finally, we discuss the light and charm quark mass dependence of the pole position, from the global fit.


\subsubsection{Padmanath24 data}\label{Padma}

\begin{figure*}
     \centering
     \includegraphics[width=0.95\linewidth]{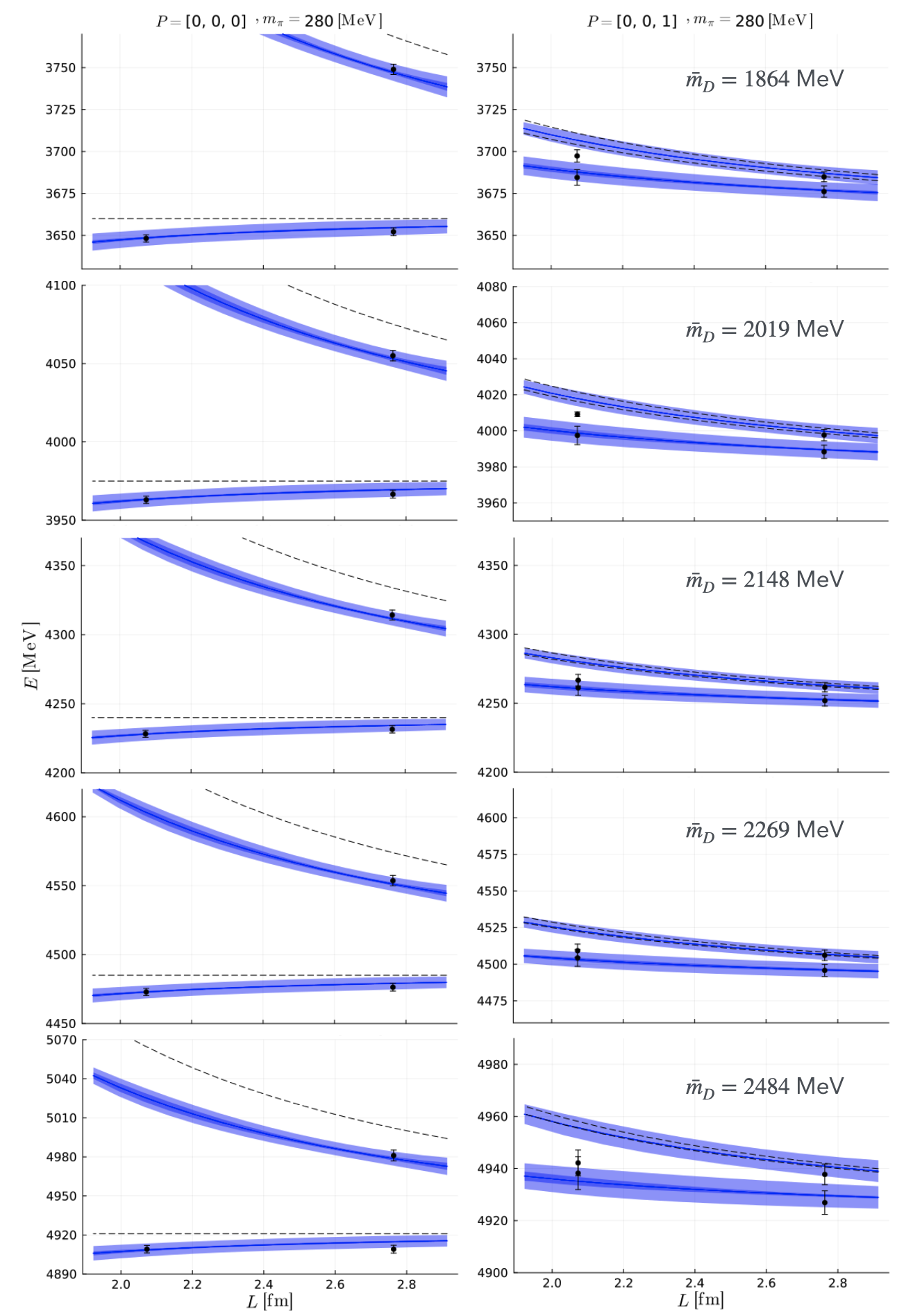}
   \caption{Results from the energy-level fit to Padmanath24 data \cite{Padmanath:2022cvl,Collins:2024sfi}. The five rows of plots correspond to the five sets of different charm quark mass from Ref. \cite{Collins:2024sfi}. Black dashed lines are the noninteracting energy levels, while blue lines are the results of our fit. Here and in the next figures, error bars include statistical and systematic errors.}
   \label{fig:PrelovElvls}
\end{figure*} 

Energy-level data from Refs.~\cite{Padmanath:2022cvl, Collins:2024sfi} are analyzed. Note that these data include ensembles at five different charm quark masses. The goodness-of-fit measure $\chi^2$ is expressed as
\begin{align}
    \chi^2=\Delta E^TC^{-1}\Delta E
\end{align}
where $\Delta E_i=E_i-E^0_i$, with $E^0_i$ the lattice energy $i$, and $C$ is the covariance matrix provided by the authors of Ref.~\cite{Padmanath:2022cvl}. In principle, the fitting parameters are the cutoff $q_\mathrm{max}$ in Eqs.~(\ref{eq:bethe})-(\ref{eq:cut}) and the coupling constant $g$ in Eq.~(\ref{g_rho}). However, when attempting to use these two variables in the fit, we find a strong correlation between these two parameters. For this reason, we fix $q_\mathrm{max}=612$~MeV, which is the value of the cutoff from the global fit explained in Sec.~\ref{sec:gl}\footnote{When the global fit is performed, we find that the correlation between these two parameters is negligible.}. The energy levels that we obtain are shown in Fig. \ref{fig:PrelovElvls} in comparison with the lattice data from Ref.~\cite{Collins:2024sfi} for different charm quark masses. For the error bands, we show, here and in the following figures, statistic and systematic errors due to the lattice spacing in a double band, with the first one shown in a darker color. As shown in Fig. \ref{fig:PrelovElvls}, the description of the data is overall good except for the second energy level at $L=2.07$~fm, which our model does not reproduce well. However, note that this point is also not well reproduced in Ref.~\cite{Padmanath:2022cvl}, and it is systematically not considered in the LQCD phase shift results~\cite{Padmanath:2022cvl,Collins:2024sfi}.\footnote{In addition, we have checked that the effect of the pion exchange is not important at the energy of this point, which lies far from the lhc, in an energy region where other points are well fitted.} We obtain a reasonable value of $\chi^2$, which is $\chi^2/dof=1.47$ with $35$ data points. The value of the coupling $g$ obtained is given in the first row of Table \ref{tab:param}.
 The values of the pole positions relative to the threshold, i. e., $\Delta E=E_{\mathrm{pole}}-(m_D+m_{D^*})$, are given in Table \ref{tab:poleslat}. In all cases, for the five ensembles, a virtual bound state is found. As shown in Table~\ref{tab:poleslat}, the interaction becomes more attractive when the charm quark mass increases, in agreement with the findings of Ref.~\cite{Collins:2024sfi}. In the quantities tabulated, the first and second numbers in parentheses denote the statistical and systematic errors, respectively. In the same table, we also include the result of the analysis of the LQCD data from Ref.~\cite{Whyte:2024ihh}  and the one of the global fit. Both results are discussed below.\footnote{We do not perform an individual analysis of the data from Ref.~\cite{Chen:2022vpo} because in this case the number of data points is too small to find a reasonable solution.}

\begin{table*}
\centering
 \setlength{\tabcolsep}{0.5em}
{\renewcommand{\arraystretch}{1.6}
\begin{tabular}{|c|c|c|}
\hline
  & $g_{(0)}$ & $g_2m_{phys}^2$  \\
 \hline
Padmanath24 \cite{Padmanath:2022cvl}\cite{Collins:2024sfi}  & $2.60\pm0.62$ & $-$ \\
 \hline
 HSC24 \cite{Whyte:2024ihh} & $3.01\pm 0.10$ & $-$ \\
 \hline
  Global & $3.13\pm0.10$ & $-0.057\pm 0.058$ \\
 \hline
 \end{tabular}}
\caption{Best parameters obtained from the several energy-level fits. The cutoff is fixed to $q_{\mathrm{max}}=612$ MeV.}
\label{tab:param}
\end{table*}

\begin{table*}
\centering
 \setlength{\tabcolsep}{0.5em}
{\renewcommand{\arraystretch}{1.6}
\begin{tabular}{|c|c|c|c|c|}
\hline
\multicolumn{2}{|c|}{Collaboration} & Fit Padmanath24 & Fit HSC24 & Global fit \\ 
 \hline
  \multicolumn{2}{|c|}{Physical} & $-$ & $-$ & $-0.06 \left(^{+1.30}_{-2.20}\right) \left(^{+0.50}_{-1.11}\right)$ \\
 \hline
\multirow{5}{*}{Padmanath24 \cite{Padmanath:2022cvl}\cite{Collins:2024sfi}} & 1 & $-19.64 \left(^{+6.63}_{-7.32}\right) \left(^{+5.14}_{-5.50}\right)$ & $-$ & $-7.67 \left(^{+7.45}_{-16.34}\right) \left(^{+4.85}_{-6.68}\right)$\\
\cline{2-5}
  & 2 & $-13.25 \left(^{+5.48}_{-6.37}\right) \left(^{+4.31}_{-4.79}\right)$ & $-$ & $-3.87 \left(^{+3.97}_{-13.19}\right) \left(^{+3.20}_{-5.16}\right)$\\
 \cline{2-5}
   & 3 & $-9.28 \left(^{+4.54}_{-5.58}\right) \left(^{+3.62}_{-4.19}\right)$ & $-$ & $-1.89 \left(^{+2.69}_{-10.72}\right) \left(^{+1.82}_{-3.98}\right)$ \\
\cline{2-5}
   & 4 & $-6.45 \left(^{+3.71}_{-4.87}\right) \left(^{+2.99}_{-3.64}\right)$ & $-$ & $-0.77 \left(^{+2.64}_{-8.59}\right) \left(^{+0.83}_{-2.96}\right)$  \\
\cline{2-5}
   & 5 & $-3.03 \left(^{+2.32}_{-3.64}\right) \left(^{+1.94}_{-2.71}\right)$ & $-$ & $-0.01 \left(^{+4.35}_{-5.18}\right) \left(^{+1.02}_{-1.34}\right)$\\
 \hline
 \multicolumn{2}{|c|}{CLQCD22
 \cite{Chen:2022vpo}} & $-$ & $-$ & $-13.09 \left(^{+12.67}_{-26.65}\right) \left(^{+7.48}_{-9.72}\right)$ \\
\hline
 \multicolumn{2}{|c|}{HSC24 \cite{Whyte:2024ihh}} & $-$ & $-7.87 \left(^{+7.75}_{-18.72}\right) \left(^{+4.08}_{-5.22}\right)$ & $ -22.00 \left(^{+20.49}_{-36.26}\right) \left(^{+10.84}_{-13.05}\right)$\\
 \hline
 \end{tabular}}
\caption{Binding energy in MeV units obtained for different collaborations from the lattice energy-level fits.}
\label{tab:poleslat}
\end{table*}

\subsubsection{HSC24}

\begin{figure}[h!]
\centering
   \begin{minipage}{0.5\textwidth}
     \centering
     \includegraphics[width=1.\linewidth]{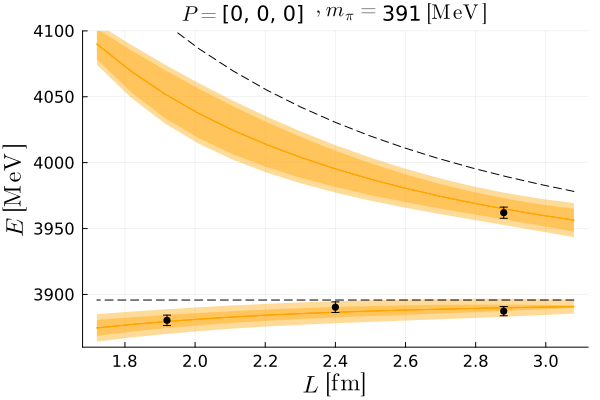}
   \end{minipage}
   \begin{minipage}{0.5\textwidth}
     \centering
     \includegraphics[width=1.\linewidth]{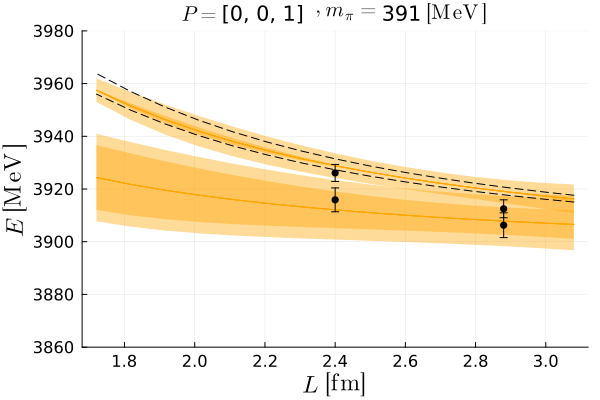}
   \end{minipage}
   \begin{minipage}{0.5\textwidth}
     \centering
     \includegraphics[width=1.\linewidth]{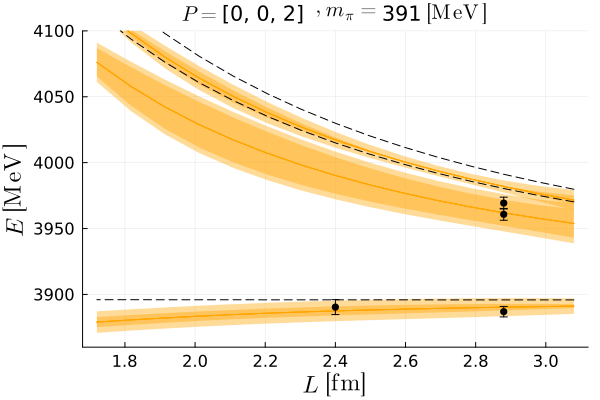}
   \end{minipage}
   \caption{Results from the energy-level fit to HSC24 \cite{Whyte:2024ihh} data. Black dashed lines are the noninteracting energy levels, while orange lines are the results of our fit.}
   \label{fig:HSCElvls}
\end{figure} 

The energy levels obtained from this fit are shown in Fig. \ref{fig:HSCElvls}. We show the statistic and systematic errors in the plot. The parameters obtained are given in Table \ref{tab:param}. The first energy level is very well described. Indeed, only with this energy level can one determine the pole position of $T_{cc}$ since this level is the closest one to the pole. We include the first two energy levels in the fit with data points below the $D^*D^*$ threshold. For the second one, the description is inside either the statistical or systematic error band.
As can be seen, the coupling is slightly larger than the one in the previous analysis with data from Ref.~\cite{Collins:2024sfi}. We also obtain a virtual bound state as in Ref.~\cite{Whyte:2024ihh}. See Table~\ref{tab:poleslat}. In this case, the state is further from the threshold, indicating that the interaction becomes less attractive as the pion mass increases, consistently with Refs.~\cite{Padmanath:2022cvl, Collins:2024sfi}. We include 12 points and obtain the $\chi^2$ minimum value for $\chi^2/dof=1.59$. 

\subsubsection{Global fit}~\label{sec:gl}

In this section, we perform a global fit, which includes the  energy levels studied in the previous sections~\cite{Padmanath:2022cvl,Collins:2024sfi,Whyte:2024ihh} together with the first two energy levels from CLQCD22 \cite{Chen:2022vpo}, and also the scattering length data from HALQCD23 \cite{Lyu:2023xro} and HALQCD14
\cite{Ikeda:2013vwa},\footnote{Note that in Refs.~\cite{Lyu:2023xro,Ikeda:2013vwa} no energy levels are computed.} collected in Table \ref{tab:a0data}.

\begin{table}
\centering
 \setlength{\tabcolsep}{0.5em}
{\renewcommand{\arraystretch}{1.6}
\begin{tabular}{|c|c|c|c|c|c|}
\hline
 Collaboration & $a$ & $m_\pi$ & $a_0^{-1}$ & $M^{c\bar{c}}_{avg}$ \\
 \hline
 HALQCD23 \cite{Lyu:2023xro} & $0.0846$ & $146$ & $0.05$ & $3097$\\
 \hline
 HALQCD14
\cite{Ikeda:2013vwa} & $0.0907$ & $411$ & $2.34$ & $3070$\\
 \hline
\end{tabular}}
\caption{Summary of the scattering length data for $DD^*$ for different lattice collaborations. The lattice spacing $a$ is given in fm, the inverse of the scattering length, $a_0^{-1}$, in fm$^{-1}$, and the masses in units of MeV.}
\label{tab:a0data}
\end{table}

\begin{figure*}
\centering
   \begin{minipage}{0.95 \textwidth}
     \centering
     \includegraphics[width=1.\linewidth]{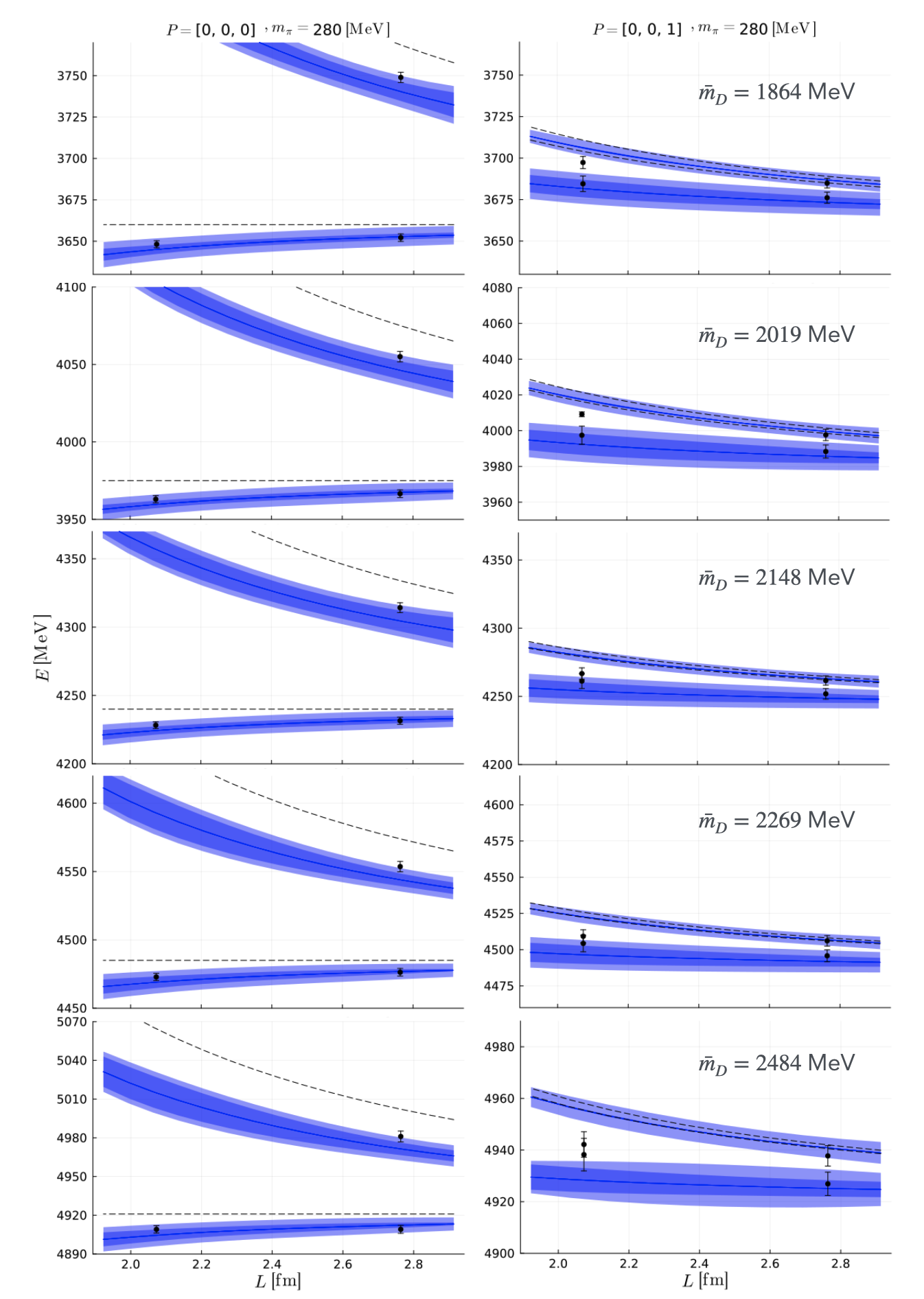}
   \end{minipage}
   \caption{Results from the global fit (blue solid lines). The five rows of plots correspond to the five sets of different charm quark mass from Ref. \cite{Collins:2024sfi}. Black dashed lines denote the noninteracting energy levels.}
   \label{fig:GLOBAL_PrelovElvls}
\end{figure*} 
\begin{figure}[h!]
\centering
   \begin{minipage}{0.5\textwidth}
     \centering
     \includegraphics[width=1.\linewidth]{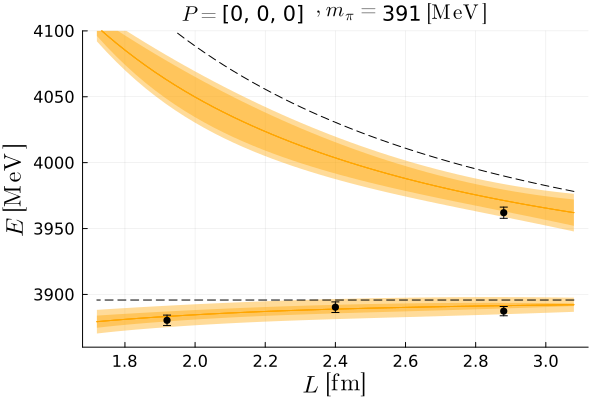}
   \end{minipage}
   \begin{minipage}{0.5\textwidth}
     \centering
     \includegraphics[width=1.\linewidth]{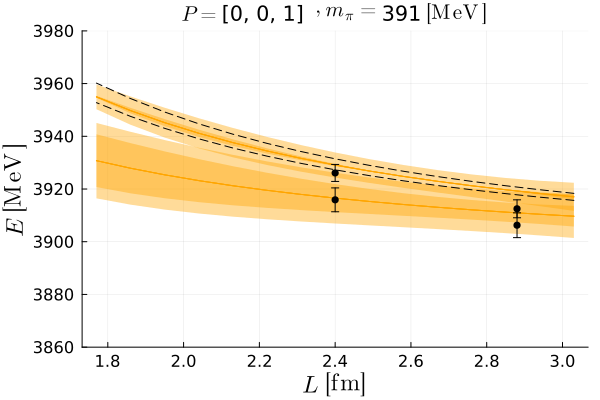}
   \end{minipage}
   \begin{minipage}{0.5\textwidth}
     \centering
     \includegraphics[width=1.\linewidth]{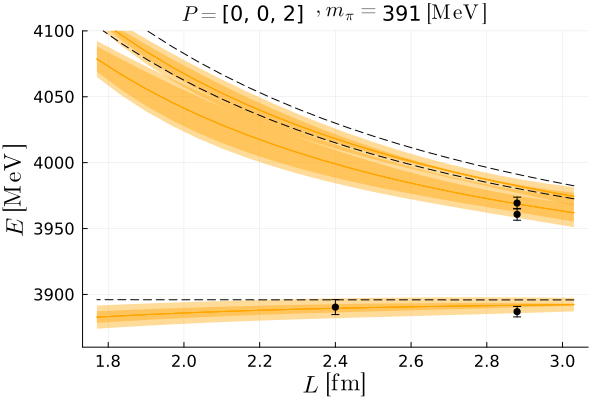}
   \end{minipage}
   \caption{Results from the global fit (orange solid lines). These plots correspond to the HSC24 \cite{Whyte:2024ihh} data. Black dashed lines denote the noninteracting energy levels.}
   \label{fig:GLOBAL_HSCElvls}
\end{figure} 
\begin{figure}[h!]
\centering
   \begin{minipage}{0.5\textwidth}
     \centering
     \includegraphics[width=1.\linewidth]{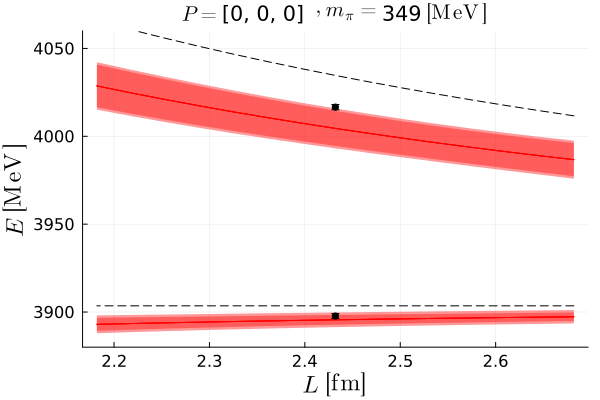}
   \end{minipage}
   \caption{Results from the global fit (red solid lines). These plots correspond to the CLQCD22 \cite{Chen:2022vpo} data. Black dashed lines are the noninteracting energy levels.}
   \label{fig:GLOBAL_ChenElvls}
\end{figure} 

\begin{figure*}
\begin{minipage}{0.47\textwidth}
     \includegraphics[width=1\linewidth]{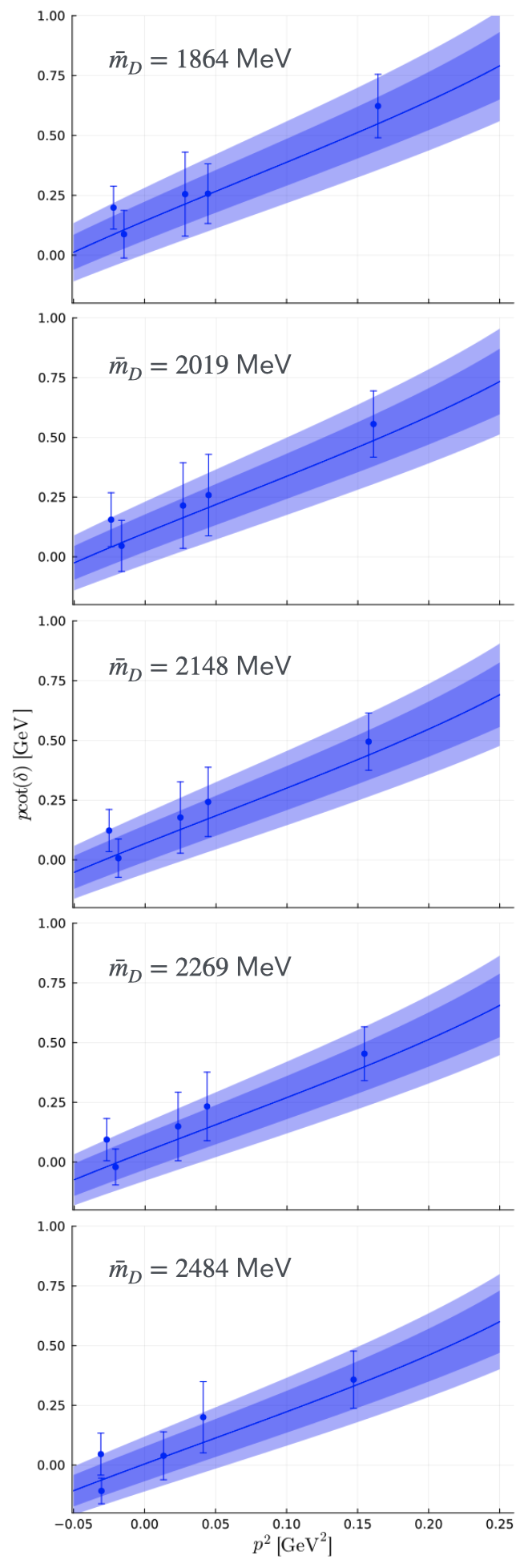}
   \end{minipage}
   \hfill
   \begin{minipage}{0.47\textwidth}
     \includegraphics[width=1\linewidth]{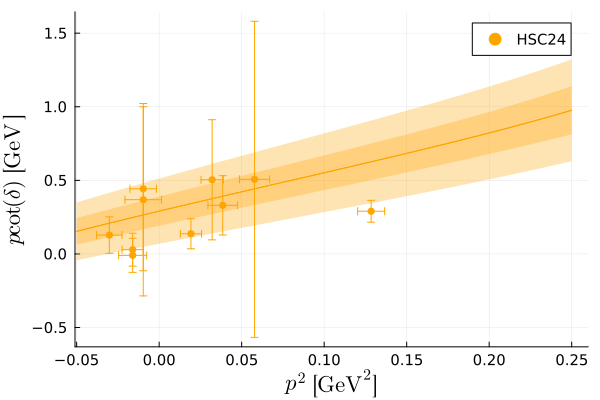}
      \includegraphics[width=1\linewidth]{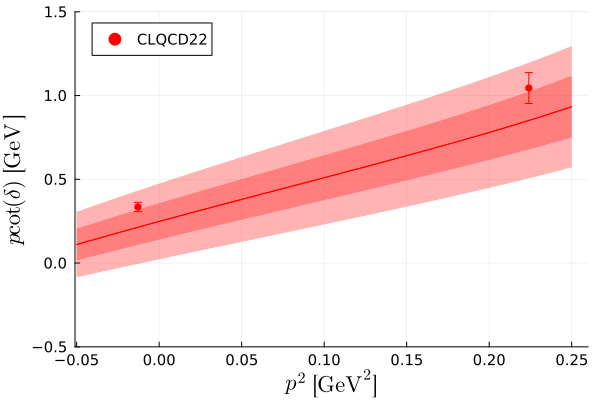}
   \end{minipage}
   \caption{Phase shifts as a function of center-of-mass momentum for each collaboration. The error bands are calculated, including the errors of the parameters from our fit shown in darker color (statistical), and the ones from the lattice spacing error in lighter color (systematic).}
   \label{fig:GLOBAL_pcotdelta}
\end{figure*}

As we have data from several pion masses, we use the parametrization of Eq. (\ref{g_rho}) with a pion-mass-dependent coupling. Hence, for the global fit, we have three parameters: the cutoff $q_\mathrm{max}$, $g_0$, and $g_2$ for the coupling. The values of the parameters that we obtain are $q_\mathrm{max}=612 \pm 29$ MeV, $ g_0 = 3.13 \pm 0.10$, and $g_2m^2_{phys}=(-0.057 \pm 0.058 )$, the latter having no dimensions. These are also shown in Table~\ref{tab:param}, together with the results from individual fits. The value of $g_0$ obtained is close to the value of $g$ obtained in individual fits, since the pion mass dependence of the coupling is mild. The results for the energy levels are depicted in Figs. \ref{fig:GLOBAL_PrelovElvls}-\ref{fig:GLOBAL_ChenElvls}. Indeed, these are very similar to the ones from individual fits. In Fig. \ref{fig:GLOBAL_pcotdelta}, we show the phase shifts obtained for the different simulations. Overall, we find very good agreement with the LQCD data analyzed. The total number of data points used is $51$, and the $\chi^2/dof$ obtained is $0.91$. We determine the pole positions from this fit. These are given in Table~\ref{tab:poleslat}. In all cases, we obtain a virtual bound state. The central value of the binding energy is smaller than in the individual analysis of the Padmanath24 data, while it is larger for the HSC24 data. In any case, the result of the global fit, once the statistical and systematic errors are included, is compatible with the individual analyses, showing consistency between the lattice data sets considered within the current errors of LQCD data.
In Appendix \ref{mpi4} we study the effects of possible higher-order effects by adding a quartic term in the pion mass in Eq. (\ref{g_rho}). As shown in Fig. \ref{fig:highorders}, the fit is slightly improved, but the results are compatible within errors with the ones presented here.

\begin{figure}
\centering
   \begin{minipage}{0.5\textwidth}
     \centering
     \includegraphics[width=1.\linewidth]{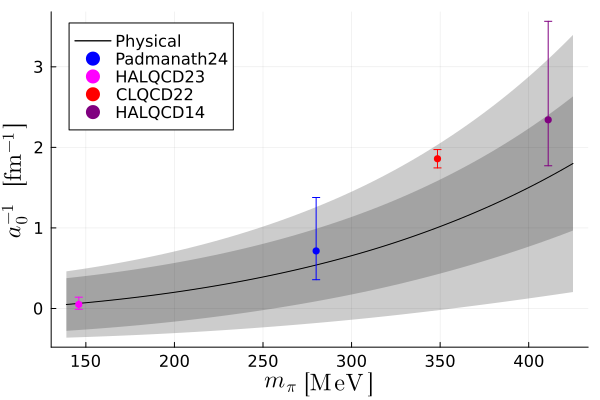}
   \end{minipage}
   \caption{Inverse of the scattering length as a function of the pion mass. The black line is the physical trajectory extracted from our global fit.}
   \label{fig:GLOBAL_a0}
\end{figure}

\begin{figure}
\centering
   \begin{minipage}{0.5\textwidth}
     \centering
     \includegraphics[width=1.\linewidth]{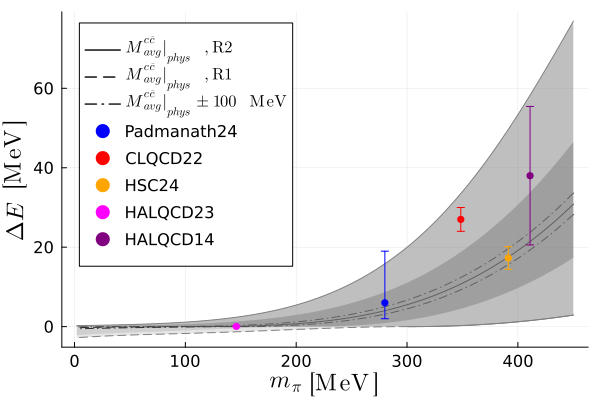}
   \end{minipage}
   \begin{minipage}{0.5\textwidth}
     \centering
     \includegraphics[width=1.\linewidth]{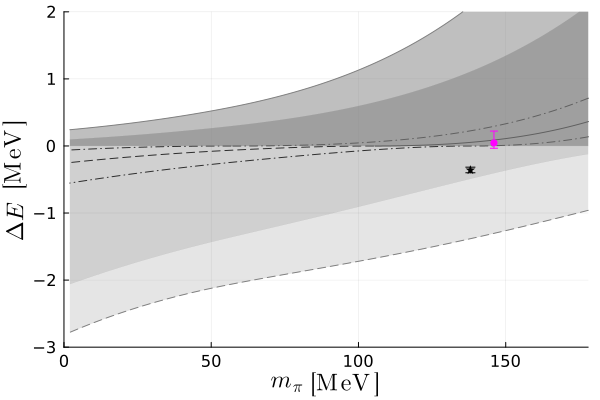}
   \end{minipage}
   \begin{minipage}{0.5\textwidth}
     \centering
     \includegraphics[width=1.\linewidth]{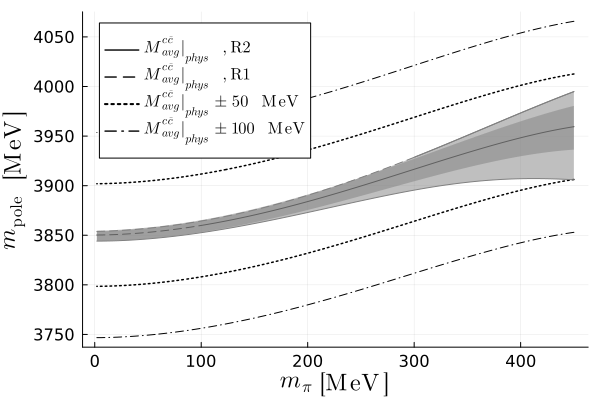}
   \end{minipage}
   \caption{Binding energy of the pole associated with $T_{cc}(3875)^+$ as a function of the pion mass. The black line denotes the physical charm quark mass trajectory extracted from the global fit. The solid line represents a virtual bound state in the second Riemann sheet, and the dashed line shows when a bound state in the first RS is found. We change the sign of the binding energy when the pole is in the second RS for better clarity. The black star represents the physical binding energy from Ref. \cite{ParticleDataGroup:2022pth}. The CLQCD22 and HALQCD14 pole positions are calculated by solving the equation $p \cot \delta=ip$.}
   \label{fig:GLOBAL_Binding}
\end{figure}

\begin{figure}
\centering
   \begin{minipage}{0.5\textwidth}
     \centering
     \includegraphics[width=1.\linewidth]{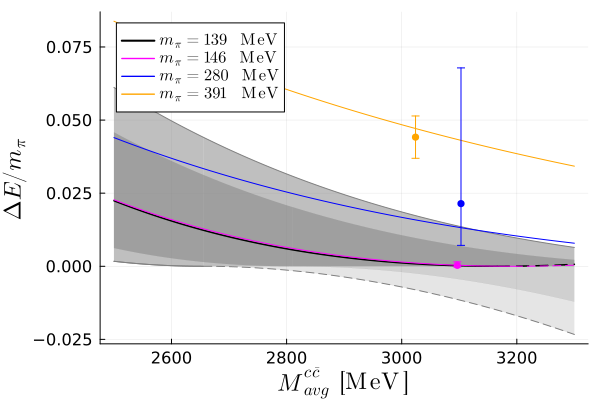}
   \end{minipage}
   \begin{minipage}{0.5\textwidth}
     \centering
     \includegraphics[width=1.\linewidth]{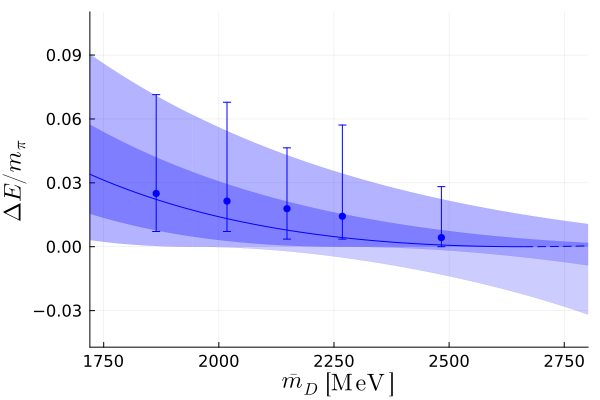}
   \end{minipage}
   \caption{Top panel: binding energy of $T_{cc}(3875)^+$ as a function of $M^{c\bar{c}}_{avg}$ for each collaboration. The black line and error band represent the physical charm quark mass trajectory extracted from our global fit. The results for the various pion masses are denoted with different colors. Bottom panel: binding energy of $T_{cc}(3875)^+$ as a function of $\bar{m}_D= 1/4(m_{D}+3m_{D^*})$ for all the ensembles of Ref. \cite{Collins:2024sfi}.As in the previous figure, we change the sign of the binding energy when the pole is in the second RS for better clarity.}
   \label{fig:GLOBAL_Binding_mcs}
\end{figure}

With the results of the global fit, we are ready to further investigate the quark mass dependence of the $T_{cc}(3875)^+$ pole position, which comes from difference sources. The Weinberg-Tomozawa (WT) term,~Eq.~(\ref{treeampli_rho}), depends explicitly on the masses of the low-lying charmed mesons: it is obtained in Ref.~\cite{Gil-Dominguez:2023eld} from one-loop HHChPT, and we consider explicitly the $\rho$-meson exchange propagator in Eq.~(\ref{treeampli_rho}). The quark mass dependence of the $\rho$ meson is taken from the one-loop NLO unitarized ChPT analysis of Ref.~\cite{Molina:2020qpw}. Figs.~\ref{fig:GLOBAL_a0} and \ref{fig:GLOBAL_Binding}, we show the result with the spin average charmonia mass at the physical point, which is $M_{avg}^{c\bar{c}}=3069$ MeV.
The pion mass dependence of the inverse of the scattering length, $a_0^{-1}$, is shown in Fig. \ref{fig:GLOBAL_a0}, in comparison with the available LQCD data.\footnote{Note that not all collaborations calculate the scattering length.} We can see that this quantity increases with the pion mass, and our prediction describes well the scattering LQCD data; i. e., the scattering length decreases with the pion mass.
The light and heavy quark mass dependence of the pole position are given in Figs.~ \ref{fig:GLOBAL_Binding} and \ref{fig:GLOBAL_Binding_mcs}. In these figures, the dashed line denotes a pole in the first Riemann sheet (bound state), while the solid line stands for the pole position in the second Riemann sheet below threshold (virtual state). In Fig. \ref{fig:GLOBAL_Binding} (top and middle panels), we plot the binding energy of $T_{cc}(3875)^+$ as a function of the pion mass.\footnote{In the case of the HALQCD14 and CLQCD22 the pole position is not given in the LQCD articles. We have estimated them by evaluating the energy such that $p\mathrm{cot}\delta=ip$. We evaluate the error through the propagation of the scattering length error.} We can see that, at $m_\pi \sim 100$ MeV, the pole switches Riemann sheets, becoming a bound state for lower pion masses. When taking into account the statistical error, the result is consistent with a bound state for the physical pion mass with a binding energy similar to the experimental one. However, note that the available LQCD data have less precision than the experimental ones. In the middle plot, we can see that region in detail. Overall, the trend found indicates that the interaction becomes less attractive as the pion mass increases. We have also shown, in these two panels of Fig.~\ref{fig:GLOBAL_Binding}, the result of the binding energy when raising or lowering the charm quark mass. Concretely, we plot the binding energy for $M_\mathrm{avg}^{c\bar{c}}=M_\mathrm{avg}^{c\bar{c}}\vert_\mathrm{phys}\pm 100$, with the lower curve corresponding to a larger spin average charm quark mass. The result is interesting-the binding energy is affected very little, by $\sim0.25$ MeV, by the change of this charm quark mass in the region near the pole. This tells us that the pole indeed follows the threshold in such a way that the binding energy is not very sensitive to the charm quark mass. In Fig.~\ref{fig:GLOBAL_Binding} (bottom panel), the mass of the pole is shown as a function of the pion mass for different charm quark masses, with the lower curve corresponding to a smaller charm quark mass. The variation of the pole of the $T_{cc}$ mass due to a change of $50$~MeV in the spin average charm quark mass is about $52$~MeV. 

In the top panel of Fig. \ref{fig:GLOBAL_Binding_mcs}, we plot the binding energy as a function of the spin average charmonium mass $M^{c\bar{c}}_{avg}$, where the gray line and error band stand for the result at the physical pion mass, while other lines denote the dependence for a fixed pion mass of the given LQCD Collaboration in comparison with data. We notice that the attraction becomes stronger as the charm quark mass increases, in agreement with Ref. \cite{Collins:2024sfi}. In the range chosen for $M^{c\bar{c}}_{avg}$, the pole of $T_{cc}(3875)^+$ becomes a bound state around the physical charm quark mass. In the bottom panel of Fig. \ref{fig:GLOBAL_Binding_mcs}, we plot the binding energy of $T_{cc}(3875)^+$ as a function of $\bar{m}_D=1/4(m_{D}+3m_{D^*})$ for all the ensembles of Ref. \cite{Collins:2024sfi}, where we see a very similar behavior with an error band consistent with the error from data. For $\bar{m}_D \gtrsim 2670$ MeV, the $T_{cc}(3875)^+$ pole transitions to the first RS and becomes a bound state for the pion mass of Ref. \cite{Collins:2024sfi}, $\sim 280$~MeV.

Before concluding this subsection, we would like to make some remarks concerning the possible systematic errors from the lattice data sets considered. First, the number of flavors used for some of the ensembles is different. However, the only lattice data set that considers only two flavors, $N_f=2$, is CLQCD22, Ref.~\cite{Chen:2022vpo}, with two data points used as input. There appears to be some tension between this simulation and the others, as can be seen from Figs.~\ref{fig:GLOBAL_a0} and~\ref{fig:GLOBAL_Binding}: however, we do not expect from the phenomenological point of view, that the number of flavors considered is responsible for the tension between the CLQCD22 data sets and others in this case since, even if the $s\bar{s}$ pair is created, the coupling to a channel containing this pair would be highly suppressed because $DD^*$ is an open flavor channel. A possible reason could be the fact that the lattice spacing for this collaboration is larger than the one used in other simulations. Second, all lattice data sets analyzed here use the same operators, i. e., $DD^*$ operators, except for HSC24~\cite{Whyte:2024ihh}, which also includes $D^*D^*$ operators. This threshold is far from the first energy level but could affect the second energy level more. We only take data points below this threshold, and we have checked that the energy level responsible for the $T_{cc}$ pole position is the first one, naturally, because it lies closer to the $DD^*$ threshold. Taking the second energy level helps us obtain the energy-dependent trend of the phase shift. Thus, it is possible that this channel can have some impact on the phase shift obtained here above the energy of the $D^*D^*$ threshold, which can be tested in a future analysis by including this channel in our basis. In addition, new simulations are being conducted which also include diquark-antidiquark interpolators~\cite{Vujmilovic:2024snz}. The effect of such operators is also interesting to study, but this is beyond our scope in the present article.
 To account for systematic errors due to the lattice spacing, in the figures throughout this work, we have shown the errors due to the lattice spacing with lighter color error bands compared to the statistical error band from the fit parameter errors.
 
\subsection{Inclusion of off-shell effects and the pion exchange}\label{infinite_volume}

In this section, we examine the effect of including an off-shell momentum-dependent framework for the interaction from the $\rho$-meson and pion exchanges on the pole positions and phase shifts in the infinite volume limit.
\begin{figure}[h!]
\centering
   \begin{minipage}{0.5\textwidth}
     \centering
     \includegraphics[width=1.\linewidth]{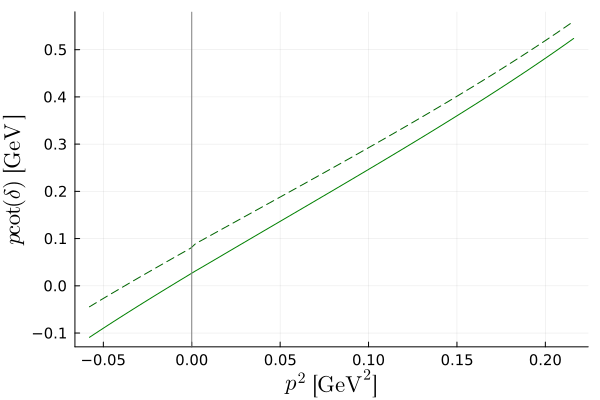}
   \end{minipage}
   \caption{Comparison of the phase shifts in the physical case between the on-shell factorization (solid line) and the momentum-dependent framework (dashed line).}
   \label{fig:pmomdepVSindep}
\end{figure} 

First, we discuss the effect from considering an off-shell interaction for the vector-meson exchange potential.
When considering the momentum-dependent framework, Eq.~(\ref{LS}), with the same parameters obtained from the global fit, one can see that the scattering amplitude and phase shift change with respect to the on-shell factorization. This effect is displayed in Fig. \ref{fig:pmomdepVSindep}, where we observe that $p \cot{\delta}$ increases in the momentum dependent framework. This variation can be absorbed in the coupling $g$. In order to obtain the best results that match the LQCD data, we accommodate this variation in the coupling constant, replacing $g_0 \to g_0 + 0.23$ and $g_2 m^2_{phys} \to g_2 m^2_{phys} -0.019$, where  $ g_0 = 3.13 \pm 0.10$ and $g_2 m^2_{phys}=-0.057 \pm 0.058 $ are the best parameters obtained from the global fit in Sec.~\ref{sec:gl}.\footnote{We obtain these values by matching the values of $p \cot{\delta}$ obtained from the momentum-dependent equation with those obtained in the on-shell factorization}

\begin{figure*}
\begin{minipage}{0.49\textwidth}
     \includegraphics[width=1\linewidth]{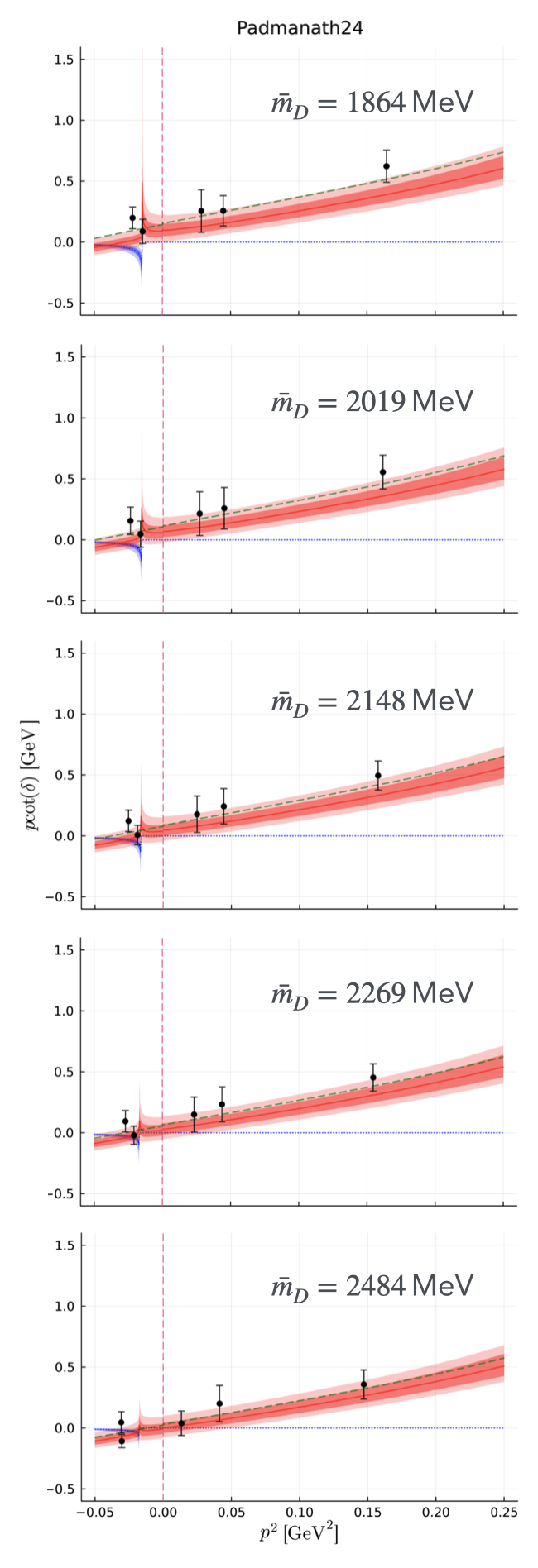}
   \end{minipage}
   \hfill
   \begin{minipage}{0.47\textwidth}
     \includegraphics[width=1\linewidth]{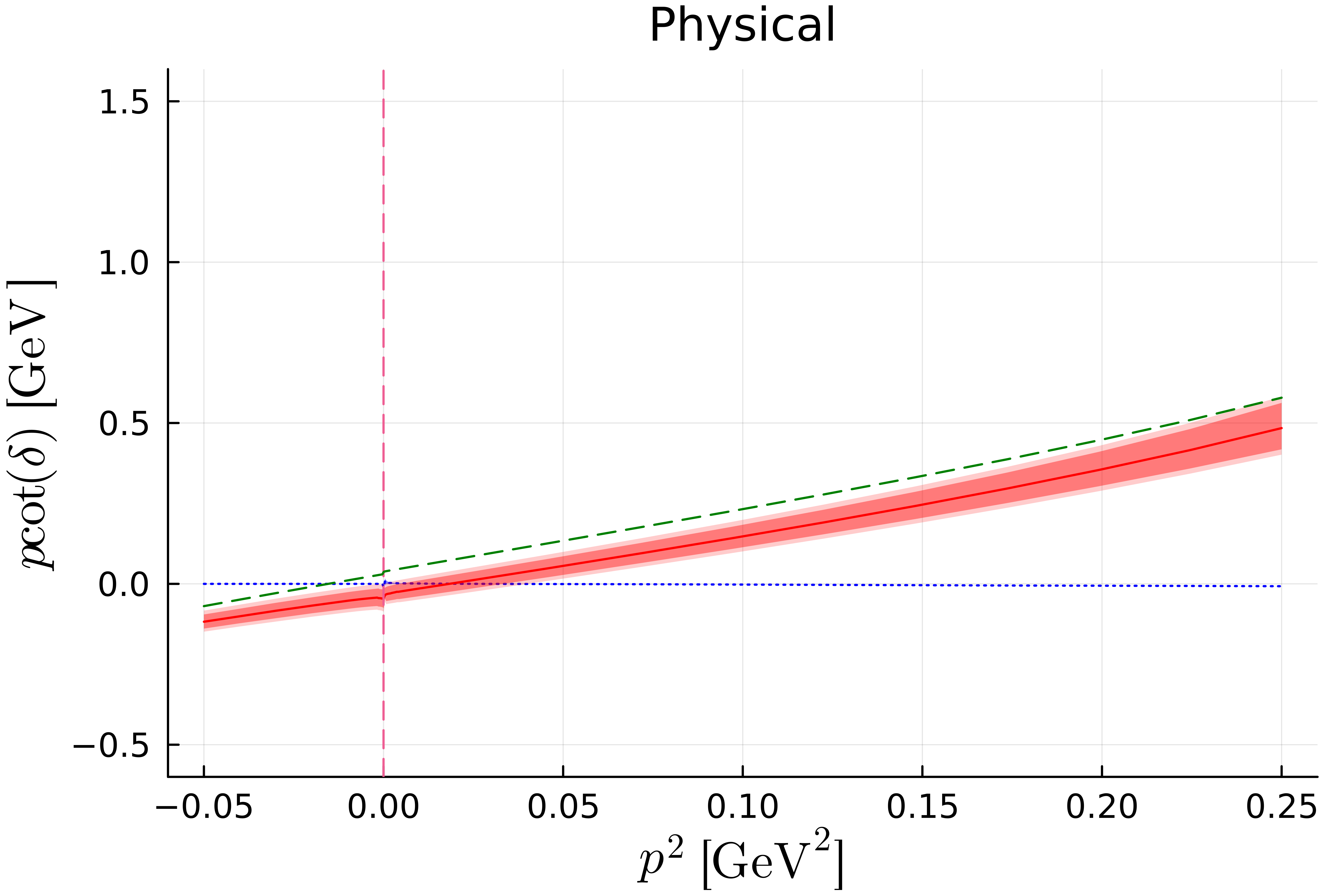}
            \includegraphics[width=1\linewidth]{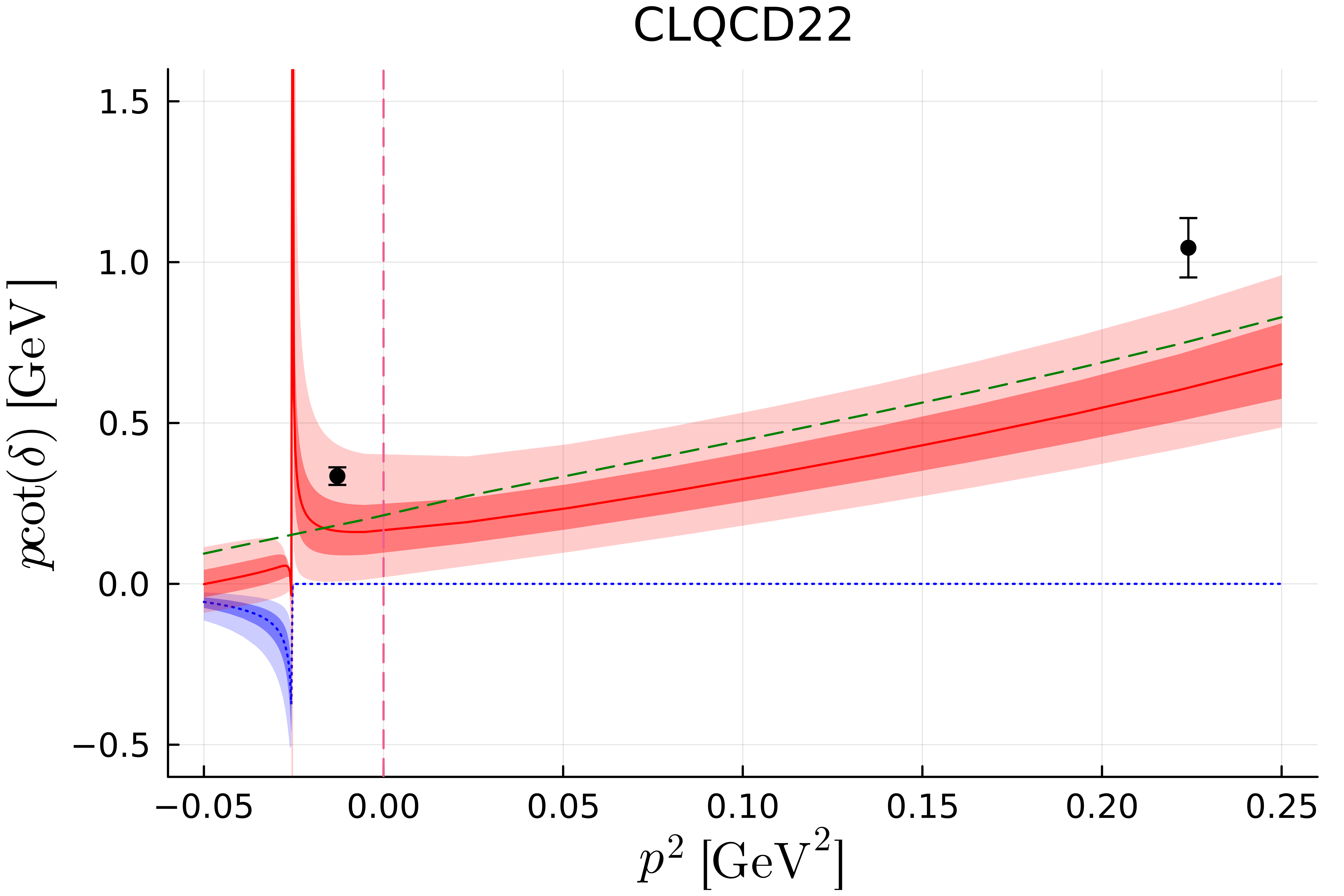}
           \includegraphics[width=1\linewidth]{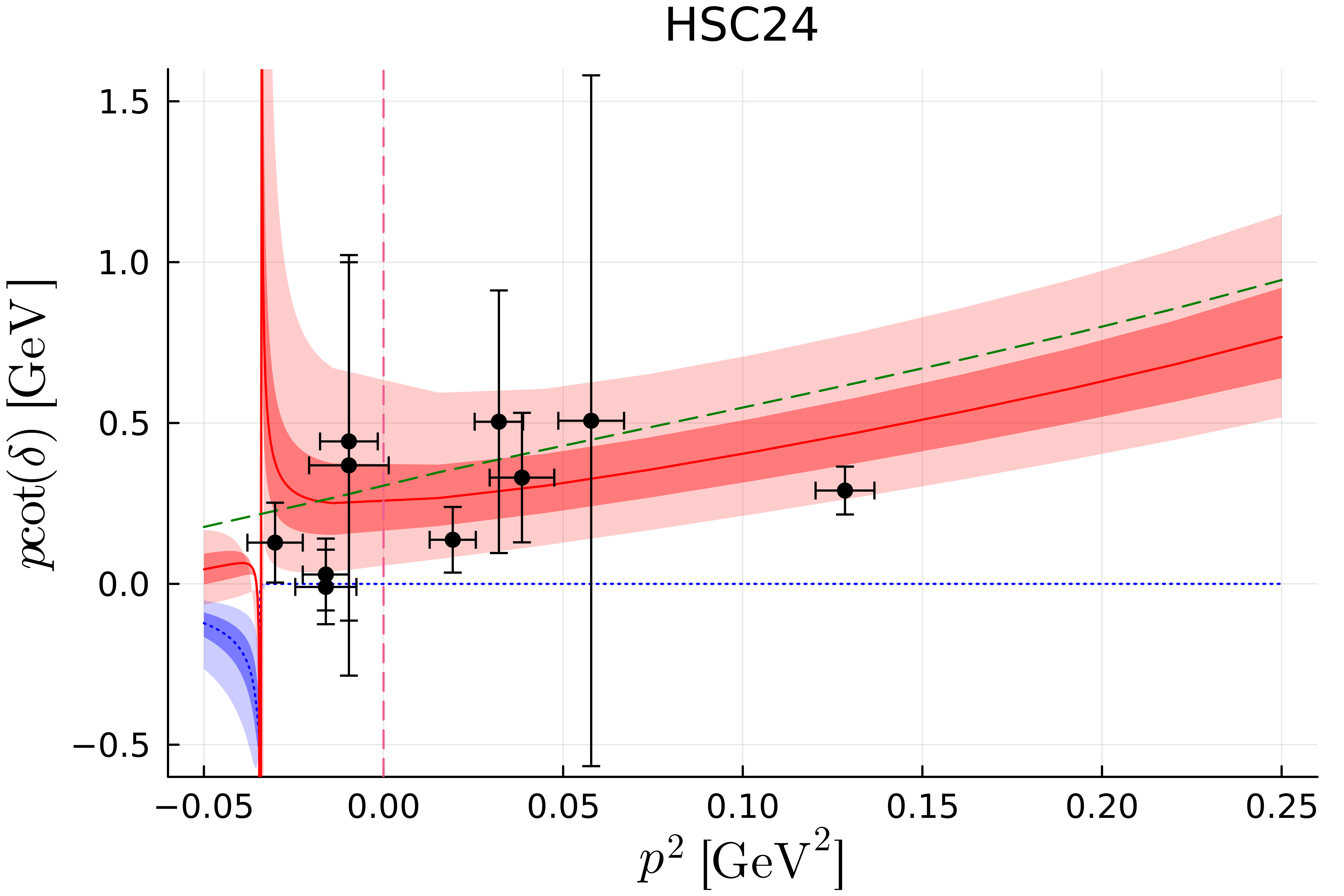}
   \end{minipage}
   \caption{Comparison between the phase shifts from LQCD \cite{Padmanath:2022cvl,Whyte:2024ihh,Chen:2022vpo} and the ones obtained here by solving Eq. (\ref{LS}) for the $\rho$-meson exchange (green dashed line) and the $\rho$+$\pi$-meson exchange (the red and blue lines stand for the real and imaginary parts). The dashed vertical line denotes the $D D^*$ threshold. See also the explanation in the text.}
   \label{fig:phase_shifts}
\end{figure*}

Now, we turn to the inclusion of the pion exchange. Using this new value for the coupling $g$ in the vector-meson exchange, we include the pion exchange and solve the momentum-dependent Bethe-Salpeter equation, Eq. (\ref{LS}). In Fig. \ref{fig:phase_shifts}, we compare our results with the lattice data \cite{Padmanath:2022cvl,Whyte:2024ihh,Chen:2022vpo} and show the physical limit extrapolation in the top-right panel. In this figure, the real and imaginary parts of $p \cot{\delta}$ after including the pion are depicted in red and blue colors, respectively, while the result considering only vector-meson exchange is shown with a dashed green line.
As one can see from Fig. \ref{fig:phase_shifts}, the inclusion of the pion produces an overall decrease of $p \cot{\delta}$ in the energy range studied, the overall trend being preserved with the exception of the region close to the left-hand cut.
In particular, below the left-hand cut, $p \cot{\delta}$ acquires an imaginary part produced by the pion exchange, while the real part diverges, in agreement with the results found in Ref.~\cite{Collins:2024sfi}. As in previous plots, the darker error band accounts for the error of the fit parameters, while the lighter one accounts for the propagation of the lattice spacing error.

\begin{figure*}
   \begin{minipage}{0.49\textwidth}
   \centering
     \includegraphics[width=1\linewidth]{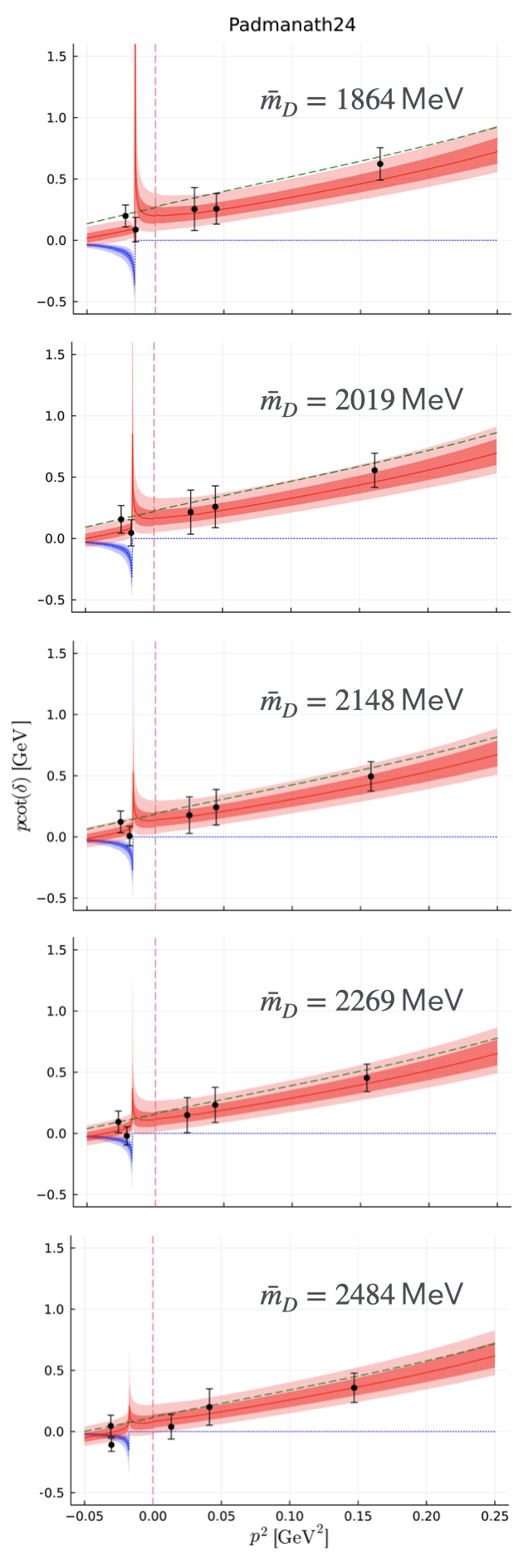}
   \end{minipage}
    \begin{minipage}{0.47\textwidth}
    \centering
     \includegraphics[width=1\linewidth]{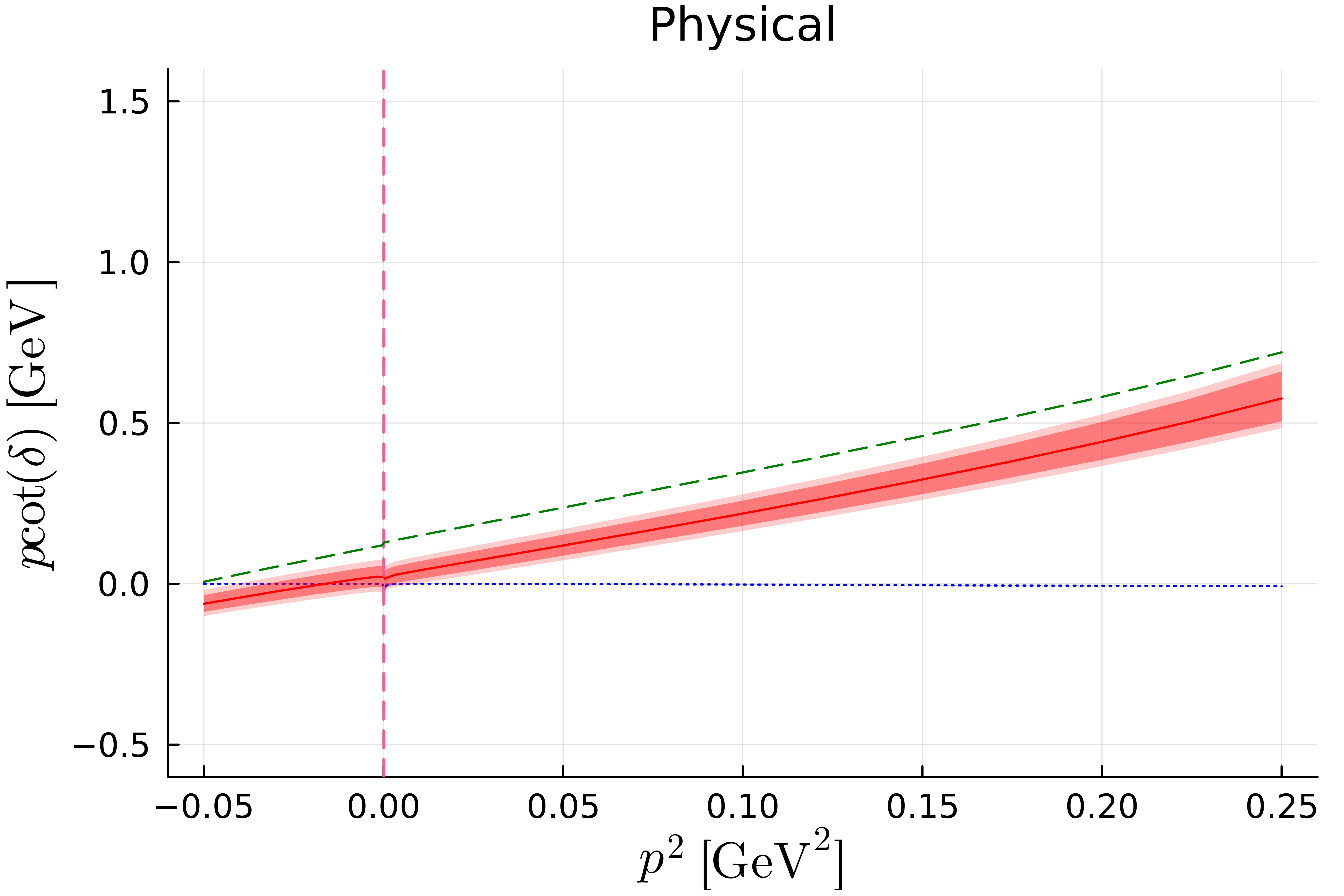}
        \includegraphics[width=1\linewidth]{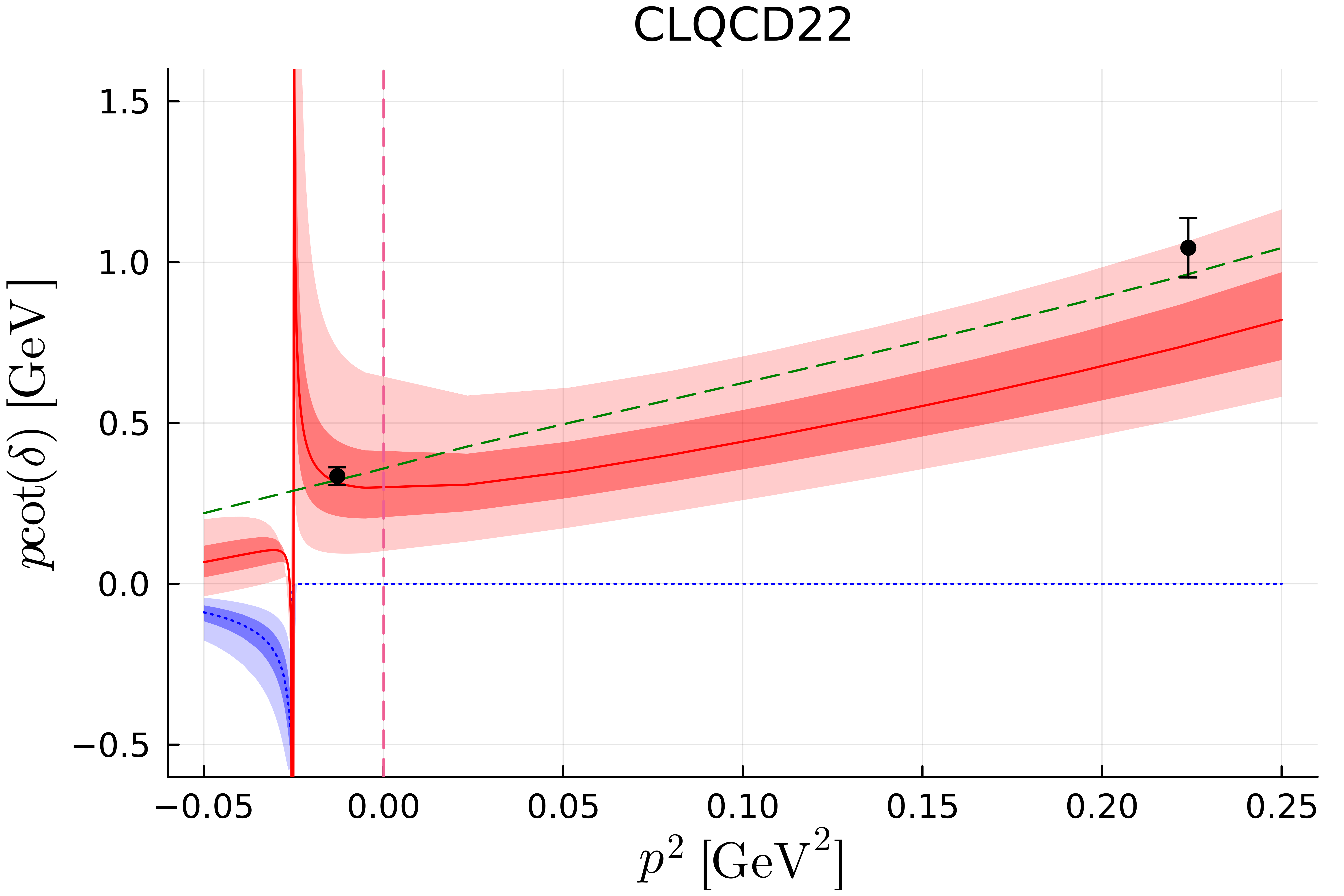}
 \includegraphics[width=1\linewidth]{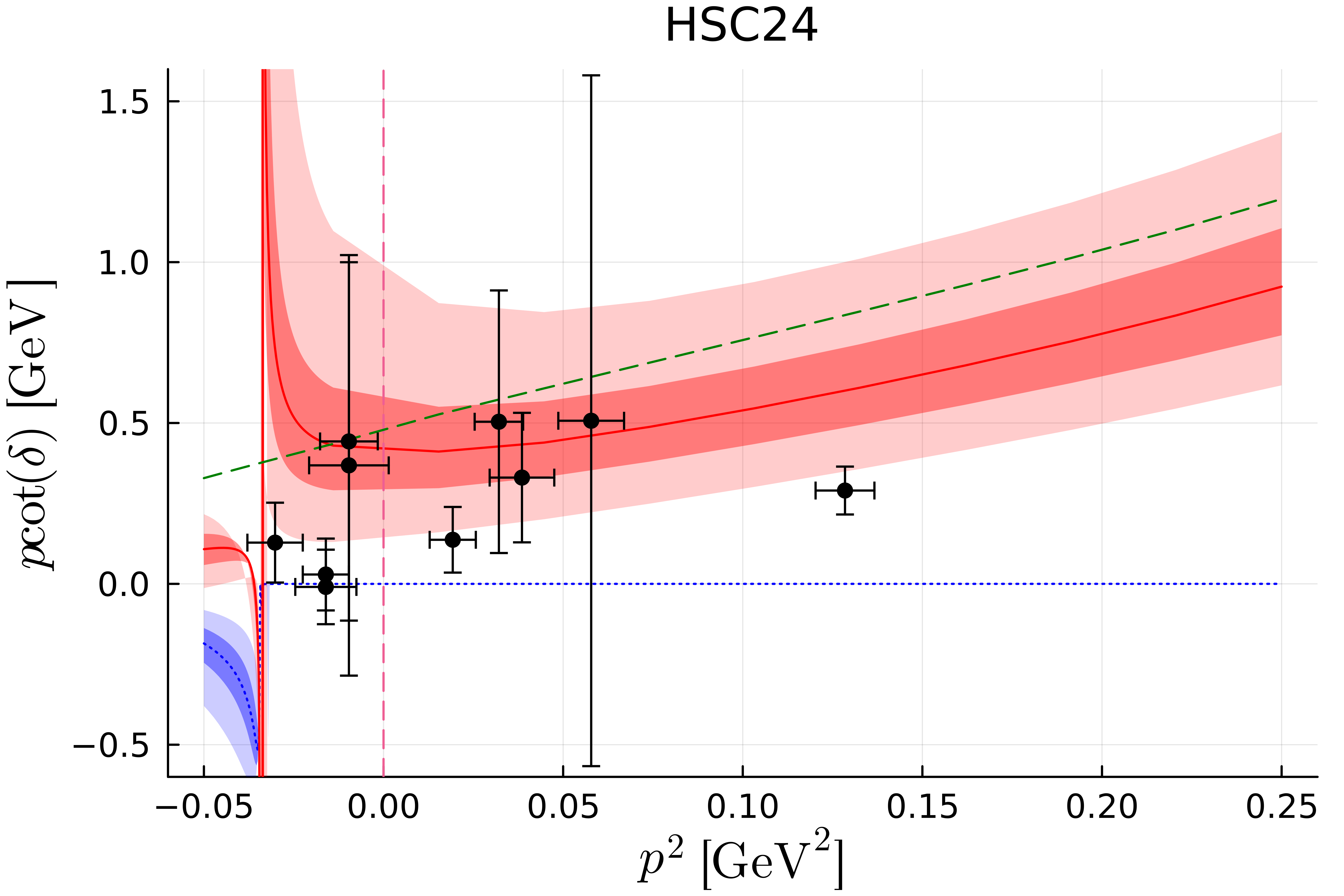}
     \end{minipage}
   \caption{Similar to Fig. \ref{fig:phase_shifts}, but readjusting the $g$ coupling to account for the pion exchange outside the lhc region as explained in the text.}
   \label{fig:phase_shifts2}
\end{figure*}

While the effect of the pion exchange is large around the lhc, this effect is smaller than the phase shift error bars from the LQCD data in the rest of the energy range. Still, we can have a better agreement of our momentum-dependent framework including $\rho+\pi$ exchange with the LQCD data, by absorbing this effect, once again, in the $g$ coupling constant. 
Thus, we tune the $g$ coupling to obtain the best possible agreement with the LQCD phase shift data outside the lhc energy region. This requires changing  $g_0 \to g_0  -0.35, g_2 m^2_{phys} \to g_2 m^2_{phys} +  0.005$, where $g_0$ and $g_2$ are given by the previously calculated parameters, which are $ g_0 = 3.36 $ and $g_2 m^2_{phys}=-0.076$. The new result is shown in Fig.~\ref{fig:phase_shifts2}, and a similar effect of the lhc is observed.
In Table \ref{tab:poles}, we present the new pole positions obtained, including only $\rho$-meson exchange (first column) or taking into account both $\rho+\pi$ (second column), in the momentum-dependent framework of Eq.~(\ref{LS}). As expected, the results from the first column are compatible within errors with the ones shown in Table~\ref{tab:poleslat}.

The inclusion of the pion has two visible effects. First, it moves the virtual state away from the threshold when the pole is near the lhc, which is a repulsive effect. This is the case of the pion mass of $280$~MeV, when it is noticeable. On the contrary, when the pole is far from the lhc, the pion exchange turns out to be attractive. See Fig.~\ref{fig:onshellV} in Appendix \ref{onshelV}. The latter occurs at the physical point. Second, the pion exchange causes the pole to acquire an imaginary part that increases as the pion mass becomes larger.

\begin{figure*}
\centering
   \begin{minipage}{0.45\textwidth}
     \centering
     \includegraphics[width=1.\linewidth]{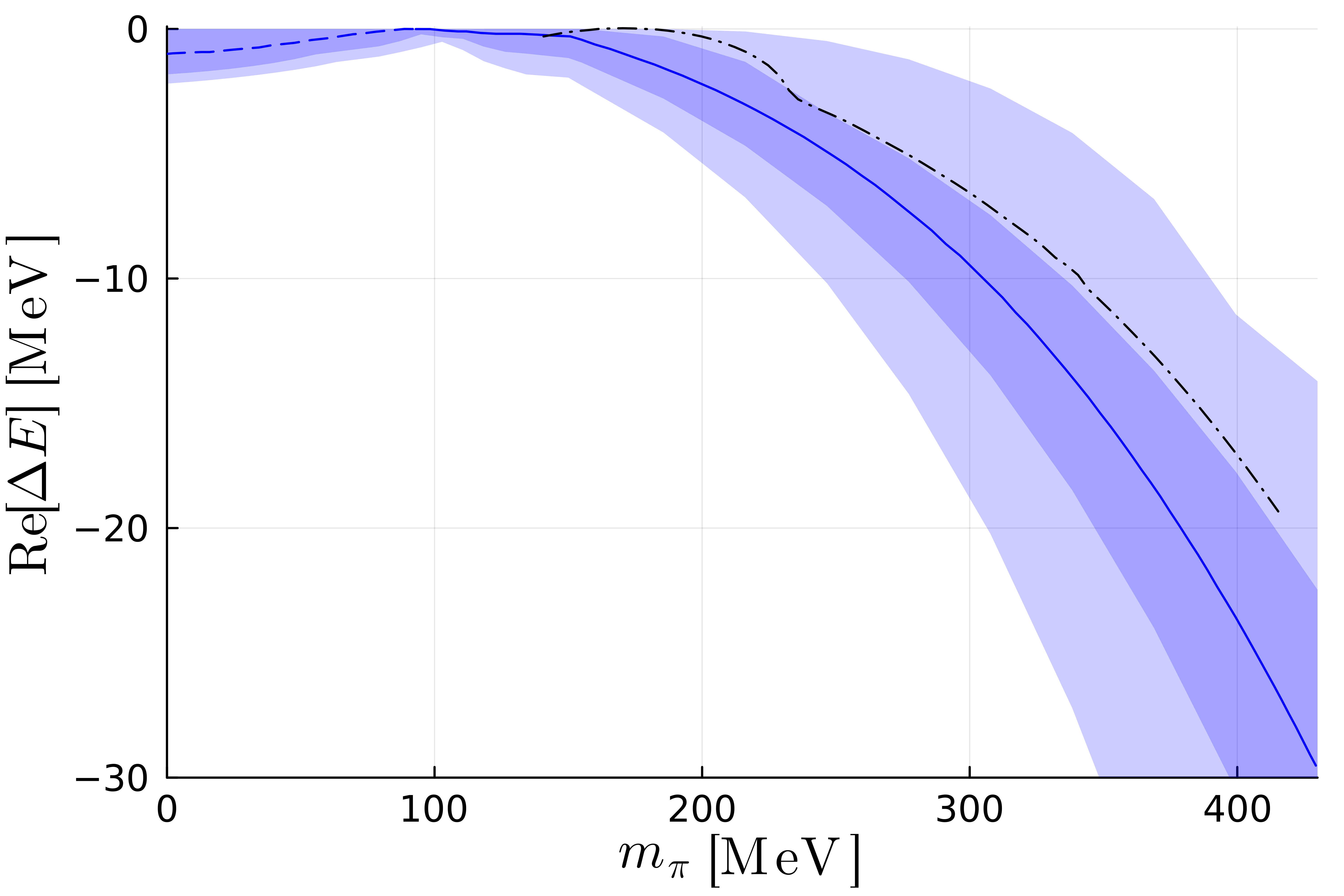}
   \end{minipage}
   \begin{minipage}{0.45\textwidth}
     \centering
     \includegraphics[width=1.\linewidth]{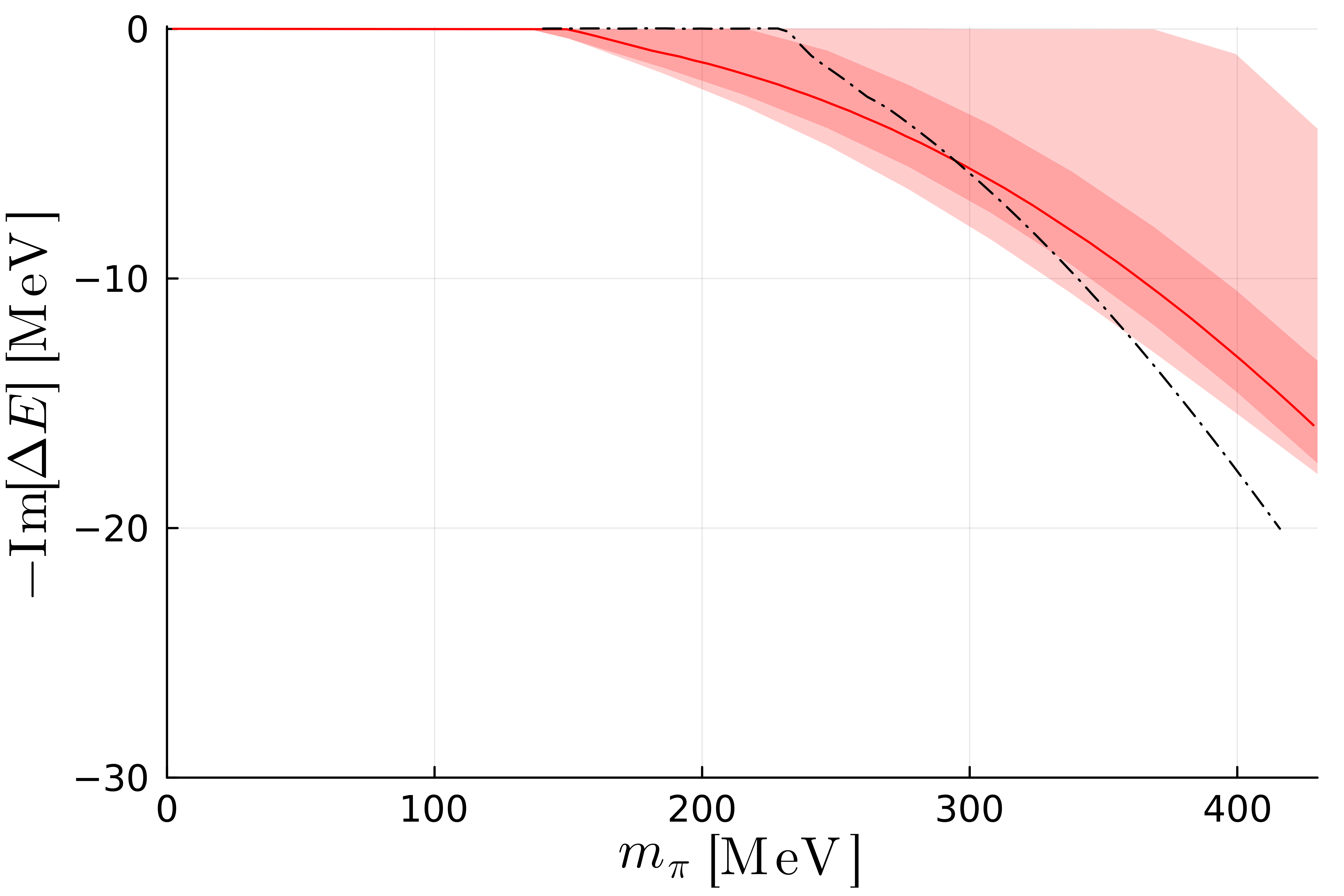}
   \end{minipage}
   \caption{Dependence of the real and imaginary parts of the pole with the pion mass for the physical charm quark mass trajectory. The error bands are calculated by including the errors of the parameters from our fit (statistical), shown in darker color, and the ones from the lattice spacing error (systematic), shown in lighter color. The dash-dotted black line corresponds to the result of Ref. \cite{Abolnikov:2024key}.}
   \label{fig:poleposition_reim}
\end{figure*}

Finally, we depict the pion mass dependence of the pole position, the real and imaginary parts, including both $\rho+\pi$-meson exchanges, in Fig.~\ref{fig:poleposition_reim}. The dashed line of the real part of the pole denotes a bound state, while the solid line stands for a virtual state. The result from Ref.~\cite{Abolnikov:2024key} is shown with a dot-dashed line. In this work, the state becomes bound around $m_\pi=100$~MeV, and it transitions from a virtual bound into a virtual resonance state around $m_\pi$$\sim 150$~MeV. The trend obtained here is similar to the one obtained in Ref. \cite{Abolnikov:2024key},\footnote{Note, however, that there are some differences between the approach followed in Ref.~\cite{Abolnikov:2024key} and here. First, in~\cite{Abolnikov:2024key}, the quark mass dependence is extracted from the data set of~\cite{Padmanath:2022cvl} at $m_\pi=280$ MeV (also included here) and the experimental point, while our analysis is based only on lattice data. Second, the interaction terms are different. In both works the pion exchange is considered (in our work only in the infinite limit case), but whereas in Ref.~\cite{Abolnikov:2024key} there is a contact plus ${\cal O}$($p^2$) term, the Weinberg-Tomozawa term explicitly accounts for the dependence on the charmed meson masses and the vector-meson exchange in the propagator. We take the $\rho$-meson quark mass dependence from Ref.~\cite{Molina:2020qpw} (one-loop NLO unitarized ChPT), and the light and heavy quark mass dependence of the low-lying charmed mesons from Ref.~\cite{Gil-Dominguez:2023eld} (one-loop HHChPT). Thus, our analysis allows for the extraction of not only the light but also the heavy quark mass dependence of $T_{cc}$.} implying that the effect of the one-pion exchange interaction in the imaginary part of the pole becomes larger as the pion mass increases. 
However, the value of the pion mass in which the state acquires an imaginary part, $m_\pi\simeq 150$~MeV, is lower in this work compared to $m_\pi=230$~MeV \cite{Abolnikov:2024key}. First, we notice that for the range of pion masses between the Padmanath24 and HSC24 data sets, 280-400~MeV, the results of both calculations are similar, the difference being more relevant for low pion masses. At low pion masses, we include the HALQCD23 data point at $146$~MeV, and we obtain a virtual bound state, as in the lattice simulation article~\cite{Lyu:2023xro}. Around this pion mass, Ref.~\cite{Abolnikov:2024key} includes the experimental result, which leads to a bound state. Thus, at these pion masses, a more attractive interaction is achieved in Ref.~\cite{Abolnikov:2024key} than here, induced by the use of the experimental point. Since the trends are very similar for higher pion masses and only differ around the physical point, we think that the most likely reason for the discrepancy is the input used. Still, the result of Ref.~\cite{Abolnikov:2024key} is close to our error band in most of the energy range. There are also other differences. Our analysis is mostly based on the quark mass dependence of the WT term, dominated by vector-meson exchange and supplemented by the pion-exchange term, as well as in a chiral extrapolation of a global fit to several LQCD simulations at different pion masses ranging from $m_\pi=146-411$~MeV. Note also that, in the recent work~\cite{Meng:2024kkp}, the lattice simulation for $DD^*$ scattering with $I = 1$ has been performed, obtaining repulsion in this sector, according to the earlier predictions of vector-meson exchange~\cite{Molina:2010tx}. More precise LQCD simulations are needed in the future to determine, with higher precision, the quark mass dependence of doubly charm mesons.

\begin{table*}
\centering
 \setlength{\tabcolsep}{0.5em}
{\renewcommand{\arraystretch}{1.6}
\begin{tabular}{|c|c|c|c|}
\hline
 \multicolumn{2}{|c|}{Collaboration} & $\Delta E_\rho$ & $\Delta E_{\rho + \pi} $  \\
 \hline
 \multicolumn{2}{|c|}{ Physical} & $-0.66 \left(^{+0.62}_{-0.82}\right) \left(^{+0.20}_{-0.95}\right)$ & $-0.48 \left(^{+0.48}_{-1.43}\right) \left(^{+0.11}_{-1.08}\right)$  \\
 \hline
\multirow{5}{*}{Padmanath24 \cite{Padmanath:2022cvl}\cite{Collins:2024sfi}} & 1 & $-7.18 \left(^{+3.72}_{-4.81}\right) \left(^{+2.67}_{-7.51}\right)$ & $-9.04 +5.67i \left(^{+2.59 - 1.83i}_{-3.37 + 1.15i}\right) \left(^{+3.45 - 3.99i}_{-5.18 + 0.89i}\right)$ \\
\cline{2-4}
 & 2 & $-3.97 \left(^{+2.60}_{-3.74}\right) \left(^{+1.31}_{-6.11}\right)$ & $-7.36 + 4.11i \left(^{+2.06 - 2.52i}_{-2.82 + 1.42i}\right) \left(^{+4.49 - 1.80i}_{-4.50 + 1.12i}\right)$ \\
\cline{2-4}
 & 3 & $-2.21 \left(^{+1.77}_{-2.95}\right) \left(^{+0.52}_{-5.08}\right)$ & $ -6.25+ 2.76i \left(^{+3.45 - 3.03i}_{-2.42 + 1.71i}\right) \left(^{+2.71 - 0.00i}_{-3.96 + 1.30i}\right)$ \\
\cline{2-4}
 & 4 & $-1.14 \left(^{+1.07}_{-2.29}\right) \left(^{+0.53}_{-4.24}\right)$ & $-5.42 +1.12i \left(^{+4.07 - 1.64i}_{-2.07 + 2.22i}\right) \left(^{+1.51 - 0.01i}_{-3.51 + 1.47i}\right)$ \\
\cline{2-4}
 & 5 & $-0.17 \left(^{+0.34}_{-1.26}\right) \left(^{+0.83}_{-2.95}\right)$ & $-2.02 \left(^{+1.80 - 0.00i}_{-3.61 + 2.20i}\right) \left(^{+0.50 - 0.00i}_{-2.78 + 1.91i}\right)$ \\
 \hline
\multicolumn{2}{|c|}{CLQCD22 \cite{Chen:2022vpo}} & $-12.44 \left(^{+5.87}_{-7.35}\right) \left(^{+5.24}_{-14.17}\right)$ & $-15.19 + 8.87i \left(^{+4.01 - 2.38i}_{-5.03 + 1.48i}\right) \left(^{+6.28 - 6.80i}_{-9.76 + 1.19i}\right)$ \\
 \hline
\multicolumn{2}{|c|}{HSC24 \cite{Whyte:2024ihh}} & $-20.86 \left(^{+8.65}_{-10.36}\right) \left(^{+9.16}_{-21.01}\right)$ & $-22.30 + 12.76i \left(^{+5.58 - 2.65i}_{-6.79 + 1.59i}\right) \left(^{+6.07 - 9.20i}_{-14.30 + 1.11i}\right)$ \\
 \hline
 \end{tabular}}
\caption{Comparison of the binding energy $\Delta E =E_{\text{pole}}-E_{\text{thr}}$ in MeV calculated for different lattice collaborations and the physical point, including or not including the OPE. All the poles found are in the second RS.}
\label{tab:poles}
\end{table*}

\section{Conclusions}\label{sec:conclusions}
 In this study, we examine data from the available LQCD simulations on $DD^*$ scattering, performing an extrapolation to the physical point using an EFT-based approach.
For the first time, the dependence of the $T_{cc}(3875)^+$ pole on both light and heavy quark masses is extracted based on a global LQCD data analysis. As expected, the mass of the $T_{cc}$ increases with the charm quark mass, the interaction being more attractive for larger charm quark masses. Contrarily, the interaction becomes less attractive when the pion mass increases, moving the pole further away from the threshold. At the physical point, the extrapolation carried out here is compatible with the experimental $T_{cc}$ mass within statistical errors. Furthermore, we investigate the role of $\rho$-meson exchange and the impact of the pion. According to our analysis, the $\rho$-meson exchange is dominant but the pion contribution is non-negligible. However, while the effect of the pion is clearly visible around the lhc in the scattering phase shifts, its small effect outside that region can be reabsorbed by slightly tuning the $DD^*\rho$ coupling. When extracting the pole position, the imaginary part of the pole caused by the pion exchange term increases with the pion mass. The real part of the pole is also affected when the pole is close to the lhc. Still, the impact of the pion exchange on the scattering phase shifts is smaller than or comparable to the statistical error of the LQCD data outside the lhc. Taking into account $\rho$- and $\pi$-meson exchanges, the pole evolves from a virtual bound to a virtual resonance state around $m_\pi=150$~MeV. The statistical and systematic uncertainties carried out in the LQCD simulations are significant in this case due to the fact that the binding energy of $T_{cc}$ is very small. More precise LQCD data are needed in order to determine the light and heavy quark mass dependence of the pole with higher accuracy. We have shown that the $T_{cc}$ pole can be generated from the $DD^*$ interaction for unphysical pion masses, consistently with the LQCD simulations and also the experimental point at the physical pion mass. The closeness of the pole to the $DD^*$ threshold for the wide range of pion masses studied here indicates its molecular nature.

\clearpage

\section{Acknowledgments}

We acknowledge useful discussions with J. Nieves, M. Pavon-Valderrama, and Pan-Pan Shi. We also thank M. Padmanath and S. Prelovsek for providing us with the data. R. M. acknowledges support from the CIDEGENT program through Grant No. CIDEGENT/2019/015 and the PROMETEU program through Grant No. CIPROM/2023/59, of the Generalitat Valenciana, and also from 
the Spanish Ministerio de Economia y Competitividad and European Union (NextGenerationEU/PRTR) through Grant No. CNS2022-13614, and from the Spanish National Grant No. PID2020-112777GB-I00. This project has received funding from the European Union’s Horizon 2020 Program No. 824093 for the STRONG-2020 project.

\appendix 

\section{The $D^* D \pi$ vertex}\label{pion-form-factor}

The $D^* D \pi$ coupling constant $g_{D^* D \pi}$ can be determined by comparing the expression of the $D^* \to D \pi$ decay width from the hidden gauge formalism (HGF) with the experimental value provided by the PDG~\cite{ParticleDataGroup:2022pth}. The experimental decay width is given by $\Gamma\left(D^{*+} \rightarrow D\pi\right) =(82.1\pm 1.8)\, \text{keV}$~\cite{ParticleDataGroup:2022pth}. We take into account a form factor in the $D^*D\pi$ vertex for an off-shell pion with four-momenta $q$ derived from QCD sum rules of exponential form, $F(q)=e^{q^2/\Lambda^2}$~\cite{Navarra:2001ju}. We set $\Lambda=1300$~MeV~\cite{Molina:2020hde,Molina:2022jcd}. When the pion is on shell, $q^2=m_\pi^2$, we obtain $g_{D^* D \pi}^{\exp} = 8.33 \pm 0.10$, which is larger than the one predicted by the HGF, $g_{D^* D \pi}^{}=\frac{m_{D^*}}{2 f_D}  \simeq 6.3 $.
In this way, the decay width obtained using the HGF, 
$\Gamma(D^{*+}\to D\pi)=\Gamma_{D^{*+}\to D^0\pi^+}+\Gamma_{D^{*0}\to D^+\pi^0}=\frac{1}{3}\frac{1}{8\pi}\frac{1}{m_{D^*}^2}4g_{D^*D\pi}^2F(q^2=m_\pi^2)^2(q^3_{\pi^+}+\frac{q^3_{\pi^0}}{2})$, 
should match the experimental one at the physical point. Thus, for the $D^*D\pi$ vertex, we take \begin{equation}g_{D^* D \pi}^{\exp}e^{q^2/\Lambda^2},\quad\Lambda=1300~\mathrm{MeV}\ .\end{equation}

\section{Polarization vectors}\label{app:polvec}

Explicitly, we use the following expressions for the polarization vectors:
\begin{align}
\label{eq:kinematics-INOUT}
   &\epsilon_0=\frac{1}{m}\begin{pmatrix} p \\  0 \\ 0 \\ E_p \end{pmatrix}\,,
   \epsilon_{\pm 1}=\frac{1}{\sqrt{2}}\begin{pmatrix} 0 \\  \mp 1 \\ -i \\ 0 \end{pmatrix}\,, \nonumber \\
   &\epsilon_0=\frac{1}{m}\begin{pmatrix} p' \\  E_{p'} \sin{\theta}\cos{\phi} \\ E_{p'} \sin{\theta}\sin{\phi} \\ E_{p'} \cos{\theta} \end{pmatrix}\,, \nonumber \\
   &\epsilon_{\pm 1}=\frac{1}{\sqrt{2}}\begin{pmatrix} 0 \\  \mp \cos{\theta}\cos{\phi}+i\sin{\phi} \\ -i\cos{\phi}\mp \cos{\theta}\sin{\phi} \\ \pm \sin{\theta} \end{pmatrix}
   \, ,
\end{align}
where $\vec{p}$  and $\vec{p}\,'$ are the three-momenta of the ingoing and the outgoing vector mesons, respectively. Here,  the ingoing three-momenta is chosen along the z axis, $\vec{p}=(0,0,p)$.
To simplify the partial-wave projection of Eq.~\eqref{eq:VmatrixPartial},  it is convenient to exploit the azimuthal symmetry, which allows one to choose a reference frame in which the three-momenta of the outgoing vector meson lies in the $xz$ plane with  $\phi=0$. 

\clearpage
\begin{widetext}
\section{On shell potential plots}\label{onshelV}
Here we present figures of the on-shell potentials for the different collaborations studied in this article. In Fig. \ref{fig:onshellV}, it can be observed how the effect of the pion in the region near the lhc causes the potential to become less attractive, as explained in the main text.
\begin{figure*}[h!]
\centering
   \begin{minipage}{0.3\textwidth}
   \centering
   physical case
     \includegraphics[width=1\linewidth]{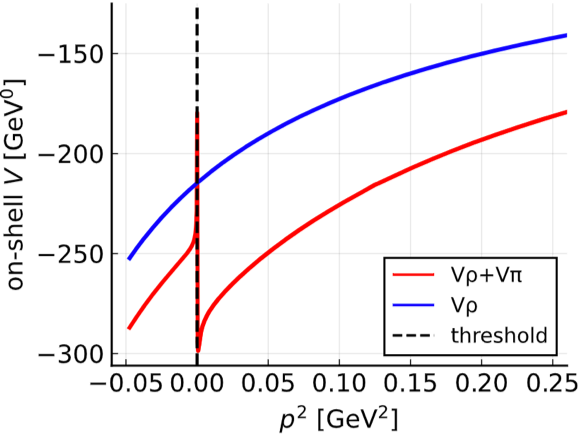}
     CLQCD22
    \includegraphics[width=1\linewidth]{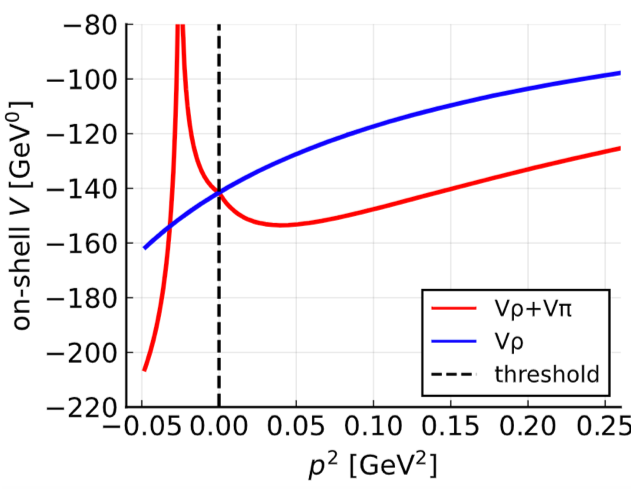}
   \end{minipage}
   \hspace{5 mm}
    \begin{minipage}{0.3\textwidth}
    \centering
    Padmanath24
	\includegraphics[width=1\linewidth]{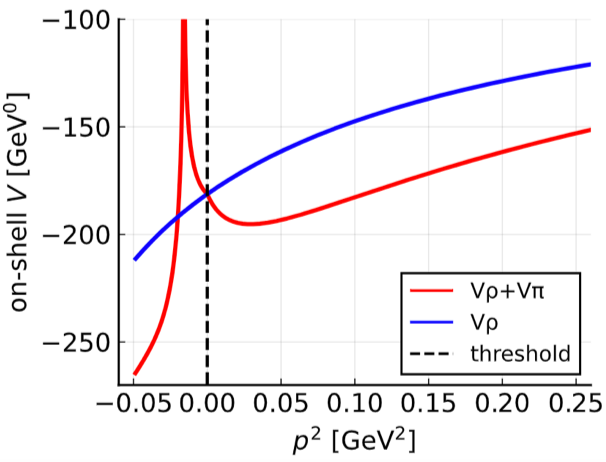}
     HSC24
   \includegraphics[width=1\linewidth]{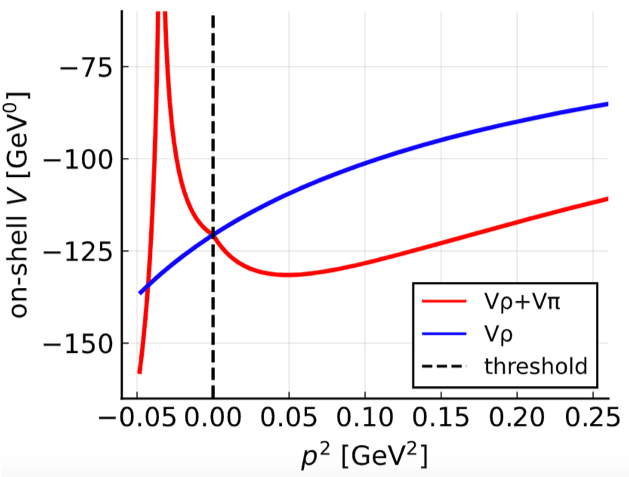}
   \end{minipage}
   \caption{On-shell potential for different pion masses.}
   \label{fig:onshellV}
\end{figure*}
\end{widetext}

\section{Higher-order effects}~\label{mpi4}

\begin{figure*}[h!]
\centering
   \begin{minipage}{0.4\textwidth}
   \centering
     \includegraphics[width=1\linewidth]{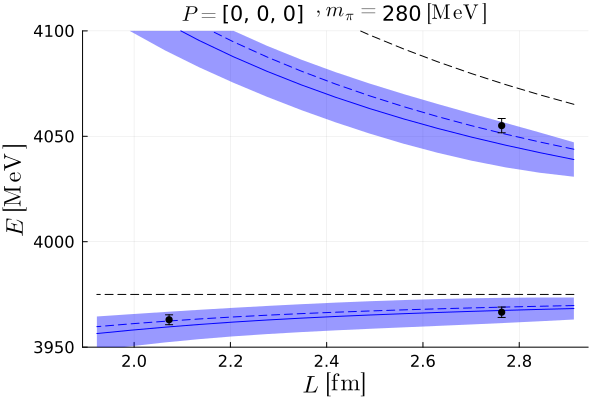}
    \includegraphics[width=1\linewidth]{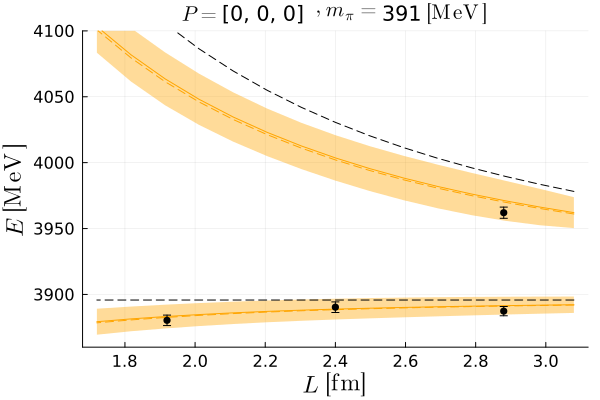}
   \end{minipage}
   \hspace{5 mm}
    \begin{minipage}{0.4\textwidth}
    \centering
	\includegraphics[width=1\linewidth]{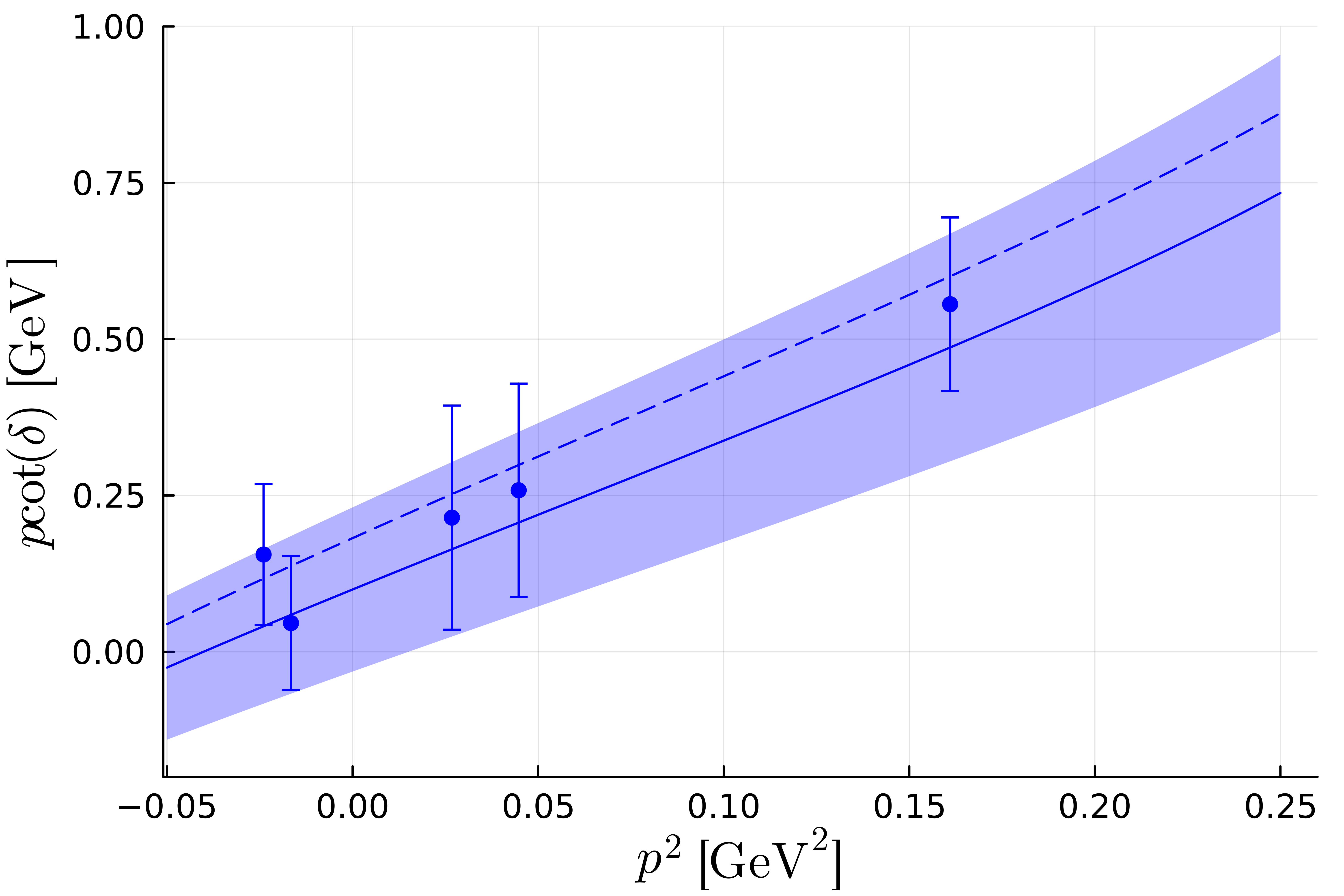}
   \includegraphics[width=1\linewidth]{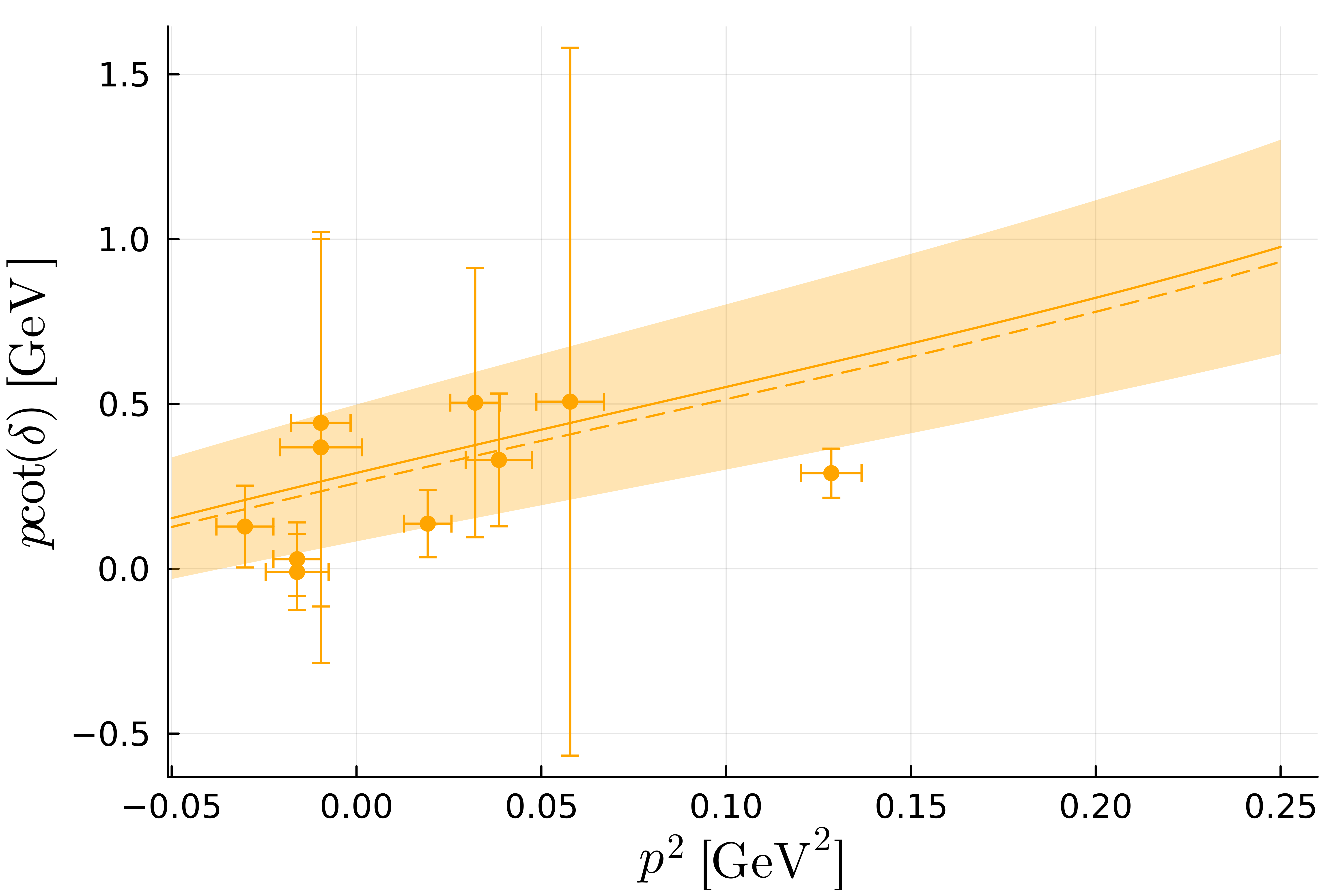}
   \end{minipage}
   \caption{Some energy levels (left) and phase shifts (right) from the Padmanath24 (top) and HSC24 (bottom) Collaborations. Solid lines and error bands represent the results of our global fit discussed in Sec. \ref{sec:gl}, while the colored dashed lines indicate the results of a fit that includes higher-order effects through Eq. (\ref{eq:mpi4}).}
   \label{fig:highorders}
\end{figure*}

Here, we check the stability of the results against higher-order terms by considering the expansion of the coupling up to ${\cal O}(m_\pi^4)$, i. e., 
\begin{equation}
g=g_0+g_2\,m_\pi^2+g_4 \,m_\pi^4\label{eq:mpi4}
\end{equation}
instead of Eq.~(\ref{g_rho}). Note that, even though the WT term is of ${\cal O}(2)$ in the momenta and in the meson masses, some effect due to ${\cal O}(4)$ is already present through the charmed meson and $\rho$-meson masses which are evaluated up to one-loop NLO~\cite{Gil-Dominguez:2023eld},~\cite{Molina:2020qpw}. We perform a new global fit using Eq.~(\ref{eq:mpi4}). The results for some of the energy levels and phase shifts are given in Fig. \ref{fig:highorders} by a dashed line, in comparison with the previous ones.  As can be seen, this term improves the result of the fit, as expected; however, the new results are inside the error bands evaluated, and therefore, the errors given in this article are of reasonable size since the calculation is reasonably stable against possible higher-order effects.

\clearpage

\bibliography{Tcc_v2}

\begin{thebibliography}{72}
\expandafter\ifx\csname natexlab\endcsname\relax\def\natexlab#1{#1}\fi
\expandafter\ifx\csname bibnamefont\endcsname\relax
  \def\bibnamefont#1{#1}\fi
\expandafter\ifx\csname bibfnamefont\endcsname\relax
  \def\bibfnamefont#1{#1}\fi
\expandafter\ifx\csname citenamefont\endcsname\relax
  \def\citenamefont#1{#1}\fi
\expandafter\ifx\csname url\endcsname\relax
  \def\url#1{\texttt{#1}}\fi
\expandafter\ifx\csname urlprefix\endcsname\relax\def\urlprefix{URL }\fi
\providecommand{\bibinfo}[2]{#2}
\providecommand{\eprint}[2][]{\url{#2}}

\bibitem[{\citenamefont{Aaij et~al.}(2022{\natexlab{a}})}]{LHCb:2021vvq}
\bibinfo{author}{\bibfnamefont{R.}~\bibnamefont{Aaij}} \bibnamefont{et~al.}
  (\bibinfo{collaboration}{LHCb}), \bibinfo{journal}{Nature Phys.}
  \textbf{\bibinfo{volume}{18}}, \bibinfo{pages}{751}
  (\bibinfo{year}{2022}{\natexlab{a}}), \eprint{2109.01038}.

\bibitem[{\citenamefont{Aaij et~al.}(2022{\natexlab{b}})}]{LHCb:2021auc}
\bibinfo{author}{\bibfnamefont{R.}~\bibnamefont{Aaij}} \bibnamefont{et~al.}
  (\bibinfo{collaboration}{LHCb}), \bibinfo{journal}{Nature Commun.}
  \textbf{\bibinfo{volume}{13}}, \bibinfo{pages}{3351}
  (\bibinfo{year}{2022}{\natexlab{b}}), \eprint{2109.01056}.

\bibitem[{\citenamefont{Janc and Rosina}(2004)}]{Janc:2004qn}
\bibinfo{author}{\bibfnamefont{D.}~\bibnamefont{Janc}} \bibnamefont{and}
  \bibinfo{author}{\bibfnamefont{M.}~\bibnamefont{Rosina}},
  \bibinfo{journal}{Few Body Syst.} \textbf{\bibinfo{volume}{35}},
  \bibinfo{pages}{175} (\bibinfo{year}{2004}), \eprint{hep-ph/0405208}.

\bibitem[{\citenamefont{Yang et~al.}(2009)\citenamefont{Yang, Deng, Ping, and
  Goldman}}]{Yang:2009zzp}
\bibinfo{author}{\bibfnamefont{Y.}~\bibnamefont{Yang}},
  \bibinfo{author}{\bibfnamefont{C.}~\bibnamefont{Deng}},
  \bibinfo{author}{\bibfnamefont{J.}~\bibnamefont{Ping}}, \bibnamefont{and}
  \bibinfo{author}{\bibfnamefont{T.}~\bibnamefont{Goldman}},
  \bibinfo{journal}{Phys. Rev. D} \textbf{\bibinfo{volume}{80}},
  \bibinfo{pages}{114023} (\bibinfo{year}{2009}).

\bibitem[{\citenamefont{Carames et~al.}(2011)\citenamefont{Carames, Valcarce,
  and Vijande}}]{Carames:2011zz}
\bibinfo{author}{\bibfnamefont{T.~F.} \bibnamefont{Carames}},
  \bibinfo{author}{\bibfnamefont{A.}~\bibnamefont{Valcarce}}, \bibnamefont{and}
  \bibinfo{author}{\bibfnamefont{J.}~\bibnamefont{Vijande}},
  \bibinfo{journal}{Phys. Lett. B} \textbf{\bibinfo{volume}{699}},
  \bibinfo{pages}{291} (\bibinfo{year}{2011}).

\bibitem[{\citenamefont{Ohkoda et~al.}(2012)\citenamefont{Ohkoda, Yamaguchi,
  Yasui, Sudoh, and Hosaka}}]{Ohkoda:2012hv}
\bibinfo{author}{\bibfnamefont{S.}~\bibnamefont{Ohkoda}},
  \bibinfo{author}{\bibfnamefont{Y.}~\bibnamefont{Yamaguchi}},
  \bibinfo{author}{\bibfnamefont{S.}~\bibnamefont{Yasui}},
  \bibinfo{author}{\bibfnamefont{K.}~\bibnamefont{Sudoh}}, \bibnamefont{and}
  \bibinfo{author}{\bibfnamefont{A.}~\bibnamefont{Hosaka}},
  \bibinfo{journal}{Phys. Rev. D} \textbf{\bibinfo{volume}{86}},
  \bibinfo{pages}{034019} (\bibinfo{year}{2012}), \eprint{1202.0760}.

\bibitem[{\citenamefont{Li et~al.}(2013)\citenamefont{Li, Sun, Liu, and
  Zhu}}]{Li:2012ss}
\bibinfo{author}{\bibfnamefont{N.}~\bibnamefont{Li}},
  \bibinfo{author}{\bibfnamefont{Z.-F.} \bibnamefont{Sun}},
  \bibinfo{author}{\bibfnamefont{X.}~\bibnamefont{Liu}}, \bibnamefont{and}
  \bibinfo{author}{\bibfnamefont{S.-L.} \bibnamefont{Zhu}},
  \bibinfo{journal}{Phys. Rev. D} \textbf{\bibinfo{volume}{88}},
  \bibinfo{pages}{114008} (\bibinfo{year}{2013}), \eprint{1211.5007}.

\bibitem[{\citenamefont{Liu et~al.}(2019)\citenamefont{Liu, Wu,
  Pavon~Valderrama, Xie, and Geng}}]{Liu:2019stu}
\bibinfo{author}{\bibfnamefont{M.-Z.} \bibnamefont{Liu}},
  \bibinfo{author}{\bibfnamefont{T.-W.} \bibnamefont{Wu}},
  \bibinfo{author}{\bibfnamefont{M.}~\bibnamefont{Pavon~Valderrama}},
  \bibinfo{author}{\bibfnamefont{J.-J.} \bibnamefont{Xie}}, \bibnamefont{and}
  \bibinfo{author}{\bibfnamefont{L.-S.} \bibnamefont{Geng}},
  \bibinfo{journal}{Phys. Rev. D} \textbf{\bibinfo{volume}{99}},
  \bibinfo{pages}{094018} (\bibinfo{year}{2019}), \eprint{1902.03044}.

\bibitem[{\citenamefont{Liu et~al.}(2020)\citenamefont{Liu, Xie, and
  Geng}}]{Liu:2020nil}
\bibinfo{author}{\bibfnamefont{M.-Z.} \bibnamefont{Liu}},
  \bibinfo{author}{\bibfnamefont{J.-J.} \bibnamefont{Xie}}, \bibnamefont{and}
  \bibinfo{author}{\bibfnamefont{L.-S.} \bibnamefont{Geng}},
  \bibinfo{journal}{Phys. Rev. D} \textbf{\bibinfo{volume}{102}},
  \bibinfo{pages}{091502} (\bibinfo{year}{2020}), \eprint{2008.07389}.

\bibitem[{\citenamefont{Ader et~al.}(1982)\citenamefont{Ader, Richard, and
  Taxil}}]{Ader:1981db}
\bibinfo{author}{\bibfnamefont{J.~P.} \bibnamefont{Ader}},
  \bibinfo{author}{\bibfnamefont{J.~M.} \bibnamefont{Richard}},
  \bibnamefont{and} \bibinfo{author}{\bibfnamefont{P.}~\bibnamefont{Taxil}},
  \bibinfo{journal}{Phys. Rev. D} \textbf{\bibinfo{volume}{25}},
  \bibinfo{pages}{2370} (\bibinfo{year}{1982}).

\bibitem[{\citenamefont{Zouzou et~al.}(1986)\citenamefont{Zouzou,
  Silvestre-Brac, Gignoux, and Richard}}]{Zouzou:1986qh}
\bibinfo{author}{\bibfnamefont{S.}~\bibnamefont{Zouzou}},
  \bibinfo{author}{\bibfnamefont{B.}~\bibnamefont{Silvestre-Brac}},
  \bibinfo{author}{\bibfnamefont{C.}~\bibnamefont{Gignoux}}, \bibnamefont{and}
  \bibinfo{author}{\bibfnamefont{J.~M.} \bibnamefont{Richard}},
  \bibinfo{journal}{Z. Phys. C} \textbf{\bibinfo{volume}{30}},
  \bibinfo{pages}{457} (\bibinfo{year}{1986}).

\bibitem[{\citenamefont{Heller and Tjon}(1987)}]{Heller:1986bt}
\bibinfo{author}{\bibfnamefont{L.}~\bibnamefont{Heller}} \bibnamefont{and}
  \bibinfo{author}{\bibfnamefont{J.~A.} \bibnamefont{Tjon}},
  \bibinfo{journal}{Phys. Rev. D} \textbf{\bibinfo{volume}{35}},
  \bibinfo{pages}{969} (\bibinfo{year}{1987}).

\bibitem[{\citenamefont{Silvestre-Brac and
  Semay}(1993)}]{Silvestre-Brac:1993zem}
\bibinfo{author}{\bibfnamefont{B.}~\bibnamefont{Silvestre-Brac}}
  \bibnamefont{and} \bibinfo{author}{\bibfnamefont{C.}~\bibnamefont{Semay}},
  \bibinfo{journal}{Z. Phys. C} \textbf{\bibinfo{volume}{57}},
  \bibinfo{pages}{273} (\bibinfo{year}{1993}).

\bibitem[{\citenamefont{Navarra et~al.}(2007)\citenamefont{Navarra, Nielsen,
  and Lee}}]{Navarra:2007yw}
\bibinfo{author}{\bibfnamefont{F.~S.} \bibnamefont{Navarra}},
  \bibinfo{author}{\bibfnamefont{M.}~\bibnamefont{Nielsen}}, \bibnamefont{and}
  \bibinfo{author}{\bibfnamefont{S.~H.} \bibnamefont{Lee}},
  \bibinfo{journal}{Phys. Lett. B} \textbf{\bibinfo{volume}{649}},
  \bibinfo{pages}{166} (\bibinfo{year}{2007}), \eprint{hep-ph/0703071}.

\bibitem[{\citenamefont{Ebert et~al.}(2007)\citenamefont{Ebert, Faustov,
  Galkin, and Lucha}}]{Ebert:2007rn}
\bibinfo{author}{\bibfnamefont{D.}~\bibnamefont{Ebert}},
  \bibinfo{author}{\bibfnamefont{R.~N.} \bibnamefont{Faustov}},
  \bibinfo{author}{\bibfnamefont{V.~O.} \bibnamefont{Galkin}},
  \bibnamefont{and} \bibinfo{author}{\bibfnamefont{W.}~\bibnamefont{Lucha}},
  \bibinfo{journal}{Phys. Rev. D} \textbf{\bibinfo{volume}{76}},
  \bibinfo{pages}{114015} (\bibinfo{year}{2007}), \eprint{0706.3853}.

\bibitem[{\citenamefont{Karliner and Rosner}(2017)}]{Karliner:2017qjm}
\bibinfo{author}{\bibfnamefont{M.}~\bibnamefont{Karliner}} \bibnamefont{and}
  \bibinfo{author}{\bibfnamefont{J.~L.} \bibnamefont{Rosner}},
  \bibinfo{journal}{Phys. Rev. Lett.} \textbf{\bibinfo{volume}{119}},
  \bibinfo{pages}{202001} (\bibinfo{year}{2017}), \eprint{1707.07666}.

\bibitem[{\citenamefont{Yang et~al.}(2020)\citenamefont{Yang, Ping, and
  Segovia}}]{Yang:2019itm}
\bibinfo{author}{\bibfnamefont{G.}~\bibnamefont{Yang}},
  \bibinfo{author}{\bibfnamefont{J.}~\bibnamefont{Ping}}, \bibnamefont{and}
  \bibinfo{author}{\bibfnamefont{J.}~\bibnamefont{Segovia}},
  \bibinfo{journal}{Phys. Rev. D} \textbf{\bibinfo{volume}{101}},
  \bibinfo{pages}{014001} (\bibinfo{year}{2020}), \eprint{1911.00215}.

\bibitem[{\citenamefont{Tang et~al.}(2020)\citenamefont{Tang, Wan, Maltman, and
  Qiao}}]{Tang:2019nwv}
\bibinfo{author}{\bibfnamefont{L.}~\bibnamefont{Tang}},
  \bibinfo{author}{\bibfnamefont{B.-D.} \bibnamefont{Wan}},
  \bibinfo{author}{\bibfnamefont{K.}~\bibnamefont{Maltman}}, \bibnamefont{and}
  \bibinfo{author}{\bibfnamefont{C.-F.} \bibnamefont{Qiao}},
  \bibinfo{journal}{Phys. Rev. D} \textbf{\bibinfo{volume}{101}},
  \bibinfo{pages}{094032} (\bibinfo{year}{2020}), \eprint{1911.10951}.

\bibitem[{\citenamefont{Dong et~al.}(2021)\citenamefont{Dong, Guo, and
  Zou}}]{Dong:2021bvy}
\bibinfo{author}{\bibfnamefont{X.-K.} \bibnamefont{Dong}},
  \bibinfo{author}{\bibfnamefont{F.-K.} \bibnamefont{Guo}}, \bibnamefont{and}
  \bibinfo{author}{\bibfnamefont{B.-S.} \bibnamefont{Zou}},
  \bibinfo{journal}{Commun. Theor. Phys.} \textbf{\bibinfo{volume}{73}},
  \bibinfo{pages}{125201} (\bibinfo{year}{2021}), \eprint{2108.02673}.

\bibitem[{\citenamefont{Chen et~al.}(2023)\citenamefont{Chen, Chen, Liu, Liu,
  and Zhu}}]{Chen:2022asf}
\bibinfo{author}{\bibfnamefont{H.-X.} \bibnamefont{Chen}},
  \bibinfo{author}{\bibfnamefont{W.}~\bibnamefont{Chen}},
  \bibinfo{author}{\bibfnamefont{X.}~\bibnamefont{Liu}},
  \bibinfo{author}{\bibfnamefont{Y.-R.} \bibnamefont{Liu}}, \bibnamefont{and}
  \bibinfo{author}{\bibfnamefont{S.-L.} \bibnamefont{Zhu}},
  \bibinfo{journal}{Rept. Prog. Phys.} \textbf{\bibinfo{volume}{86}},
  \bibinfo{pages}{026201} (\bibinfo{year}{2023}), \eprint{2204.02649}.

\bibitem[{\citenamefont{Choi et~al.}(2003)}]{Belle:2003nnu}
\bibinfo{author}{\bibfnamefont{S.~K.} \bibnamefont{Choi}} \bibnamefont{et~al.}
  (\bibinfo{collaboration}{Belle}), \bibinfo{journal}{Phys. Rev. Lett.}
  \textbf{\bibinfo{volume}{91}}, \bibinfo{pages}{262001}
  (\bibinfo{year}{2003}), \eprint{hep-ex/0309032}.

\bibitem[{\citenamefont{Wang et~al.}(2023)\citenamefont{Wang, Yang, Wu, Oka,
  and Zhu}}]{Wang:2023ovj}
\bibinfo{author}{\bibfnamefont{G.-J.} \bibnamefont{Wang}},
  \bibinfo{author}{\bibfnamefont{Z.}~\bibnamefont{Yang}},
  \bibinfo{author}{\bibfnamefont{J.-J.} \bibnamefont{Wu}},
  \bibinfo{author}{\bibfnamefont{M.}~\bibnamefont{Oka}}, \bibnamefont{and}
  \bibinfo{author}{\bibfnamefont{S.-L.} \bibnamefont{Zhu}}
  (\bibinfo{year}{2023}), \eprint{2306.12406}.

\bibitem[{\citenamefont{Padmanath and Prelovsek}(2022)}]{Padmanath:2022cvl}
\bibinfo{author}{\bibfnamefont{M.}~\bibnamefont{Padmanath}} \bibnamefont{and}
  \bibinfo{author}{\bibfnamefont{S.}~\bibnamefont{Prelovsek}},
  \bibinfo{journal}{Phys. Rev. Lett.} \textbf{\bibinfo{volume}{129}},
  \bibinfo{pages}{032002} (\bibinfo{year}{2022}), \eprint{2202.10110}.

\bibitem[{\citenamefont{Chen et~al.}(2022)\citenamefont{Chen, Shi, Chen, Gong,
  Liu, Sun, and Zhang}}]{Chen:2022vpo}
\bibinfo{author}{\bibfnamefont{S.}~\bibnamefont{Chen}},
  \bibinfo{author}{\bibfnamefont{C.}~\bibnamefont{Shi}},
  \bibinfo{author}{\bibfnamefont{Y.}~\bibnamefont{Chen}},
  \bibinfo{author}{\bibfnamefont{M.}~\bibnamefont{Gong}},
  \bibinfo{author}{\bibfnamefont{Z.}~\bibnamefont{Liu}},
  \bibinfo{author}{\bibfnamefont{W.}~\bibnamefont{Sun}}, \bibnamefont{and}
  \bibinfo{author}{\bibfnamefont{R.}~\bibnamefont{Zhang}},
  \bibinfo{journal}{Phys. Lett. B} \textbf{\bibinfo{volume}{833}},
  \bibinfo{pages}{137391} (\bibinfo{year}{2022}), \eprint{2206.06185}.

\bibitem[{\citenamefont{Lyu et~al.}(2023)\citenamefont{Lyu, Aoki, Doi, Hatsuda,
  Ikeda, and Meng}}]{Lyu:2023xro}
\bibinfo{author}{\bibfnamefont{Y.}~\bibnamefont{Lyu}},
  \bibinfo{author}{\bibfnamefont{S.}~\bibnamefont{Aoki}},
  \bibinfo{author}{\bibfnamefont{T.}~\bibnamefont{Doi}},
  \bibinfo{author}{\bibfnamefont{T.}~\bibnamefont{Hatsuda}},
  \bibinfo{author}{\bibfnamefont{Y.}~\bibnamefont{Ikeda}}, \bibnamefont{and}
  \bibinfo{author}{\bibfnamefont{J.}~\bibnamefont{Meng}},
  \bibinfo{journal}{Phys. Rev. Lett.} \textbf{\bibinfo{volume}{131}},
  \bibinfo{pages}{161901} (\bibinfo{year}{2023}), \eprint{2302.04505}.

\bibitem[{\citenamefont{Collins et~al.}(2024)\citenamefont{Collins, Nefediev,
  Padmanath, and Prelovsek}}]{Collins:2024sfi}
\bibinfo{author}{\bibfnamefont{S.}~\bibnamefont{Collins}},
  \bibinfo{author}{\bibfnamefont{A.}~\bibnamefont{Nefediev}},
  \bibinfo{author}{\bibfnamefont{M.}~\bibnamefont{Padmanath}},
  \bibnamefont{and}
  \bibinfo{author}{\bibfnamefont{S.}~\bibnamefont{Prelovsek}},
  \bibinfo{journal}{Phys. Rev. D} \textbf{\bibinfo{volume}{109}},
  \bibinfo{pages}{094509} (\bibinfo{year}{2024}), \eprint{2402.14715}.

\bibitem[{\citenamefont{Whyte et~al.}(2024)\citenamefont{Whyte, Wilson, and
  Thomas}}]{Whyte:2024ihh}
\bibinfo{author}{\bibfnamefont{T.}~\bibnamefont{Whyte}},
  \bibinfo{author}{\bibfnamefont{D.~J.} \bibnamefont{Wilson}},
  \bibnamefont{and} \bibinfo{author}{\bibfnamefont{C.~E.} \bibnamefont{Thomas}}
  (\bibinfo{year}{2024}), \eprint{2405.15741}.

\bibitem[{\citenamefont{Feijoo et~al.}(2021)\citenamefont{Feijoo, Liang, and
  Oset}}]{Feijoo:2021ppq}
\bibinfo{author}{\bibfnamefont{A.}~\bibnamefont{Feijoo}},
  \bibinfo{author}{\bibfnamefont{W.~H.} \bibnamefont{Liang}}, \bibnamefont{and}
  \bibinfo{author}{\bibfnamefont{E.}~\bibnamefont{Oset}},
  \bibinfo{journal}{Phys. Rev. D} \textbf{\bibinfo{volume}{104}},
  \bibinfo{pages}{114015} (\bibinfo{year}{2021}), \eprint{2108.02730}.

\bibitem[{\citenamefont{Albaladejo}(2022)}]{Albaladejo:2021vln}
\bibinfo{author}{\bibfnamefont{M.}~\bibnamefont{Albaladejo}},
  \bibinfo{journal}{Phys. Lett. B} \textbf{\bibinfo{volume}{829}},
  \bibinfo{pages}{137052} (\bibinfo{year}{2022}), \eprint{2110.02944}.

\bibitem[{\citenamefont{Ling et~al.}(2022)\citenamefont{Ling, Liu, Geng, Wang,
  and Xie}}]{Ling:2021bir}
\bibinfo{author}{\bibfnamefont{X.-Z.} \bibnamefont{Ling}},
  \bibinfo{author}{\bibfnamefont{M.-Z.} \bibnamefont{Liu}},
  \bibinfo{author}{\bibfnamefont{L.-S.} \bibnamefont{Geng}},
  \bibinfo{author}{\bibfnamefont{E.}~\bibnamefont{Wang}}, \bibnamefont{and}
  \bibinfo{author}{\bibfnamefont{J.-J.} \bibnamefont{Xie}},
  \bibinfo{journal}{Phys. Lett. B} \textbf{\bibinfo{volume}{826}},
  \bibinfo{pages}{136897} (\bibinfo{year}{2022}), \eprint{2108.00947}.

\bibitem[{\citenamefont{Du et~al.}(2022)\citenamefont{Du, Baru, Dong, Filin,
  Guo, Hanhart, Nefediev, Nieves, and Wang}}]{Du:2021zzh}
\bibinfo{author}{\bibfnamefont{M.-L.} \bibnamefont{Du}},
  \bibinfo{author}{\bibfnamefont{V.}~\bibnamefont{Baru}},
  \bibinfo{author}{\bibfnamefont{X.-K.} \bibnamefont{Dong}},
  \bibinfo{author}{\bibfnamefont{A.}~\bibnamefont{Filin}},
  \bibinfo{author}{\bibfnamefont{F.-K.} \bibnamefont{Guo}},
  \bibinfo{author}{\bibfnamefont{C.}~\bibnamefont{Hanhart}},
  \bibinfo{author}{\bibfnamefont{A.}~\bibnamefont{Nefediev}},
  \bibinfo{author}{\bibfnamefont{J.}~\bibnamefont{Nieves}}, \bibnamefont{and}
  \bibinfo{author}{\bibfnamefont{Q.}~\bibnamefont{Wang}},
  \bibinfo{journal}{Phys. Rev. D} \textbf{\bibinfo{volume}{105}},
  \bibinfo{pages}{014024} (\bibinfo{year}{2022}), \eprint{2110.13765}.

\bibitem[{\citenamefont{Aaron et~al.}(1968)\citenamefont{Aaron, Amado, and
  Young}}]{Aaron:1968aoz}
\bibinfo{author}{\bibfnamefont{R.}~\bibnamefont{Aaron}},
  \bibinfo{author}{\bibfnamefont{R.~D.} \bibnamefont{Amado}}, \bibnamefont{and}
  \bibinfo{author}{\bibfnamefont{J.~E.} \bibnamefont{Young}},
  \bibinfo{journal}{Phys. Rev.} \textbf{\bibinfo{volume}{174}},
  \bibinfo{pages}{2022} (\bibinfo{year}{1968}).

\bibitem[{\citenamefont{Mai et~al.}(2017)\citenamefont{Mai, Hu, Doring,
  Pilloni, and Szczepaniak}}]{Mai:2017vot}
\bibinfo{author}{\bibfnamefont{M.}~\bibnamefont{Mai}},
  \bibinfo{author}{\bibfnamefont{B.}~\bibnamefont{Hu}},
  \bibinfo{author}{\bibfnamefont{M.}~\bibnamefont{Doring}},
  \bibinfo{author}{\bibfnamefont{A.}~\bibnamefont{Pilloni}}, \bibnamefont{and}
  \bibinfo{author}{\bibfnamefont{A.}~\bibnamefont{Szczepaniak}},
  \bibinfo{journal}{Eur. Phys. J. A} \textbf{\bibinfo{volume}{53}},
  \bibinfo{pages}{177} (\bibinfo{year}{2017}), \eprint{1706.06118}.

\bibitem[{\citenamefont{Dai et~al.}(2023)\citenamefont{Dai, Fleming, Hodges,
  and Mehen}}]{Dai:2023mxm}
\bibinfo{author}{\bibfnamefont{L.}~\bibnamefont{Dai}},
  \bibinfo{author}{\bibfnamefont{S.}~\bibnamefont{Fleming}},
  \bibinfo{author}{\bibfnamefont{R.}~\bibnamefont{Hodges}}, \bibnamefont{and}
  \bibinfo{author}{\bibfnamefont{T.}~\bibnamefont{Mehen}},
  \bibinfo{journal}{Phys. Rev. D} \textbf{\bibinfo{volume}{107}},
  \bibinfo{pages}{076001} (\bibinfo{year}{2023}), \eprint{2301.11950}.

\bibitem[{\citenamefont{Fleming et~al.}(2007)\citenamefont{Fleming, Kusunoki,
  Mehen, and van Kolck}}]{Fleming:2007rp}
\bibinfo{author}{\bibfnamefont{S.}~\bibnamefont{Fleming}},
  \bibinfo{author}{\bibfnamefont{M.}~\bibnamefont{Kusunoki}},
  \bibinfo{author}{\bibfnamefont{T.}~\bibnamefont{Mehen}}, \bibnamefont{and}
  \bibinfo{author}{\bibfnamefont{U.}~\bibnamefont{van Kolck}},
  \bibinfo{journal}{Phys. Rev. D} \textbf{\bibinfo{volume}{76}},
  \bibinfo{pages}{034006} (\bibinfo{year}{2007}), \eprint{hep-ph/0703168}.

\bibitem[{\citenamefont{Dai et~al.}(2020)\citenamefont{Dai, Guo, and
  Mehen}}]{Dai:2019hrf}
\bibinfo{author}{\bibfnamefont{L.}~\bibnamefont{Dai}},
  \bibinfo{author}{\bibfnamefont{F.-K.} \bibnamefont{Guo}}, \bibnamefont{and}
  \bibinfo{author}{\bibfnamefont{T.}~\bibnamefont{Mehen}},
  \bibinfo{journal}{Phys. Rev. D} \textbf{\bibinfo{volume}{101}},
  \bibinfo{pages}{054024} (\bibinfo{year}{2020}), \eprint{1912.04317}.

\bibitem[{\citenamefont{Luscher}(1986)}]{Luscher:1986pf}
\bibinfo{author}{\bibfnamefont{M.}~\bibnamefont{Luscher}},
  \bibinfo{journal}{Commun. Math. Phys.} \textbf{\bibinfo{volume}{105}},
  \bibinfo{pages}{153} (\bibinfo{year}{1986}).

\bibitem[{\citenamefont{Luscher}(1991)}]{Luscher:1990ux}
\bibinfo{author}{\bibfnamefont{M.}~\bibnamefont{Luscher}},
  \bibinfo{journal}{Nucl. Phys. B} \textbf{\bibinfo{volume}{354}},
  \bibinfo{pages}{531} (\bibinfo{year}{1991}).

\bibitem[{\citenamefont{Du et~al.}(2023)\citenamefont{Du, Filin, Baru, Dong,
  Epelbaum, Guo, Hanhart, Nefediev, Nieves, and Wang}}]{Du:2023hlu}
\bibinfo{author}{\bibfnamefont{M.-L.} \bibnamefont{Du}},
  \bibinfo{author}{\bibfnamefont{A.}~\bibnamefont{Filin}},
  \bibinfo{author}{\bibfnamefont{V.}~\bibnamefont{Baru}},
  \bibinfo{author}{\bibfnamefont{X.-K.} \bibnamefont{Dong}},
  \bibinfo{author}{\bibfnamefont{E.}~\bibnamefont{Epelbaum}},
  \bibinfo{author}{\bibfnamefont{F.-K.} \bibnamefont{Guo}},
  \bibinfo{author}{\bibfnamefont{C.}~\bibnamefont{Hanhart}},
  \bibinfo{author}{\bibfnamefont{A.}~\bibnamefont{Nefediev}},
  \bibinfo{author}{\bibfnamefont{J.}~\bibnamefont{Nieves}}, \bibnamefont{and}
  \bibinfo{author}{\bibfnamefont{Q.}~\bibnamefont{Wang}},
  \bibinfo{journal}{Phys. Rev. Lett.} \textbf{\bibinfo{volume}{131}},
  \bibinfo{pages}{131903} (\bibinfo{year}{2023}), \eprint{2303.09441}.

\bibitem[{\citenamefont{Meng et~al.}(2024{\natexlab{a}})\citenamefont{Meng,
  Baru, Epelbaum, Filin, and Gasparyan}}]{Meng:2023bmz}
\bibinfo{author}{\bibfnamefont{L.}~\bibnamefont{Meng}},
  \bibinfo{author}{\bibfnamefont{V.}~\bibnamefont{Baru}},
  \bibinfo{author}{\bibfnamefont{E.}~\bibnamefont{Epelbaum}},
  \bibinfo{author}{\bibfnamefont{A.~A.} \bibnamefont{Filin}}, \bibnamefont{and}
  \bibinfo{author}{\bibfnamefont{A.~M.} \bibnamefont{Gasparyan}},
  \bibinfo{journal}{Phys. Rev. D} \textbf{\bibinfo{volume}{109}},
  \bibinfo{pages}{L071506} (\bibinfo{year}{2024}{\natexlab{a}}),
  \eprint{2312.01930}.

\bibitem[{\citenamefont{Raposo and Hansen}(2024)}]{Raposo:2023oru}
\bibinfo{author}{\bibfnamefont{A.~B.~a.} \bibnamefont{Raposo}}
  \bibnamefont{and} \bibinfo{author}{\bibfnamefont{M.~T.}
  \bibnamefont{Hansen}}, \bibinfo{journal}{JHEP} \textbf{\bibinfo{volume}{08}},
  \bibinfo{pages}{075} (\bibinfo{year}{2024}), \eprint{2311.18793}.

\bibitem[{\citenamefont{Bubna et~al.}(2024)\citenamefont{Bubna, Hammer,
  M\"uller, Pang, Rusetsky, and Wu}}]{Bubna:2024izx}
\bibinfo{author}{\bibfnamefont{R.}~\bibnamefont{Bubna}},
  \bibinfo{author}{\bibfnamefont{H.-W.} \bibnamefont{Hammer}},
  \bibinfo{author}{\bibfnamefont{F.}~\bibnamefont{M\"uller}},
  \bibinfo{author}{\bibfnamefont{J.-Y.} \bibnamefont{Pang}},
  \bibinfo{author}{\bibfnamefont{A.}~\bibnamefont{Rusetsky}}, \bibnamefont{and}
  \bibinfo{author}{\bibfnamefont{J.-J.} \bibnamefont{Wu}},
  \bibinfo{journal}{JHEP} \textbf{\bibinfo{volume}{05}}, \bibinfo{pages}{168}
  (\bibinfo{year}{2024}), \eprint{2402.12985}.

\bibitem[{\citenamefont{Hansen et~al.}(2024)\citenamefont{Hansen,
  Romero-L\'opez, and Sharpe}}]{Hansen:2024ffk}
\bibinfo{author}{\bibfnamefont{M.~T.} \bibnamefont{Hansen}},
  \bibinfo{author}{\bibfnamefont{F.}~\bibnamefont{Romero-L\'opez}},
  \bibnamefont{and} \bibinfo{author}{\bibfnamefont{S.~R.}
  \bibnamefont{Sharpe}}, \bibinfo{journal}{JHEP} \textbf{\bibinfo{volume}{06}},
  \bibinfo{pages}{051} (\bibinfo{year}{2024}), \eprint{2401.06609}.

\bibitem[{\citenamefont{Du et~al.}(2024)\citenamefont{Du, Guo, and
  Wu}}]{Du:2024snq}
\bibinfo{author}{\bibfnamefont{M.-L.} \bibnamefont{Du}},
  \bibinfo{author}{\bibfnamefont{F.-K.} \bibnamefont{Guo}}, \bibnamefont{and}
  \bibinfo{author}{\bibfnamefont{B.}~\bibnamefont{Wu}} (\bibinfo{year}{2024}),
  \eprint{2408.09375}.

\bibitem[{\citenamefont{Abolnikov et~al.}(2024)\citenamefont{Abolnikov, Baru,
  Epelbaum, Filin, Hanhart, and Meng}}]{Abolnikov:2024key}
\bibinfo{author}{\bibfnamefont{M.}~\bibnamefont{Abolnikov}},
  \bibinfo{author}{\bibfnamefont{V.}~\bibnamefont{Baru}},
  \bibinfo{author}{\bibfnamefont{E.}~\bibnamefont{Epelbaum}},
  \bibinfo{author}{\bibfnamefont{A.~A.} \bibnamefont{Filin}},
  \bibinfo{author}{\bibfnamefont{C.}~\bibnamefont{Hanhart}}, \bibnamefont{and}
  \bibinfo{author}{\bibfnamefont{L.}~\bibnamefont{Meng}}
  (\bibinfo{year}{2024}), \eprint{2407.04649}.

\bibitem[{\citenamefont{Molina et~al.}(2010)\citenamefont{Molina, Branz, and
  Oset}}]{Molina:2010tx}
\bibinfo{author}{\bibfnamefont{R.}~\bibnamefont{Molina}},
  \bibinfo{author}{\bibfnamefont{T.}~\bibnamefont{Branz}}, \bibnamefont{and}
  \bibinfo{author}{\bibfnamefont{E.}~\bibnamefont{Oset}},
  \bibinfo{journal}{Phys. Rev. D} \textbf{\bibinfo{volume}{82}},
  \bibinfo{pages}{014010} (\bibinfo{year}{2010}), \eprint{1005.0335}.

\bibitem[{\citenamefont{Dai et~al.}(2022)\citenamefont{Dai, Molina, and
  Oset}}]{Dai:2021vgf}
\bibinfo{author}{\bibfnamefont{L.~R.} \bibnamefont{Dai}},
  \bibinfo{author}{\bibfnamefont{R.}~\bibnamefont{Molina}}, \bibnamefont{and}
  \bibinfo{author}{\bibfnamefont{E.}~\bibnamefont{Oset}},
  \bibinfo{journal}{Phys. Rev. D} \textbf{\bibinfo{volume}{105}},
  \bibinfo{pages}{016029} (\bibinfo{year}{2022}), \bibinfo{note}{[Erratum:
  Phys.Rev.D 106, 099902 (2022)]}, \eprint{2110.15270}.

\bibitem[{\citenamefont{Bando et~al.}(1988)\citenamefont{Bando, Kugo, and
  Yamawaki}}]{Bando:1987br}
\bibinfo{author}{\bibfnamefont{M.}~\bibnamefont{Bando}},
  \bibinfo{author}{\bibfnamefont{T.}~\bibnamefont{Kugo}}, \bibnamefont{and}
  \bibinfo{author}{\bibfnamefont{K.}~\bibnamefont{Yamawaki}},
  \bibinfo{journal}{Phys. Rept.} \textbf{\bibinfo{volume}{164}},
  \bibinfo{pages}{217} (\bibinfo{year}{1988}).

\bibitem[{\citenamefont{Harada and Yamawaki}(2003)}]{Harada:2003jx}
\bibinfo{author}{\bibfnamefont{M.}~\bibnamefont{Harada}} \bibnamefont{and}
  \bibinfo{author}{\bibfnamefont{K.}~\bibnamefont{Yamawaki}},
  \bibinfo{journal}{Phys. Rept.} \textbf{\bibinfo{volume}{381}},
  \bibinfo{pages}{1} (\bibinfo{year}{2003}), \eprint{hep-ph/0302103}.

\bibitem[{\citenamefont{Meissner}(1988)}]{Meissner:1987ge}
\bibinfo{author}{\bibfnamefont{U.~G.} \bibnamefont{Meissner}},
  \bibinfo{journal}{Phys. Rept.} \textbf{\bibinfo{volume}{161}},
  \bibinfo{pages}{213} (\bibinfo{year}{1988}).

\bibitem[{\citenamefont{Nagahiro et~al.}(2009)\citenamefont{Nagahiro, Roca,
  Hosaka, and Oset}}]{Nagahiro:2008cv}
\bibinfo{author}{\bibfnamefont{H.}~\bibnamefont{Nagahiro}},
  \bibinfo{author}{\bibfnamefont{L.}~\bibnamefont{Roca}},
  \bibinfo{author}{\bibfnamefont{A.}~\bibnamefont{Hosaka}}, \bibnamefont{and}
  \bibinfo{author}{\bibfnamefont{E.}~\bibnamefont{Oset}},
  \bibinfo{journal}{Phys. Rev. D} \textbf{\bibinfo{volume}{79}},
  \bibinfo{pages}{014015} (\bibinfo{year}{2009}), \eprint{0809.0943}.

\bibitem[{\citenamefont{Molina et~al.}(2009)\citenamefont{Molina, Nagahiro,
  Hosaka, and Oset}}]{Molina:2009eb}
\bibinfo{author}{\bibfnamefont{R.}~\bibnamefont{Molina}},
  \bibinfo{author}{\bibfnamefont{H.}~\bibnamefont{Nagahiro}},
  \bibinfo{author}{\bibfnamefont{A.}~\bibnamefont{Hosaka}}, \bibnamefont{and}
  \bibinfo{author}{\bibfnamefont{E.}~\bibnamefont{Oset}},
  \bibinfo{journal}{Phys. Rev. D} \textbf{\bibinfo{volume}{80}},
  \bibinfo{pages}{014025} (\bibinfo{year}{2009}), \eprint{0903.3823}.

\bibitem[{\citenamefont{Molina and Oset}(2009)}]{Molina:2009ct}
\bibinfo{author}{\bibfnamefont{R.}~\bibnamefont{Molina}} \bibnamefont{and}
  \bibinfo{author}{\bibfnamefont{E.}~\bibnamefont{Oset}},
  \bibinfo{journal}{Phys. Rev. D} \textbf{\bibinfo{volume}{80}},
  \bibinfo{pages}{114013} (\bibinfo{year}{2009}), \eprint{0907.3043}.

\bibitem[{\citenamefont{Xiao et~al.}(2013)\citenamefont{Xiao, Nieves, and
  Oset}}]{Xiao:2013yca}
\bibinfo{author}{\bibfnamefont{C.~W.} \bibnamefont{Xiao}},
  \bibinfo{author}{\bibfnamefont{J.}~\bibnamefont{Nieves}}, \bibnamefont{and}
  \bibinfo{author}{\bibfnamefont{E.}~\bibnamefont{Oset}},
  \bibinfo{journal}{Phys. Rev. D} \textbf{\bibinfo{volume}{88}},
  \bibinfo{pages}{056012} (\bibinfo{year}{2013}), \eprint{1304.5368}.

\bibitem[{\citenamefont{Molina and Oset}(2020)}]{Molina:2020hde}
\bibinfo{author}{\bibfnamefont{R.}~\bibnamefont{Molina}} \bibnamefont{and}
  \bibinfo{author}{\bibfnamefont{E.}~\bibnamefont{Oset}},
  \bibinfo{journal}{Phys. Lett. B} \textbf{\bibinfo{volume}{811}},
  \bibinfo{pages}{135870} (\bibinfo{year}{2020}), \bibinfo{note}{[Erratum:
  Phys.Lett.B 837, 137645 (2023)]}, \eprint{2008.11171}.

\bibitem[{\citenamefont{Molina and Oset}(2023)}]{Molina:2022jcd}
\bibinfo{author}{\bibfnamefont{R.}~\bibnamefont{Molina}} \bibnamefont{and}
  \bibinfo{author}{\bibfnamefont{E.}~\bibnamefont{Oset}},
  \bibinfo{journal}{Phys. Rev. D} \textbf{\bibinfo{volume}{107}},
  \bibinfo{pages}{056015} (\bibinfo{year}{2023}), \eprint{2211.01302}.

\bibitem[{\citenamefont{Sadasivan et~al.}(2022)\citenamefont{Sadasivan,
  Alexandru, Akdag, Amorim, Brett, Culver, D\"oring, Lee, and
  Mai}}]{Sadasivan:2021emk}
\bibinfo{author}{\bibfnamefont{D.}~\bibnamefont{Sadasivan}},
  \bibinfo{author}{\bibfnamefont{A.}~\bibnamefont{Alexandru}},
  \bibinfo{author}{\bibfnamefont{H.}~\bibnamefont{Akdag}},
  \bibinfo{author}{\bibfnamefont{F.}~\bibnamefont{Amorim}},
  \bibinfo{author}{\bibfnamefont{R.}~\bibnamefont{Brett}},
  \bibinfo{author}{\bibfnamefont{C.}~\bibnamefont{Culver}},
  \bibinfo{author}{\bibfnamefont{M.}~\bibnamefont{D\"oring}},
  \bibinfo{author}{\bibfnamefont{F.~X.} \bibnamefont{Lee}}, \bibnamefont{and}
  \bibinfo{author}{\bibfnamefont{M.}~\bibnamefont{Mai}},
  \bibinfo{journal}{Phys. Rev. D} \textbf{\bibinfo{volume}{105}},
  \bibinfo{pages}{054020} (\bibinfo{year}{2022}), \eprint{2112.03355}.

\bibitem[{\citenamefont{Chung}(1971)}]{Chung:1971ri}
\bibinfo{author}{\bibfnamefont{S.~U.} \bibnamefont{Chung}}
  (\bibinfo{year}{1971}).

\bibitem[{\citenamefont{Bayar et~al.}(2024)\citenamefont{Bayar, Molina, Oset,
  Liu, and Geng}}]{Bayar:2023azy}
\bibinfo{author}{\bibfnamefont{M.}~\bibnamefont{Bayar}},
  \bibinfo{author}{\bibfnamefont{R.}~\bibnamefont{Molina}},
  \bibinfo{author}{\bibfnamefont{E.}~\bibnamefont{Oset}},
  \bibinfo{author}{\bibfnamefont{M.-Z.} \bibnamefont{Liu}}, \bibnamefont{and}
  \bibinfo{author}{\bibfnamefont{L.-S.} \bibnamefont{Geng}},
  \bibinfo{journal}{Phys. Rev. D} \textbf{\bibinfo{volume}{109}},
  \bibinfo{pages}{076027} (\bibinfo{year}{2024}), \eprint{2312.12004}.

\bibitem[{\citenamefont{Workman et~al.}(2022)}]{ParticleDataGroup:2022pth}
\bibinfo{author}{\bibfnamefont{R.~L.} \bibnamefont{Workman}}
  \bibnamefont{et~al.} (\bibinfo{collaboration}{Particle Data Group}),
  \bibinfo{journal}{PTEP} \textbf{\bibinfo{volume}{2022}},
  \bibinfo{pages}{083C01} (\bibinfo{year}{2022}).

\bibitem[{\citenamefont{Doring et~al.}(2012{\natexlab{a}})\citenamefont{Doring,
  Meissner, Oset, and Rusetsky}}]{Doring:2012eu}
\bibinfo{author}{\bibfnamefont{M.}~\bibnamefont{Doring}},
  \bibinfo{author}{\bibfnamefont{U.~G.} \bibnamefont{Meissner}},
  \bibinfo{author}{\bibfnamefont{E.}~\bibnamefont{Oset}}, \bibnamefont{and}
  \bibinfo{author}{\bibfnamefont{A.}~\bibnamefont{Rusetsky}},
  \bibinfo{journal}{Eur. Phys. J. A} \textbf{\bibinfo{volume}{48}},
  \bibinfo{pages}{114} (\bibinfo{year}{2012}{\natexlab{a}}),
  \eprint{1205.4838}.

\bibitem[{\citenamefont{Gil-Dom\'\i{}nguez and
  Molina}(2024)}]{Gil-Dominguez:2023huq}
\bibinfo{author}{\bibfnamefont{F.}~\bibnamefont{Gil-Dom\'\i{}nguez}}
  \bibnamefont{and} \bibinfo{author}{\bibfnamefont{R.}~\bibnamefont{Molina}},
  \bibinfo{journal}{Phys. Rev. D} \textbf{\bibinfo{volume}{109}},
  \bibinfo{pages}{096002} (\bibinfo{year}{2024}), \eprint{2306.01848}.

\bibitem[{\citenamefont{Martinez~Torres
  et~al.}(2012)\citenamefont{Martinez~Torres, Dai, Koren, Jido, and
  Oset}}]{MartinezTorres:2011pr}
\bibinfo{author}{\bibfnamefont{A.}~\bibnamefont{Martinez~Torres}},
  \bibinfo{author}{\bibfnamefont{L.~R.} \bibnamefont{Dai}},
  \bibinfo{author}{\bibfnamefont{C.}~\bibnamefont{Koren}},
  \bibinfo{author}{\bibfnamefont{D.}~\bibnamefont{Jido}}, \bibnamefont{and}
  \bibinfo{author}{\bibfnamefont{E.}~\bibnamefont{Oset}},
  \bibinfo{journal}{Phys. Rev. D} \textbf{\bibinfo{volume}{85}},
  \bibinfo{pages}{014027} (\bibinfo{year}{2012}), \eprint{1109.0396}.

\bibitem[{\citenamefont{Oller}(2020)}]{Oller:2019opk}
\bibinfo{author}{\bibfnamefont{J.~A.} \bibnamefont{Oller}},
  \bibinfo{journal}{Prog. Part. Nucl. Phys.} \textbf{\bibinfo{volume}{110}},
  \bibinfo{pages}{103728} (\bibinfo{year}{2020}), \eprint{1909.00370}.

\bibitem[{\citenamefont{Doring et~al.}(2012{\natexlab{b}})\citenamefont{Doring,
  Meissner, Oset, and Rusetsky}}]{doringframe}
\bibinfo{author}{\bibfnamefont{M.}~\bibnamefont{Doring}},
  \bibinfo{author}{\bibfnamefont{U.~G.} \bibnamefont{Meissner}},
  \bibinfo{author}{\bibfnamefont{E.}~\bibnamefont{Oset}}, \bibnamefont{and}
  \bibinfo{author}{\bibfnamefont{A.}~\bibnamefont{Rusetsky}},
  \bibinfo{journal}{Eur. Phys. J. A} \textbf{\bibinfo{volume}{48}},
  \bibinfo{pages}{114} (\bibinfo{year}{2012}{\natexlab{b}}),
  \eprint{1205.4838}.

\bibitem[{\citenamefont{Chen et~al.}(2021)\citenamefont{Chen, Huang, Liu, and
  Zhu}}]{Chen:2021vhg}
\bibinfo{author}{\bibfnamefont{R.}~\bibnamefont{Chen}},
  \bibinfo{author}{\bibfnamefont{Q.}~\bibnamefont{Huang}},
  \bibinfo{author}{\bibfnamefont{X.}~\bibnamefont{Liu}}, \bibnamefont{and}
  \bibinfo{author}{\bibfnamefont{S.-L.} \bibnamefont{Zhu}},
  \bibinfo{journal}{Phys. Rev. D} \textbf{\bibinfo{volume}{104}},
  \bibinfo{pages}{114042} (\bibinfo{year}{2021}), \eprint{2108.01911}.

\bibitem[{\citenamefont{Ikeda et~al.}(2014)\citenamefont{Ikeda, Charron, Aoki,
  Doi, Hatsuda, Inoue, Ishii, Murano, Nemura, and Sasaki}}]{Ikeda:2013vwa}
\bibinfo{author}{\bibfnamefont{Y.}~\bibnamefont{Ikeda}},
  \bibinfo{author}{\bibfnamefont{B.}~\bibnamefont{Charron}},
  \bibinfo{author}{\bibfnamefont{S.}~\bibnamefont{Aoki}},
  \bibinfo{author}{\bibfnamefont{T.}~\bibnamefont{Doi}},
  \bibinfo{author}{\bibfnamefont{T.}~\bibnamefont{Hatsuda}},
  \bibinfo{author}{\bibfnamefont{T.}~\bibnamefont{Inoue}},
  \bibinfo{author}{\bibfnamefont{N.}~\bibnamefont{Ishii}},
  \bibinfo{author}{\bibfnamefont{K.}~\bibnamefont{Murano}},
  \bibinfo{author}{\bibfnamefont{H.}~\bibnamefont{Nemura}}, \bibnamefont{and}
  \bibinfo{author}{\bibfnamefont{K.}~\bibnamefont{Sasaki}},
  \bibinfo{journal}{Phys. Lett. B} \textbf{\bibinfo{volume}{729}},
  \bibinfo{pages}{85} (\bibinfo{year}{2014}), \eprint{1311.6214}.

\bibitem[{\citenamefont{Vujmilovic et~al.}(2024)\citenamefont{Vujmilovic,
  Collins, Leskovec, Ortiz-Pacheco, Padmanath, and
  Prelovsek}}]{Vujmilovic:2024snz}
\bibinfo{author}{\bibfnamefont{I.}~\bibnamefont{Vujmilovic}},
  \bibinfo{author}{\bibfnamefont{S.}~\bibnamefont{Collins}},
  \bibinfo{author}{\bibfnamefont{L.}~\bibnamefont{Leskovec}},
  \bibinfo{author}{\bibfnamefont{E.}~\bibnamefont{Ortiz-Pacheco}},
  \bibinfo{author}{\bibfnamefont{M.}~\bibnamefont{Padmanath}},
  \bibnamefont{and}
  \bibinfo{author}{\bibfnamefont{S.}~\bibnamefont{Prelovsek}}, in
  \emph{\bibinfo{booktitle}{{41st International Symposium on Lattice Field
  Theory}}} (\bibinfo{year}{2024}), \eprint{2411.08646}.

\bibitem[{\citenamefont{Gil-Dom\'\i{}nguez and
  Molina}(2023)}]{Gil-Dominguez:2023eld}
\bibinfo{author}{\bibfnamefont{F.}~\bibnamefont{Gil-Dom\'\i{}nguez}}
  \bibnamefont{and} \bibinfo{author}{\bibfnamefont{R.}~\bibnamefont{Molina}},
  \bibinfo{journal}{Phys. Lett. B} \textbf{\bibinfo{volume}{843}},
  \bibinfo{pages}{137997} (\bibinfo{year}{2023}), \eprint{2302.12861}.

\bibitem[{\citenamefont{Molina and Ruiz~de Elvira}(2020)}]{Molina:2020qpw}
\bibinfo{author}{\bibfnamefont{R.}~\bibnamefont{Molina}} \bibnamefont{and}
  \bibinfo{author}{\bibfnamefont{J.}~\bibnamefont{Ruiz~de Elvira}},
  \bibinfo{journal}{JHEP} \textbf{\bibinfo{volume}{11}}, \bibinfo{pages}{017}
  (\bibinfo{year}{2020}), \eprint{2005.13584}.

\bibitem[{\citenamefont{Meng et~al.}(2024{\natexlab{b}})\citenamefont{Meng,
  Ortiz-Pacheco, Baru, Epelbaum, Padmanath, and Prelovsek}}]{Meng:2024kkp}
\bibinfo{author}{\bibfnamefont{L.}~\bibnamefont{Meng}},
  \bibinfo{author}{\bibfnamefont{E.}~\bibnamefont{Ortiz-Pacheco}},
  \bibinfo{author}{\bibfnamefont{V.}~\bibnamefont{Baru}},
  \bibinfo{author}{\bibfnamefont{E.}~\bibnamefont{Epelbaum}},
  \bibinfo{author}{\bibfnamefont{M.}~\bibnamefont{Padmanath}},
  \bibnamefont{and} \bibinfo{author}{\bibfnamefont{S.}~\bibnamefont{Prelovsek}}
  (\bibinfo{year}{2024}{\natexlab{b}}), \eprint{2411.06266}.

\bibitem[{\citenamefont{Navarra et~al.}(2002)\citenamefont{Navarra, Nielsen,
  and Bracco}}]{Navarra:2001ju}
\bibinfo{author}{\bibfnamefont{F.~S.} \bibnamefont{Navarra}},
  \bibinfo{author}{\bibfnamefont{M.}~\bibnamefont{Nielsen}}, \bibnamefont{and}
  \bibinfo{author}{\bibfnamefont{M.~E.} \bibnamefont{Bracco}},
  \bibinfo{journal}{Phys. Rev. D} \textbf{\bibinfo{volume}{65}},
  \bibinfo{pages}{037502} (\bibinfo{year}{2002}), \eprint{hep-ph/0109188}.

\end{thebibliography}

\end{document}